\documentclass[lettersize,journal]{IEEEtran}
\usepackage{amsmath,amsfonts}
\usepackage{array}

\usepackage[caption=false, font=footnotesize, labelformat=empty]{subfig}
\usepackage{textcomp}
\usepackage{stfloats}
\usepackage{url}
\usepackage{verbatim}
\usepackage{graphicx}
\usepackage{cite}
\usepackage{algorithm}
\usepackage{algorithmicx}
\usepackage{algpseudocode}
\algnewcommand{\LineComment}[1]{\State \(\triangleright\) #1}

\usepackage{stmaryrd}

\newcolumntype{C}[1]{>{\centering\arraybackslash}m{#1}}
\newcolumntype{R}[1]{>{\raggedright\arraybackslash}m{#1}}
\newcolumntype{L}[1]{>{\raggedleft\arraybackslash}m{#1}}

\makeatletter
\def\bstctlcite{\@ifnextchar[{\@bstctlcite}{\@bstctlcite[@auxout]}}
\def\@bstctlcite[#1]#2{\@bsphack
  \@for\@citeb:=#2\do{%
    \edef\@citeb{\expandafter\@firstofone\@citeb}%
    \if@filesw\immediate\write\csname #1\endcsname{\string\citation{\@citeb}}\fi}%
  \@esphack}
\makeatother

\DeclareMathOperator*{\argmin}{arg\,min}

\usepackage{booktabs}
\usepackage{multirow}

\usepackage[T1]{fontenc}

\usepackage{calc} 

\usepackage[pdfstartview=XYZ,
bookmarks=true,
colorlinks=true,
linkcolor=blue,
urlcolor=blue,
citecolor=blue,
bookmarks=true,
linktocpage=true, 
hyperindex=true
]{hyperref}

\usepackage{orcidlink}

\begin{document}
\bstctlcite{IEEEexample:BSTcontrol} 

\title{Tell-Tale Watermarks for Explanatory Reasoning in Synthetic Media Forensics}

\author{Ching-Chun Chang and Isao Echizen

\thanks{C.-C. Chang and I. Echizen are with the Information and Society Research Division, National Institute of Informatics, Tokyo, Japan. I. Echizen is also with the Graduate School of Information Science and Technology, University of Tokyo, Tokyo, Japan.
}
\thanks{Correspondence: C.-C. Chang (email: ccchang@nii.ac.jp)
}
}

\maketitle

\begin{abstract}
The rise of synthetic media has blurred the boundary between reality and fabrication under the evolving power of artificial intelligence, fueling an infodemic that erodes public trust in cyberspace. For digital imagery, a multitude of editing applications further complicates the forensic analysis, including semantic edits that alter content, photometric adjustments that recalibrate colour characteristics, and geometric projections that reshape viewpoints. Collectively, these transformations manipulate and control perceptual interpretation of digital imagery. This susceptibility calls for forensic enquiry into reconstructing the chain of events, thereby revealing deeper evidential insight into the presence or absence of criminal intent. This study seeks to address an inverse problem of tracing the underlying generation chain that gives rise to the observed synthetic media. A tell-tale watermarking system is developed for explanatory reasoning over the nature and extent of transformations across the lifecycle of synthetic media. Tell-tale watermarks are tailored to different classes of transformations, responding in a manner that is neither strictly robust nor fragile but instead interpretable. These watermarks function as reference clues that evolve under the same transformation dynamics as the carrier media, leaving interpretable traces when subjected to transformations. Explanatory reasoning is then performed to infer the most plausible account across the combinatorial parameter space of composite transformations. Experimental evaluations demonstrate the validity of tell-tale watermarking with respect to fidelity, synchronicity and traceability.
\end{abstract}

\section{Introduction}
\IEEEPARstart{S}{ynthetic} media has come to blur the boundary between real and fabricated information, under the ever-evolving power of artificial intelligence (AI)~\cite{1276112, Nightingale:2017aa, Chesney:2019ab, 10.1145/3371409, 10.1145/3422622, 10.1145/3425780}. When falling into the wrong hands, these technologies are capable of creating persuasive visual, auditory and linguistic content that deceives human perceptual sensories and even manipulates collective memory~\cite{10.1145/2816795.2818056, Brock:2017aa, 10.1145/3197517.3201283, 10.1145/3292039, 10.1007/978-3-030-01261-8_41, 10.1145/3333002, 10.1007/978-3-030-58517-4_42, NEURIPS2022_ec795aea}. Over time, the very foundation of public trust in digital media erodes under the deluge of misinformation that sweeps through cyberspace, culminating in an infodemic~\cite{Del-Vicario:2016aa, Lewandowsky:2017aa, Lazer:2018aa, Vosoughi:2018aa}.

In response, synthetic media forensics has emerged as a branch of science dedicated to the analysis of media authenticity, often in connection with cybercrimes~\cite{10.1145/1113034.1113074, 4806203, 4806202, 10.1145/1978802.1978805, Chen_Xie_Lin_Liu_Wang_2025, 11128946}. Forensic paradigms can be broadly classified as either reactive, addressing problems after they arise, or proactive, anticipating and preventing them. Reactive forensics often relies on perceptual artefacts or statistical anomalies to identify synthetic media~\cite{5487389, 8578214, NEURIPS2019_3e9f0fc9, 8695364, 9786832, pmlr-v202-mitchell23a}. However, its reliability is undermined by the challenge of keeping pace with the growing sophistication of forgeries~\cite{8630787, 8683164, 9010912, 9141516, 9157215, 9578592, 9578910, 9880195, 10205460}. In contrast, proactive forensics takes prior precautions to protect media at the time of creation~\cite{267415}. It is typically associated with digital watermarking, the practice of imperceptibly embedding information into carrier media to serve a range of forensic purposes. The design philosophies underlying digital watermarking are shaped by the application context, which determines the expected response of watermarks to attacks on their carriers~\cite{687830, 771065, 771066, 771072, 771068, 10238689}. At one extreme, robust watermarks are designed to remain resilient against noise and distortion during transmission and dissemination, serving the purpose of source verification~\cite{650120, BARNI1998357, 985560, Luo:2020aa, pmlr-v202-kirchenbauer23a, 10377226, 10.5555/3692070.3693829, Dathathri:2024aa}. At the other extreme, fragile watermarks are engineered to disappear or be destroyed under modifications, serving the purpose of tamper detection~\cite{638587, 723401, 723413, 951543, 10030248, 10.1145/3640466, Zhao:2024aa}.

Beyond basic forensic aims, a deeper quest lies in tracing the generation chain of synthetic media, uncovering the causal link between observable evidence and the latent sequence of operations from which it arose. In the case of visual media, the generation chain may encompass a wide spectrum of transformations, including semantic edits that alter objects within digital imagery, photometric adjustments that recalibrate colour characteristics, and geometric projections that reshape viewpoints. Such content synthesis and image enhancement processes can collectively control the perceptual interpretation of images. Establishing traceability allows forensic investigation to uncover deeper evidential insight into the presence or absence of criminal intent~\cite{9107445, Jin:2022ab, Breitinger:2025aa}.

This task is inherently an inverse problem, where consequences are observable but causes are to be inferred. In classical logic, deduction infers the consequence from a known rule and an observed cause, whereas induction infers the rule from observed causes and their corresponding consequences. In contrast, the reasoning applied in this context is abduction, which infers the most plausible cause from an observed consequence along with a known rule~\cite{Niiniluoto:2011aa}. It is a form of explanatory reasoning, also referred to as inference to the best explanation~\cite{Harman:1965aa}. This concept is eloquently captured in Sir Arthur Conan Doyle’s \emph{A Study in Scarlet}, voiced through Sherlock Holmes:
\begin{quote}
	\emph{Most people, if you describe a train of events to them will tell you what the result will be. There are few people, however, that if you told them a result, would be able to evolve from their own inner consciousness what the steps were that led to that result. This power is what I mean when I talk of reasoning backwards, or analytically.}
\end{quote}
 
 In this study, we develop tell-tale watermarks for explanatory reasoning over the generation chain of synthetic images, encompassing semantic, photometric and geometric transformations. Distinct from robust and fragile watermarks, tell-tale watermarks are designed to respond in a manner that is neither strictly robust nor fragile, but instead interpretable~\cite{771070}. Transformations applied to the carrier media leave interpretable traces within the watermarks, which can subsequently be used to infer the nature and the extent of the transformations. We design tell-tale watermarks tailored to different classes of transformations and construct neural networks for encoding and decoding them into and from the carrier media. Explanatory reasoning is performed via combinatorial optimisation over the parameter space, yielding the most plausible hypothesis that explains the traces in the tell-tale watermarks. The validity of the proposed tell-tale watermarking system is verified through experimental evaluations with respect to fidelity, synchronicity and traceability.
 
The remainder of this paper is organised as follows. Section~\ref{sec:pre} formalises the problem and scope of this study. Section~\ref{sec:method} presents the methodology of tell-tale watermarking and explanatory reasoning. Section~\ref{sec:eval} provides the implementation details and evaluates the system performance in terms of fidelity, synchronicity and traceability. Section~\ref{sec:con} concludes the paper and outlines potential directions for future research.

\begin{figure*}[t!]
    \centering
    \includegraphics[width=1.5\columnwidth]{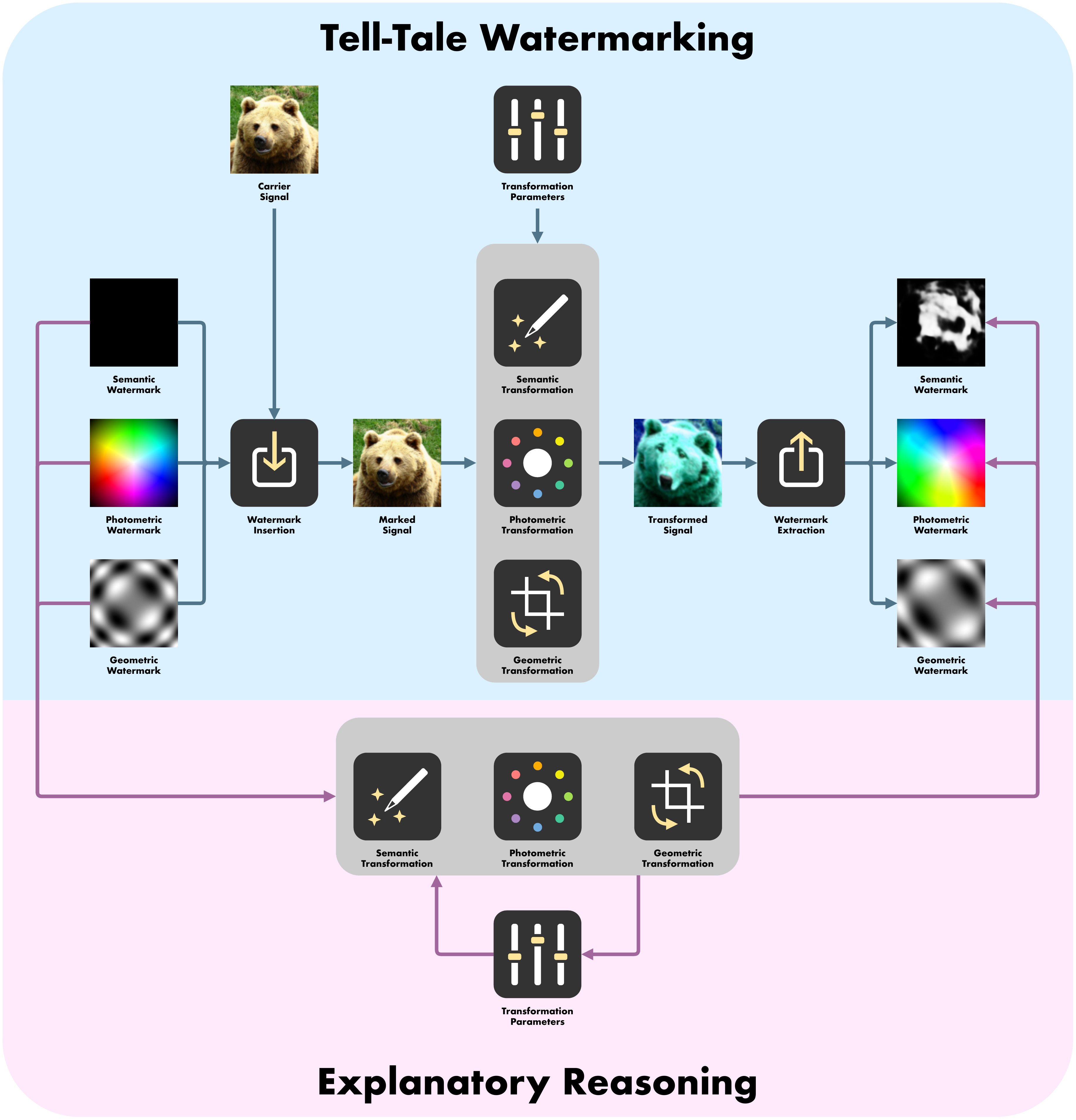}
    \caption{Overview of tell-tale watermarking and explanatory reasoning in response to semantic, photometric and geometric transformations.}
    \label{fig:overview}
\end{figure*}

\section{Preliminaries}\label{sec:pre}

The inverse problem of inferring parameterised transformation sequences from observed synthetic media is formalised with a note on intractability, followed by the scope of semantic, photometric and geometric transformations considered in this study.

\subsection{Problem Formulation}

Let $\mathcal{X}$ denote the space of digital images. We consider a set of parameterised transformation functions defined as
\begin{equation}
	\mathcal{F} = \{f_{\vartheta_i}: \mathcal{X} \rightarrow \mathcal{X} \mid \vartheta_i \in \Theta_i \} ,
\end{equation}
where each function $f_{\vartheta_i}$ represents a distinct type of image transformation and its effect is governed by a parameter $\vartheta_i$ drawn from a specific parameter space $\Theta_i$. In practice, real-world manipulations seldom result from a single isolated transformation; rather, they often arise through the composite layering of multiple transformations, which collectively induce complex and diverse effects on digital images. To model such compositions, we define $\mathcal{F}^*$ as the set of all finite ordered sequences of zero or more transformations from $\mathcal{F}$ (the free monoid over $\mathcal{F}$), given by
\begin{equation}
	\mathcal{F}^* = \bigcup _{n \geq 0} \mathcal{F}^{n} ,
\end{equation}
where $\mathcal{F}^n$ denotes the set of all sequences of $n$ transformations drawn from $\mathcal{F}$ (with repetition allowed) and $\mathcal{F}^0$ contains the identity transformation which leaves images unchanged. Each sequence $\mathbf{f} = (f_1, \dots, f_n) \in \mathcal{F}^*$ represents a composition of $n$ transformation such that
\begin{equation}
\mathbf{f}(\mathbf{x}) = f_n \circ \dots \circ f_1 (\mathbf{x}) .
\end{equation}
Note that the transformations are not necessarily commutative, meaning the order in which they are applied matters and changes the final outcome. Our objective is to reason inversely: given an observed image $\tilde{\mathbf{x}}$, we aim to infer a plausible sequence of parameterised transformations that may have been applied to an original image $\mathbf{x} \in \mathcal{X}$. Formally, we seek a hypothetical composite transformation $\hat{\mathbf{f}} \in \mathcal{F}^*$ that best explains the given observation $\tilde{\mathbf{x}}$, by solving the following inverse problem
\begin{equation}
	\hat{\mathbf{f}} = \argmin_{\mathbf{f} \in \mathcal{F}^*} \mathcal{L}(\mathbf{f}(\mathbf{x}), \tilde{\mathbf{x}}) ,
\end{equation}
where $\mathcal{L}$ denotes a loss function evaluating how well the hypothesised reconstruction approximates the observation. However, the inverse problem defined above is generally ill-posed due to the unavailability of the original source $\mathbf{x}$. This renders the optimisation problem intractable in its current form, as the loss cannot be evaluated without access to the source. To address this limitation, we develop tell-tale watermarks, serving as reference clues that evolve under the same transformation dynamics as the original source. The inverse problem can thus be reformulated as
\begin{equation}
	\hat{\mathbf{f}} = \argmin_{\mathbf{f} \in \mathcal{F}^*} \mathcal{L}(\mathbf{f}(\mathbf{w}), \tilde{\mathbf{w}} + \boldsymbol{\epsilon} ),
\end{equation}
where $\tilde{\mathbf{w}}$ denotes the ideally transformed reference watermarks under the same parameterised transformations applied to the source, and $\boldsymbol{\epsilon}$ accounts for discrepancies arising from imperfect extraction or from incomplete synchronisation with the transformation dynamics. This mirroring behaviour enables explanatory reasoning in the absence of the original source.

\subsection{Transformations}
Digital images, originating either from camera sensors or generative models, are often subjected to a range of transformations as they propagate through cyberspace. Such transformations can be applied manually by humans or autonomously by modern AI agents. Given high-level language instructions, an agent can select and execute transformation procedures via in-built tools or generated codes~\cite{Li:2022aa, 10.1145/3520312.3534862, nijkamp2023codegen}. The scope of this study focuses on three broad classes of image editing: semantic, photometric, and geometric transformations.

\paragraph*{Semantic Transformations} Semantic transformations synthesise content in the designated regions of an image. In this study, we consider semantic editing in the form of inpainting with generative models. Conditioned on a mask and a text prompt, the designated regions are filled with synthesised content.

\paragraph*{Photometric Transformations} Photometric transformations adjust how content is perceived in terms of colour characteristics. They are commonly used for image enhancement, recalibrating colour balances to improve visual interpretation for particular applications. In this study, we consider adjustments of typical components, including brightness, contrast, hue and saturation.

\paragraph*{Geometric Transformations} Geometric transformations change how content is perceived in terms of the viewpoint. They reshape the spatial arrangement of pixels, aligning images to a desired projection surface or shape. In this study, we consider classic transformations from the affine family, including rotation, translation, scaling and shear.

\section{Methodology}\label{sec:method}
The methodology of our study is organised as follows. Watermark creation establishes tell-tale patterns whose responses to transformations are interpretable and measurable. Watermark insertion and extraction are performed by a pair of encoder and decoder, jointly optimised under simulated transformations, to embed watermarks with minimal perceptual distortion of the carrier signal and maximal synchronicity with the applied transformations. Explanatory reasoning interprets the extracted watermarks, inferring the parameters of the applied transformations. An overview of the proposed method is outlined in Figure~\ref{fig:overview}.

\begin{algorithm}[t]
\caption{Watermark Creation}\label{alg:watermark}
\begin{algorithmic}

\LineComment{\textit{initialisation (polar coordinates)}}
\For{each $(x, y)$}
	\State $\bar{x}\gets 2x/\mathrm{width} -1$
	\State $\bar{y}\gets 2y/\mathrm{height} -1$
    \State $\varrho(x,y)\gets \sqrt{\bar{x}^2+\bar{y}^2}$
    \State $\varphi(x,y)\gets \operatorname{atan2}(\bar{y},\bar{x})$
\EndFor
\State

\LineComment{\textit{semantic watermark (blank canvas)}}
\For{each $(x, y)$}
	\State $\mathbf{w}_{\mathrm{sem}}(x,y)\gets 0$
\EndFor
\State

\LineComment{\textit{photometric watermark (colour wheel)}}
\For{each $(x, y)$}
	\State $h\gets ((\varphi(x,y)+\Delta\varphi)/2\pi ) \bmod 1$
    \State $l\gets 1 - \varrho(x,y)/\sqrt{2}$
    \State $s\gets 1$
    \State $\mathbf{w}_{\mathrm{pho}}(x,y) = (r,g,b) \mapsfrom (h,l,s)$
\EndFor
\State

\LineComment{\textit{geometric watermark (wave interference)}}
\For{each $(x, y)$}
    \State $\xi(x,y)\gets \xi_{\min}+(\xi_{\max}-\xi_{\min})\cdot \varrho(x,y)/\sqrt{2}$
    \State $\omega(x,y)\gets 2\pi\cdot \xi(x,y)$
    \State $\Psi(x,y)\gets \sin\!\big(\omega(x,y)\cdot x\big)\cdot \sin\!\big(\omega(x,y)\cdot y\big)$
    \State $\mathbf{w}_{\mathrm{geo}}(x,y)\gets (1+\Psi(x,y))/2$
\EndFor

\end{algorithmic}
\end{algorithm}

\subsection{Watermark Creation}
The creation of tell-tale watermarks requires tailored reference patterns whose responses to transformations are interpretable and measurable. Their design varies by transformation type: semantic watermarks as a blank canvas sensitive to content edits, photometric watermarks as a colour wheel reflecting colour adjustments, and geometric watermarks as wave interference pattern exposing affine distortions, as formulated in Algorithm~\ref{alg:watermark}.

\paragraph*{Semantic Watermark}
A semantic watermark is constructed as a blank canvas. Its simplicity allows any non-trivial semantic alteration to leave a discernible trace upon it. Let $\mathrm{height}$ and $\mathrm{width}$ denote the dimensions of the watermark and $(x, y)$ denote pixel coordinates, where $x\in\{1,\dots, \mathrm{height}\}$ and $y\in\{1,\dots, \mathrm{width}\}$. The semantic watermark is specified by
\begin{equation}
	\mathbf{w}_{\mathrm{sem}} (x, y) = 0 \quad\forall (x, y) .
\end{equation}

\paragraph*{Photometric Watermark}
A photometric watermark is constructed as a colour wheel whose behaviour under photometric transformations (brightness, contrast, saturation, hue) is predictable and straightforward to measure. In particular, it is designed as a circular spectrum in HLS colour space such that hue varies azimuthally with the angle around the centre, lightness decays radially with distance from the centre, and saturation remains spatially constant. Let $(\bar{x}, \bar{y})$ denote the normalised pixel coordinates, centred at $(0,0)$ and ranging within $[-1,1]$, defined by
\begin{equation}
\bar{x} = \frac{2x}{\mathrm{width}} - 1, 
\qquad 
\bar{y} = \frac{2y}{\mathrm{height}} - 1 .
\end{equation}
The corresponding polar coordinates are derived as
\begin{equation}
\varrho(x, y) = \sqrt{\bar{x}^2 + \bar{y}^2}, 
\qquad 
\varphi(x, y) = \operatorname{atan2}(\bar{y}, \bar{x}) .
\end{equation}
where $\varrho$ measures the distance from the centre and $\varphi$ measures the angle around the centre. The hue channel is defined as
\begin{equation}
h (x, y) = \left( \frac{\varphi(x, y) + \Delta\varphi}{2\pi} \right) \bmod 1,
\end{equation}
where $\Delta\varphi \in \mathbb{R}$ is a user-defined angular shift specifying the default hue rotation. The division by $2\pi$ converts an angle measured in radians into a fraction of a full cycle, whereas the modulo operation restricts the hue value to the interval $[0, 1]$. The lightness channel is defined as
\begin{equation}
l (x, y) = 1- \frac{\varrho(x, y)}{\sqrt{2}} .
\end{equation}
The division by $\sqrt{2}$ scales $\varrho$ from $[0, \sqrt{2}]$ to $[0, 1]$, so that the center ($\varrho=0$) attains maximum brightness ($l=1$) and the corners ($\varrho=\sqrt{2}$) approach minimum brightness ($l=0$). The saturation channel is set as a constant value $s (\bar{x}, \bar{y}) = 1$. Finally, the photometric watermark is obtained by combining the hue, lightness and saturation channels and converting them into RGB colour space
\begin{equation}
\mathbf{w}_{\mathrm{pho}}(x, y) = (r, g, b) \mapsfrom (h, l, s).
\end{equation}

\paragraph*{Geometric Watermark}
A geometric watermark is constructed as a wave interference pattern whose behaviour under geometric transformations (rotation, translation, scaling, shear) is regular and quantifiable. In particular, it is formed by the interference of two sinusoidal waves along the horizontal and vertical axes, with oscillatory frequency increasing radially from the centre. Let $\varrho(x, y)$ denote the radial distance from the centre, obtained by converting the normalised coordinates into polar form. The spatial frequency is specified as
\begin{equation}
	\xi(x, y) = \xi_{\min} + (\xi_{\max} - \xi_{\min}) \cdot \frac{\varrho(x, y)}{\sqrt{2}},
\end{equation}
where $\xi_{\min}$ and $\xi_{\max}$ are user-defined minimum and maximum frequencies, respectively. The division by $\sqrt{2}$ scales $\varrho$ from $[0, \sqrt{2}]$ to $[0, 1]$, thereby assigning the minimum frequency to the centre ($\varrho=0$) and the maximum frequency to the corners ($\varrho=\sqrt{2}$). The angular frequency is derived by multiplying the spatial frequency with a full cycle ($2\pi$ radians)
\begin{equation}
	\omega(x, y) = 2\pi \xi(x, y)
\end{equation}
The wave interference pattern is then generated by the product of two orthogonal sinusoidal gratings
\begin{equation}
	\Psi(x, y) = \sin(\omega (x, y) x) \cdot \sin(\omega (x, y) y) .
\end{equation}
Finally, the geometric watermark is obtained by normalising the wave interference pattern from $[-1, 1]$ to $[0, 1]$
\begin{equation}
	\mathbf{w}_{\mathrm{geo}}(x, y) = \frac{ 1 + \Psi(x, y) }{2} .
\end{equation}

\begin{algorithm}[t]
\caption{Watermark Insertion \& Extraction}\label{alg:insert-extract}
\begin{algorithmic}

\LineComment{\textit{Initialisation (learning \& inference)}}
\State (learning) for each $\mathbf{x} \in \mathcal{X}_{\mathrm{learn}}$ do the following
\State (inference) for each $\mathbf{x} \in \mathcal{X}_{\mathrm{infer}}$ do the following
\State

\LineComment{\textit{watermark insertion (learning \& inference)}}
\State $\mathbf{x}_{\mathbf{w}} \gets \mathcal{E}(\mathbf{x}, \mathbf{w})$
\State where $\mathbf{w} = [\mathbf{w}_{\mathrm{sem}}, \mathbf{w}_{\mathrm{pho}}, \mathbf{w}_{\mathrm{geo}}]$
\State

\LineComment{\textit{random transformations (learning \& inference)}}
\State sample $(\mathbf{f}_{\mathrm{sem}},\mathbf{f}_{\mathrm{pho}},\mathbf{f}_{\mathrm{geo}})$ randomly
\State (with random selection, intra-class ordering and parameters)
\State $\tilde{\mathbf{x}}_{\mathbf{w}} \gets \mathbf{f}_{\mathrm{geo}}( \mathbf{f}_{\mathrm{pho}}( \mathbf{f}_{\mathrm{sem}}( \mathbf{x}_{\mathbf{w}} )))$
\State (learning) $\mathbf{f}_{\mathrm{sem}}(\mathbf{x}_{\mathbf{w}} ) 
= (1-\mathbf{m})\cdot \mathbf{x}_{\mathbf{w}} + \mathbf{m}\cdot \tilde{\mathbf{x}}_{\mathrm{gen}}$
\State (inference) $\mathbf{f}_{\mathrm{sem}}(\mathbf{x}_{\mathbf{w}} ) 
= (1-\mathbf{m})\cdot \mathbf{x}_{\mathbf{w}} + \mathbf{m}\cdot \mathbf{x}_{\mathrm{gen}}$
\State

\LineComment{\textit{watermark extraction (learning \& inference)}}
\State $\hat{\mathbf{w}} \gets \mathcal{D}(\tilde{\mathbf{x}}_{\mathbf{w}})$
\State where $\hat{\mathbf{w}} = [\hat{\mathbf{w}}_{\mathrm{sem}}, \hat{\mathbf{w}}_{\mathrm{pho}}, \hat{\mathbf{w}}_{\mathrm{geo}}]$
\State

\LineComment{\textit{ground-truth watermarks (learning)}}
\State $\tilde{\mathbf{w}}_{\mathrm{sem}} \gets \mathbf{f}_{\mathrm{geo}}( \mathbf{f}_{\mathrm{sem}}( \mathbf{w}_{\mathrm{sem}} ) ) = \mathbf{f}_{\mathrm{geo}}( \mathbf{m} )$
\State $\tilde{\mathbf{w}}_{\mathrm{pho}} \gets \mathbf{f}_{\mathrm{geo}}( \mathbf{f}_{\mathrm{pho}}( \mathbf{w}_{\mathrm{pho}} ) )$
\State $\tilde{\mathbf{w}}_{\mathrm{geo}} \gets \mathbf{f}_{\mathrm{geo}}( \mathbf{w}_{\mathrm{geo}} )$
\State $\tilde{\mathbf{w}} \gets [ \tilde{\mathbf{w}}_{\mathrm{sem}}, \tilde{\mathbf{w}}_{\mathrm{pho}}, \tilde{\mathbf{w}}_{\mathrm{geo}} ]$
\State

\LineComment{\textit{loss functions (learning)}}
\State $\mathcal{L}_{\mathcal{E}}(\mathbf{x}_{\mathbf{w}},\mathbf{x}) \gets \|\mathbf{x}_{\mathbf{w}}-\mathbf{x}\|_{1} + \|\mathbf{x}_{\mathbf{w}}-\mathbf{x}\|_{\infty}$
\State $\mathcal{L}_{\mathcal{D}}(\hat{\mathbf{w}},\tilde{\mathbf{w}}) \gets \sum_{\mathrm{xform}} \| \hat{\mathbf{w}}_{\mathrm{xform}}-\tilde{\mathbf{w}}_{\mathrm{xform}} \|_1 $
\State where $\mathrm{xform} \in \{\mathrm{sem}, \mathrm{pho}, \mathrm{geo}\}$
\State $\mathcal{L} \gets \mathcal{L}_{\mathcal{E}} + \mathcal{L}_{\mathcal{D}}$
\State

\LineComment{\textit{joint optimisation (learning)}}
\State update $\mathcal{E}$ via backpropagation of gradient $\nabla_{\mathcal{E}}\mathcal{L}$
\State update $\mathcal{D}$ via backpropagation of gradient $\nabla_{\mathcal{D}}\mathcal{L}$
\end{algorithmic}
\end{algorithm}

\subsection{Watermark Insertion \& Extraction}
Watermark insertion and extraction are performed by two respective neural networks trained jointly, with simulated transformations applied in between, as presented in Algorithm~\ref{alg:insert-extract}. Given a carrier image $\mathbf{x}$ and a collection of tell-tale watermarks $\mathbf{w} = (\mathbf{w}_{\mathrm{sem}}, \mathbf{w}_{\mathrm{pho}}, \mathbf{w}_{\mathrm{geo}})$, the encoder $\mathcal{E}$ embeds $\mathbf{w}$ into $\mathbf{x}$, producing a marked image 
\begin{equation}
  \mathbf{x}_{\mathbf{w}} = \mathcal{E}(\mathbf{x}, \mathbf{w}).
\end{equation}
A transformation sequence $\mathbf{f} \in \mathcal{F}^*$ is then randomly instanciated and applied to the marked images, yielding a transformed image
\begin{equation}
  \tilde{\mathbf{x}}_{\mathbf{w}} = \mathbf{f}(\mathbf{x}_{\mathbf{w}}).
\end{equation}
We restrict attention to an ordered transformation chain that mirrors common practice in generative pipelines, defined as
\begin{equation}
  \mathbf{f} = \mathbf{f}_{\mathrm{geo}} ( \mathbf{f}_{\mathrm{pho}} (\mathbf{f}_{\mathrm{sem}}(\mathbf{x}_{\mathbf{w}}) ) ) ,
\end{equation}
where semantic editing $\mathbf{f}_{\mathrm{sem}}$ is performed first to modify content, photometric adjustment $\mathbf{f}_{\mathrm{pho}}$ follows to harmonise colour and suppress artefacts, and geometric projection $\mathbf{f}_{\mathrm{geo}}$ is performed last to control scale, orientation and viewpoint. This chain reduces the combinatorial complexity of parameter reasoning, albeit at the cost of generality. Nevertheless, full randomness is preserved within each class of transformations: brightness, contrast, hue and saturation are applied in random order for photometric transformations, while rotation, translation, scaling and shear are likewise applied in random order for geometric transformations. For semantic editing, we formulate it as inpainting based on generative models tasked with filling the designated missing regions, conditioned on a mask $\mathbf{m}$ and a text prompt. Direct semantic editing with generative models, however, is computationally intensive in the learning phase, as joint optimisation of encoder and decoder requires repeated iterations. To achieve efficiency, we replace it with a surrogate scheme that masks random regions and fills them with simulated alternatives, expressed as
\begin{equation}
\begin{split}
	\mathbf{f}_{\mathrm{sem}} ( \mathbf{x}_{\mathbf{w}} \mid \mathbf{m}, \mathrm{prompt} ) &= (1-\mathbf{m})\cdot \mathbf{x}_{\mathbf{w}} + \mathbf{m}\cdot \mathbf{x}_{\mathrm{gen}}, \\
	&\approx (1-\mathbf{m})\cdot \mathbf{x}_{\mathbf{w}} + \mathbf{m}\cdot \tilde{\mathbf{x}}_{\mathrm{gen}},
\end{split}
\end{equation}
where $\mathbf{x}{\mathrm{gen}}$ is generated content conditioned on a mask $\mathbf{m}$ and a text prompt, and $\tilde{\mathbf{x}}_{\mathrm{gen}}$ denotes its simulated counterpart. In practice, we simulate the generated content by applying random photogeometric transformations to the image $\mathbf{x}_{\mathbf{w}}$, such that the masked regions are filled with transformed variants of the image itself. From the transformed image $\tilde{\mathbf{x}}_{\mathbf{w}}$, the decoder $\mathcal{D}$ attempts to extract the watermark
\begin{equation}
  \hat{\mathbf{w}} = \mathcal{D}(\mathbf{x}_{\mathbf{w}}).
\end{equation}
To preserve perceptual fidelity after watermark insertion, we define an encoding loss $\mathcal{L}_{\mathcal{E}}(\mathbf{x}_{\mathbf{w}}, \mathbf{x})$ that penalises deviations between the carrier image and its marked counterpart, given by
\begin{equation}
  \mathcal{L}_{\mathcal{E}} (\mathbf{x}_\mathbf{w}, \mathbf{x}) = \| \mathbf{x}_\mathbf{w} - \mathbf{x} \|_1 + \| \mathbf{x}_\mathbf{w} - \mathbf{x} \|_{\infty} .
\end{equation}
where $\ell_1$ norm captures global deviations and the $\ell{\infty}$ norm suppresses localised residual spikes. To support synchronicity with the applied transformations, we define a decoding loss that measures discrepancies between the extracted watermark and its ground truth counterpart, given by
\begin{equation}
  \mathcal{L}_{\mathcal{D}} (\hat{\mathbf{w}}, \tilde{\mathbf{w}}) = \sum_{\mathrm{xform}} \| \hat{\mathbf{w}}_{\mathrm{xform}} - \tilde{\mathbf{w}}_{\mathrm{xform}} \|_1 ,
\end{equation}
where $\mathrm{xform} \in \{\mathrm{sem}, \mathrm{pho}, \mathrm{geo}\}$ indicates the transformation types, $\hat{\mathbf{w}}_{\mathrm{xform}}$ is the extracted watermark and $\tilde{\mathbf{w}}_{\mathrm{xform}}$ denotes the target ground truth. We obtain the ground truth by transforming the reference watermarks with identical parameters to those applied to the carrier image, namely
\begin{equation}
\begin{split}
	\tilde{\mathbf{w}}_{\mathrm{sem}} &= \mathbf{f}_{\mathrm{geo}} ( \mathbf{f}_{\mathrm{sem}} (\mathbf{w}_{\mathrm{sem}}) ) , \\
	\tilde{\mathbf{w}}_{\mathrm{pho}} &= \mathbf{f}_{\mathrm{geo}} ( \mathbf{f}_{\mathrm{pho}} (\mathbf{w}_{\mathrm{pho}}) ) , \\
	\tilde{\mathbf{w}}_{\mathrm{geo}} &= \mathbf{f}_{\mathrm{geo}} ( \mathbf{w}_{\mathrm{geo}} ) .
\end{split}
\end{equation}
For the semantic transformation, the filled regions are flagged with $1$s against the reference watermark initialised with $0$s, and the resulting expression reduces to the editing mask
\begin{equation}
\begin{split}
		\mathbf{f}_{\mathrm{sem}} (\mathbf{w}_{\mathrm{sem}} )
		&= (1-\mathbf{m})\cdot \mathbf{w}_{\mathrm{sem}} + \mathbf{m}\cdot \mathbf{1} \\
		&= (1-\mathbf{m})\cdot \mathbf{0} + \mathbf{m}\cdot \mathbf{1} = \mathbf{m}.
\end{split}
\end{equation}
The overall watermarking loss $\mathcal{L}_{\mathcal{W}} = \mathcal{L}_{\mathcal{E}} + \mathcal{L}_{\mathcal{D}}$ is backpropagated to update both $\mathcal{E}$ and $\mathcal{D}$ jointly. This joint formulation encourages the inserted watermark to be minimally disruptive to the carrier image yet maximally reflective of the applied transformations.

\begin{algorithm}[t]
\caption{Explanatory Reasoning}\label{alg:reasoning}
\begin{algorithmic}

\LineComment{\textit{geometric reasoning}}
\State $\mathbf{w}_{\mathrm{geo}}$ and $\hat{\mathbf{w}}_{\mathrm{geo}}$ are given
\State $\boldsymbol{\vartheta}_{\mathrm{geo}} \gets \texttt{default}$
\State $(\hat{\varpi}_{\mathrm{geo}},\hat{\boldsymbol{\vartheta}}_{\mathrm{geo}})\gets \texttt{none}; \;\mathrm{iter }\gets 0; \; L_{\min}\gets \infty$

\For{each $\varpi_{\mathrm{geo}} \in \operatorname{Sym}(\{ \mathrm{ro},\mathrm{tr}, \mathrm{sc} , \mathrm{sh} \})$}
\While{$\mathrm{iter} < \mathrm{iter}_{\max}$}
	\State $\tilde{\mathbf{w}}_{\mathrm{geo}} \gets \mathbf{f}_{\mathrm{geo}} (\mathbf{w}_{\mathrm{geo}} | \varpi_{\mathrm{geo}}, \boldsymbol{\vartheta}_{\mathrm{geo}}) $
	\State $\mathcal{L}_{\mathrm{geo}} \gets \| \tilde{\mathbf{w}}_{\mathrm{geo}} - \hat{\mathbf{w}}_{\mathrm{geo}} \|_1$
	\State update $(\varpi_{\mathrm{geo}}, \boldsymbol{\vartheta}_{\mathrm{geo}})$ via backpropagation 
	\If {$\mathcal{L}_{\mathrm{geo}} < \mathcal{L}_{\min}$}
	\State $(\hat{\varpi}_{\mathrm{geo}}, \hat{\boldsymbol{\vartheta}}_{\mathrm{geo}}) \gets (\varpi_{\mathrm{geo}}, \boldsymbol{\vartheta}_{\mathrm{geo}})$
	\EndIf
	\State $\mathrm{iter} \gets \mathrm{iter} + 1$
\EndWhile
\EndFor
\State\Return $(\hat{\varpi}_{\mathrm{geo}},\hat{\boldsymbol{\vartheta}}_{\mathrm{geo}})$
\State

\LineComment{\textit{photometric reasoning}}
\State $\mathbf{w}_{\mathrm{pho}}$ and $\hat{\mathbf{w}}_{\mathrm{pho}}$ are given
\State $\boldsymbol{\vartheta}_{\mathrm{pho}} \gets \texttt{default}$
\State $(\hat{\varpi}_{\mathrm{pho}},\hat{\boldsymbol{\vartheta}}_{\mathrm{pho}})\gets \texttt{none}; \;\mathrm{iter }\gets 0; \; L_{\min}\gets \infty$

\For{each $\varpi_{\mathrm{pho}} \in \operatorname{Sym}(\{ \mathrm{b},\mathrm{c}, \mathrm{h} , \mathrm{s} \})$}
\While{$\mathrm{iter} < \mathrm{iter}_{\max}$}
	\State $\tilde{\mathbf{w}}_{\mathrm{pho}} \gets \mathbf{f}_{\mathrm{geo}}( \mathbf{f}_{\mathrm{pho}} (\mathbf{w}_{\mathrm{geo}} | \varpi_{\mathrm{pho}}, \boldsymbol{\vartheta}_{\mathrm{pho}}) | \hat{\varpi}_{\mathrm{geo}}, \hat{\boldsymbol{\vartheta}}_{\mathrm{geo}})$
	\State $\mathcal{L}_{\mathrm{pho}} \gets \| \tilde{\mathbf{w}}_{\mathrm{pho}} - \hat{\mathbf{w}}_{\mathrm{pho}} \|_1$
	\State update $(\varpi_{\mathrm{pho}}, \boldsymbol{\vartheta}_{\mathrm{pho}})$ via backpropagation 
	\If {$\mathcal{L}_{\mathrm{pho}} < \mathcal{L}_{\min}$}
	\State $(\hat{\varpi}_{\mathrm{pho}}, \hat{\boldsymbol{\vartheta}}_{\mathrm{pho}}) \gets (\varpi_{\mathrm{pho}}, \boldsymbol{\vartheta}_{\mathrm{pho}})$
	\EndIf
	\State $\mathrm{iter} \gets \mathrm{iter} + 1$
\EndWhile
\EndFor
\State\Return $(\hat{\varpi}_{\mathrm{pho}},\hat{\boldsymbol{\vartheta}}_{\mathrm{pho}})$
\State

\LineComment{\textit{semantic reasoning}}
\State $\mathbf{w}_{\mathrm{sem}}$ and $\hat{\mathbf{w}}_{\mathrm{sem}}$ are given
\State $\hat{\boldsymbol{\vartheta}}_{\mathrm{sem}} \gets \hat{\mathbf{w}}_{\mathrm{sem}}$
\State\Return $\hat{\boldsymbol{\vartheta}}_{\mathrm{sem}}$
\end{algorithmic}
\end{algorithm}

\subsection{Explanatory Reasoning}
Explanatory reasoning concerns the estimation of transformation parameters based on the tell-tale traces left within the extracted watermarks. The procedures and objectives vary by transformation type: geometric reasoning determines affine distortions, photometric reasoning infers colour adjustments, and semantic reasoning identifies edited regions, as detailed in Algorithm~\ref{alg:reasoning}.

\paragraph*{Geometric Reasoning}
Geometric reasoning aims to determine the ordering and parameters for which the composite geometric transformation $\mathbf{f}_{\mathrm{geo}}$ maps the reference watermark $\mathbf{w}_{\mathrm{geo}}$ as closely as possible to the extracted watermark $\hat{\mathbf{w}}_{\mathrm{geo}}$. Let the hypothesised transformed watermark be denoted by
\begin{equation}
\tilde{\mathbf{w}}_{\mathrm{geo}} = \mathbf{f}_{\mathrm{geo}} (\mathbf{w}_{\mathrm{geo}} \mid \varpi_{\mathrm{geo}}, \boldsymbol{\vartheta}_{\mathrm{geo}})
\end{equation}
where $\varpi_{\mathrm{geo}}$ specifies the permutation ordering of rotation, translation, scaling and shear, and $\boldsymbol{\vartheta}_{\mathrm{geo}}$ contains their respective parameters, defined as
\begin{equation}
\begin{split}
	\varpi_{\mathrm{geo}} &\in \operatorname{Sym}(\{ \mathrm{ro}, \mathrm{tr}, \mathrm{sc}, \mathrm{sh}\} ) ,\\
	\boldsymbol{\vartheta}_{\mathrm{geo}} &= [\vartheta_{\mathrm{ro}}, \vartheta_{\mathrm{tr}}^{x}, \vartheta_{\mathrm{tr}}^{y}, \vartheta_{\mathrm{sc}}, \vartheta_{\mathrm{sh}}^{x}, \vartheta_{\mathrm{sh}}^{y}] .
\end{split}
\end{equation}
The reasoning task is therefore to jointly resolve both the ordering and the magnitudes of the underlying geometric distortions, expressed as a nested optimisation
\begin{equation}
\hat{\varpi}_{\mathrm{geo}}, \hat{\boldsymbol{\vartheta}}_{\mathrm{geo}} = \operatorname{arg\,} \min_{\varpi_{\mathrm{geo}}} \min_{\boldsymbol{\vartheta}_{\mathrm{geo}}}
\| \tilde{\mathbf{w}}_{\mathrm{geo}} - \hat{\mathbf{w}}_{\mathrm{geo}} \|_1 .
\end{equation}
Each affine transformation has an inverse affine matrix, allowing the transformed coordinates to be computed directly as
\begin{equation}
\begin{bmatrix}
u \\ v \\ 1
\end{bmatrix} = 
\mathbf{M}
\begin{bmatrix}
x \\ y \\ 1
\end{bmatrix}
\end{equation}
with the rotation, translation, scaling and shear matrices given respectively by
\begin{equation}
\begin{split}
	\mathbf{M}_{\mathrm{ro}} &=
	\begin{bmatrix}
	\cos \vartheta_{\mathrm{ro}} & \sin \vartheta_{\mathrm{ro}} & 0 \\
	-\sin \vartheta_{\mathrm{ro}} & \cos \vartheta_{\mathrm{ro}} & 0 \\
	0 & 0 & 1 \\
	\end{bmatrix} ,\\
	\mathbf{M}_{\mathrm{tr}} &=
	\begin{bmatrix}
	1 & 0 & -\vartheta_{\mathrm{tr}}^{x} \\
	0 & 1 & -\vartheta_{\mathrm{tr}}^{y} \\
	0 & 0 & 1 \\
	\end{bmatrix} ,\\
	\mathbf{M}_{\mathrm{sc}} &=
	\begin{bmatrix}
	1/\vartheta_{\mathrm{sc}} & 0 & 0 \\
	0 & 1/\vartheta_{\mathrm{sc}} & 0 \\
	0 & 0 & 1 \\
	\end{bmatrix} ,\\
	\mathbf{M}_{\mathrm{sh}} &=
	\frac{1}{1 - \vartheta_{\mathrm{sh}}^{x} \vartheta_{\mathrm{sh}}^{y}}
	\begin{bmatrix}
	1 & - \vartheta_{\mathrm{sh}}^{x} & 0 \\
	-\vartheta_{\mathrm{sh}}^{y} & 1 & 0 \\
	0 & 0 & 1 \\
	\end{bmatrix} .
\end{split}
\end{equation}
As $(u, v)$ is not generally integer-valued, bilinear interpolation is applied and the transformed watermark is given by
\begin{equation}
	\tilde{\mathbf{w}}_{\mathrm{geo}}(x, y) = \sum_{ \mathrm{cor} } \beta_{\mathrm{cor}} \cdot \mathbf{w}_{\mathrm{geo}}(u_{\mathrm{cor}}, v_{\mathrm{cor}}) ,
\end{equation}
where $\mathrm{cor} \in \{\mathtt{nw}, \mathtt{ne}, \mathtt{sw}, \mathtt{se}\}$ is the set of four corners (northwest, northeast, southwest and southeast), $(u_{\mathrm{cor}}, v_{\mathrm{cor}})$ are integer-valued corner coordinates
\begin{equation}
	\begin{split}
	(u_{\mathtt{nw}}, v_{\mathtt{nw}}) &= (\lfloor{u}\rfloor, \lfloor{v}\rfloor) ,\\
	(u_{\mathtt{ne}}, v_{\mathtt{ne}}) &= (\lceil{u}\rceil, \lfloor{v}\rfloor) ,\\
	(u_{\mathtt{sw}}, v_{\mathtt{sw}}) &= (\lfloor{u}\rfloor, \lceil{v}\rceil) ,\\
	(u_{\mathtt{se}}, v_{\mathtt{se}}) &= (\lceil{u}\rceil, \lceil{v}\rceil) ,
	\end{split}
\end{equation}
and $\beta_{\mathrm{cor}}$ are the corresponding bilinear interpolation weights
\begin{equation}
	\begin{split}
	\beta_{\mathtt{nw}} &= (u_{\mathtt{se}} - u) (v_{\mathtt{se}} - v) , \\
	\beta_{\mathtt{ne}} &= (u - u_{\mathtt{sw}}) (v_{\mathtt{sw}} - v) ,\\
	\beta_{\mathtt{sw}} &= (u_{\mathtt{ne}} - u) (v - v_{\mathtt{ne}}) ,\\
	\beta_{\mathtt{se}} &= (u - u_{\mathtt{nw}}) (v - v_{\mathtt{nw}}) .\\
	\end{split} .
\end{equation}

\begin{figure*}[t!]
    \centering
    \includegraphics[width=1.9\columnwidth]{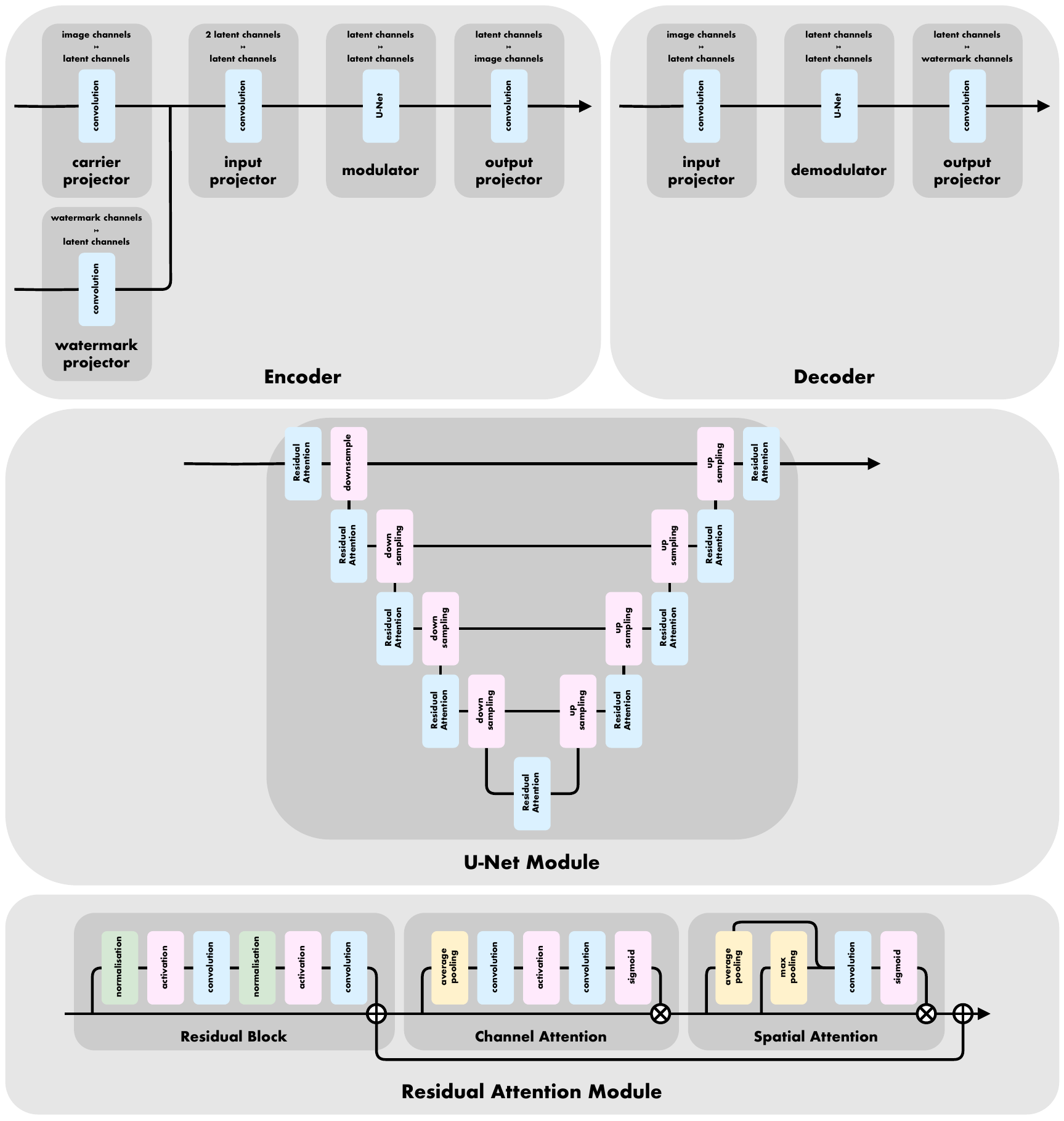}
    \caption{Architecture of the encoder and decoder neural networks, with implementation details for the U-Net module and the residual attention module.}
    \label{fig:neural_networks}
\end{figure*}

\begin{figure*}[t!]
    \centering
    \includegraphics[width=1.999\columnwidth]{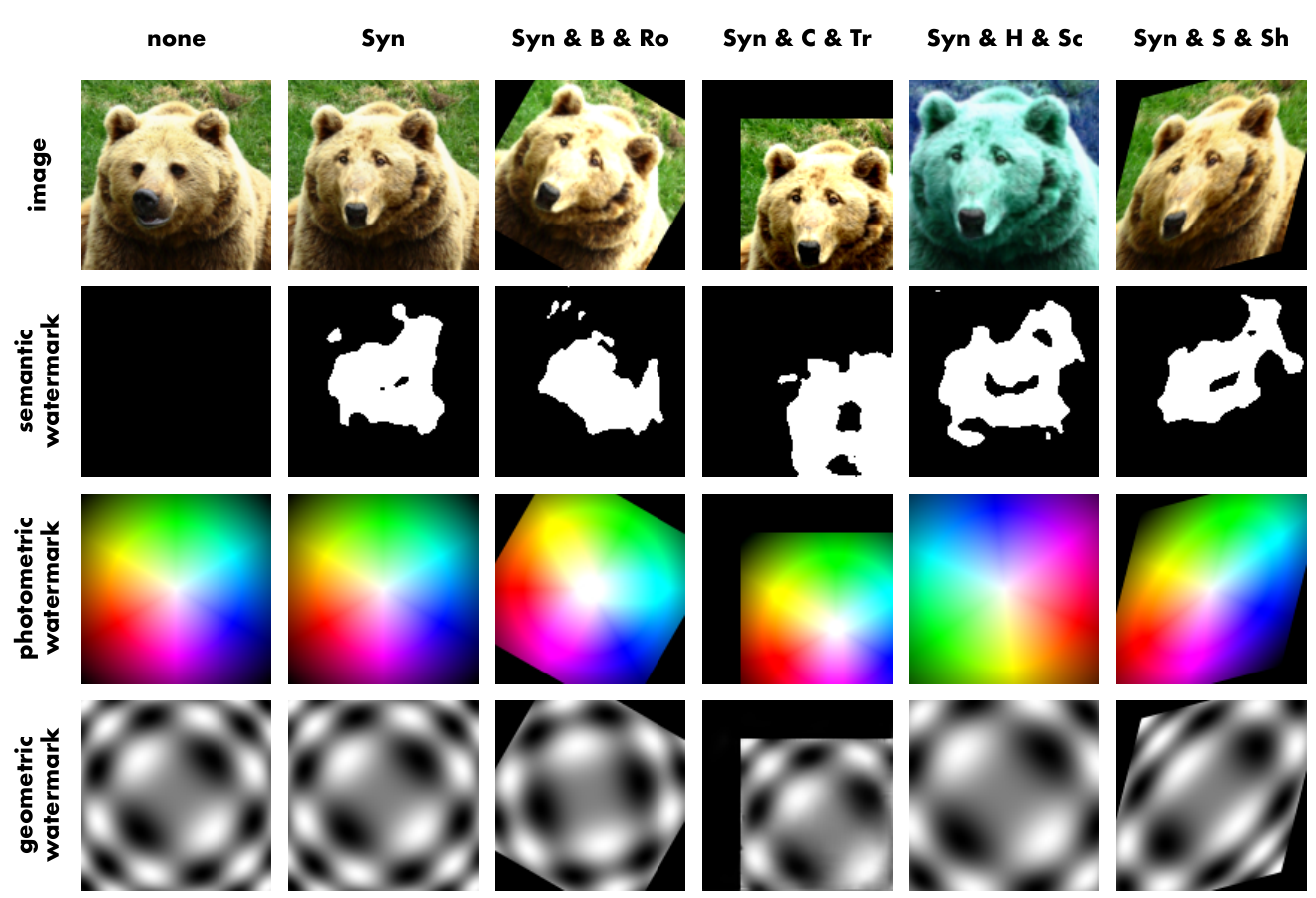}
    \caption{Demonstration of tell-tale watermarks under different transformation chains.}
    \label{fig:demo}
\end{figure*}

\begin{figure}[t!]
    \centering
    \subfloat[$\ell_1$ norm]{
    	\includegraphics[width=0.99\columnwidth]{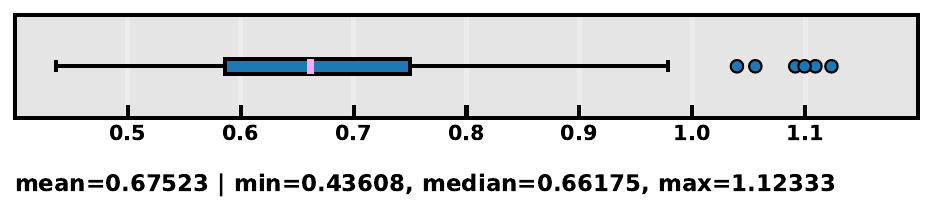}
    }
    \\
    \subfloat[$\ell_{\infty}$ norm]{
        \includegraphics[width=0.99\columnwidth]{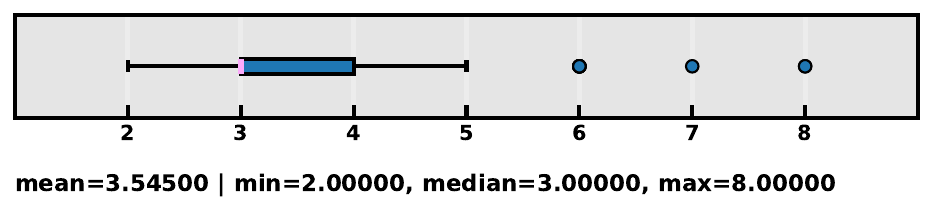}
    }
    \\
    \subfloat[PSNR]{
        \includegraphics[width=0.99\columnwidth]{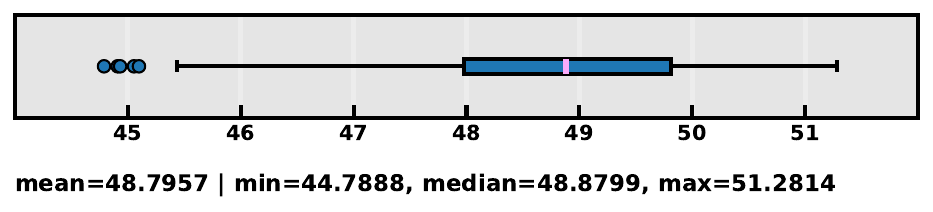}
    }
    \\
    \subfloat[SSIM]{
        \includegraphics[width=0.99\columnwidth]{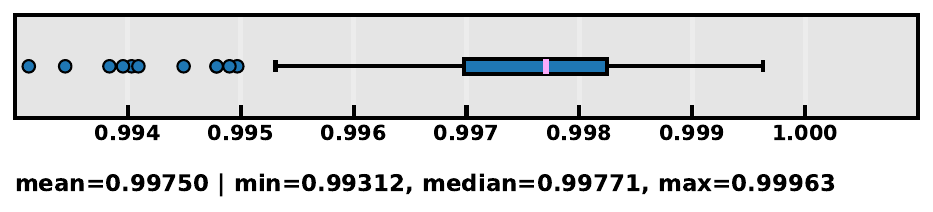}
    }
    \\
    \subfloat[LPIPS]{
        \includegraphics[width=0.99\columnwidth]{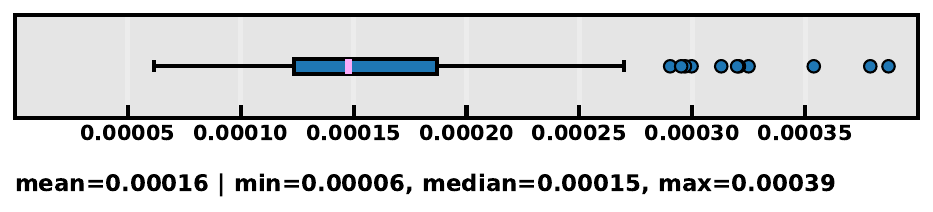}
    }
    \caption{Fidelity evaluation of watermark insertion based on distortion metrics between carrier and marked images.}
    \label{fig:fidelity}
\end{figure}

\begin{figure}[t!]
    \centering
    \subfloat[semantic and photometric transformations]{
    	\includegraphics[width=0.99\columnwidth]{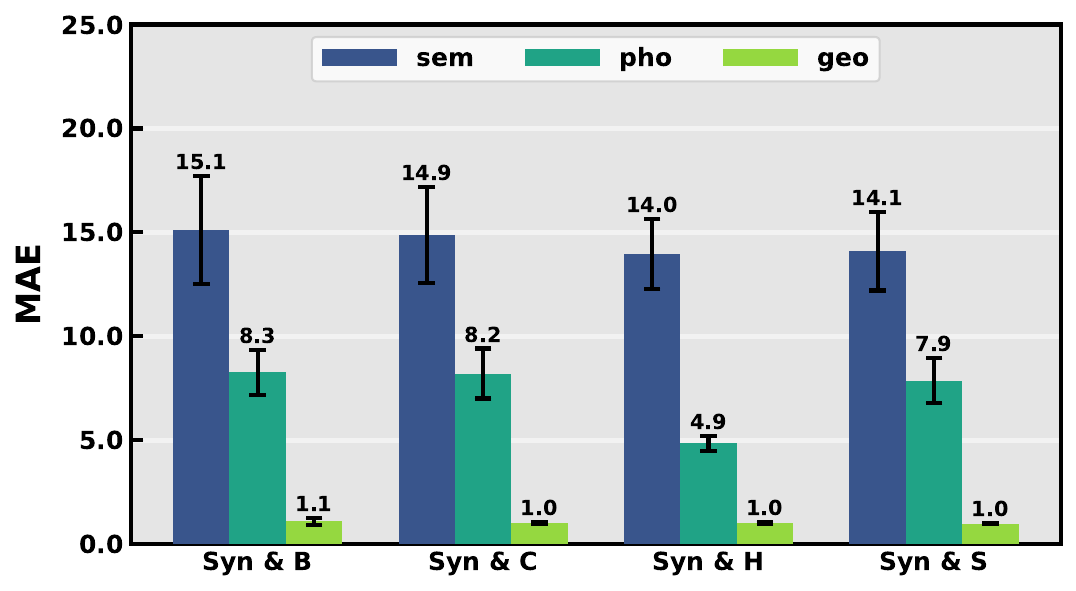}
    }
    \\
    \subfloat[semantic and geometric transformations]{
        \includegraphics[width=0.99\columnwidth]{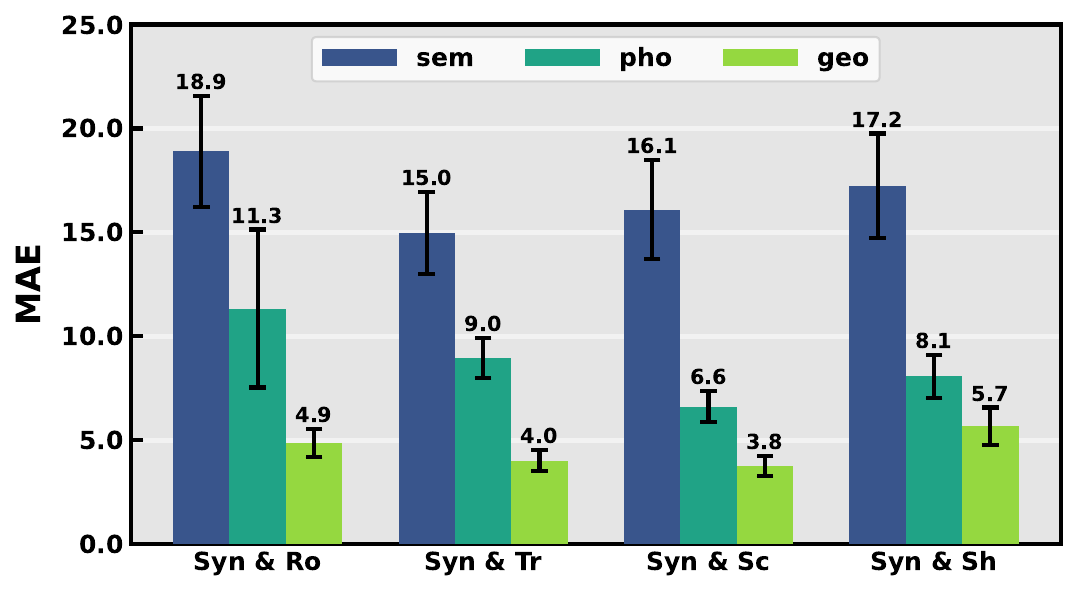}
    }
    \\
    \subfloat[composite transformations]{
        \includegraphics[width=0.99\columnwidth]{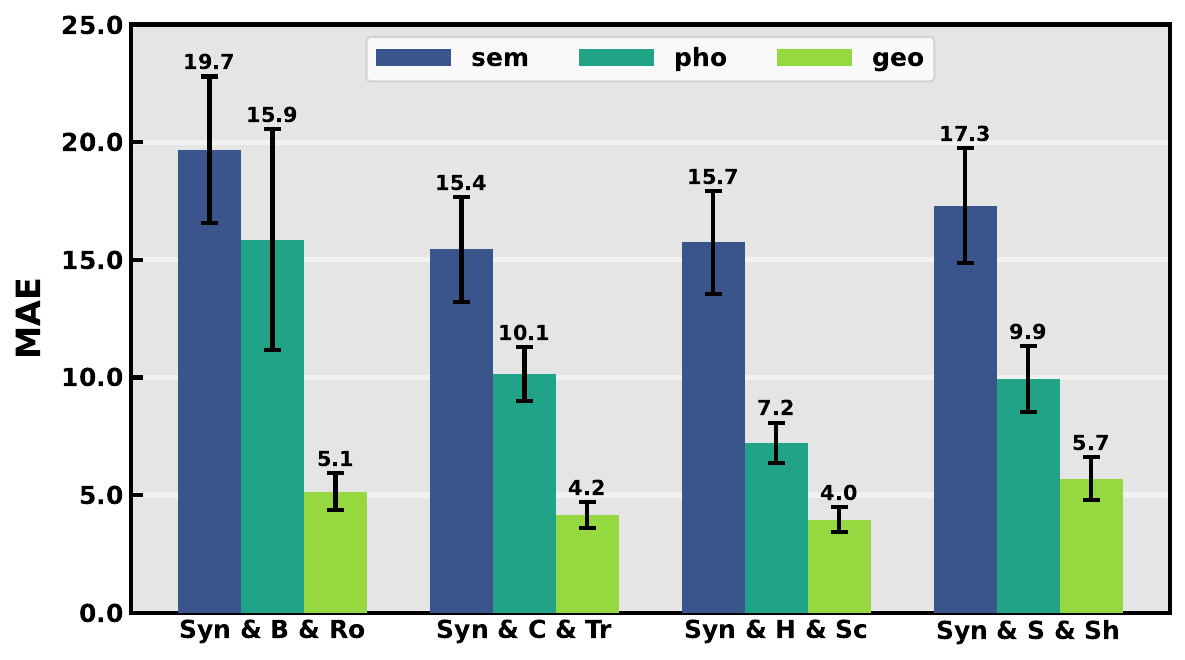}
    }
    \caption{Synchronicity evaluation of watermark extraction based on MAE between extracted and ground-truth watermarks.}
    \label{fig:wm_sync}
\end{figure}

\paragraph*{Photometric Reasoning}

Photometric reasoning seeks the ordering and parameters such that the composite photometric transformation $\mathbf{f}_{\mathrm{pho}}$, when applied to the reference watermark $\mathbf{w}_{\mathrm{pho}}$ and subsequently calibrated by the composite geometric transformation $\mathbf{f}_{\mathrm{geo}}$ (with pre-estimated ordering $\hat{\varpi}_{\mathrm{geo}}$ and parameters $\hat{\boldsymbol{\vartheta}}_{\mathrm{geo}}$), produce a result that best aligns with the extracted watermark $\hat{\mathbf{w}}_{\mathrm{pho}}$. The hypothesised transformed watermark is expressed as
\begin{equation}
\tilde{\mathbf{w}}_{\mathrm{pho}} = \mathbf{f}_{\mathrm{geo}} (\mathbf{f}_{\mathrm{pho}} (\mathbf{w}_{\mathrm{pho}} \mid \varpi_{\mathrm{pho}}, \boldsymbol{\vartheta}_{\mathrm{pho}}) \mid \hat{\varpi}_{\mathrm{geo}}, \hat{\boldsymbol{\vartheta}}_{\mathrm{geo}}),
\end{equation}
where $\varpi_{\mathrm{pho}}$ represents the permutation ordering of brightness, contrast, hue and saturation adjustments, and $\boldsymbol{\vartheta}_{\mathrm{pho}}$ comprises their respective parameters, defined as
\begin{equation}
\begin{split}
	\varpi_{\mathrm{pho}} &\in \operatorname{Sym}(\{ \mathrm{b}, \mathrm{c}, \mathrm{h}, \mathrm{s}\} ) , \\
	\boldsymbol{\vartheta}_{\mathrm{pho}} &= [\vartheta_{\mathrm{b}}, \vartheta_{\mathrm{c}}, \vartheta_{\mathrm{s}}, \vartheta_{\mathrm{h}}] .
\end{split}
\end{equation}
The reasoning task is thus to simultaneously infer the ordering and the magnitudes of the underlying photometric distortions, formulated as a nested optimisation
\begin{equation}
\hat{\varpi}_{\mathrm{pho}}, \hat{\boldsymbol{\vartheta}}_{\mathrm{pho}} = \operatorname{arg\,} \min_{\varpi_{\mathrm{pho}}} \min_{\boldsymbol{\vartheta}_{\mathrm{pho}}} 
\| \tilde{\mathbf{w}}_{\mathrm{pho}} - \hat{\mathbf{w}}_{\mathrm{pho}} \|_1 .
\end{equation}
The photometric transformations associated with brightness, contrast, hue and saturation are defined respectively as
\begin{equation}
	\begin{split}
	f_{\vartheta_{\mathrm{b}}}(\mathbf{w}_{\mathrm{pho}}) &= \vartheta_{\mathrm{b}} \mathbf{w}_{\mathrm{pho}} + (1 - \vartheta_{\mathrm{b}}) \mathbf{0}, \\
	f_{\vartheta_{\mathrm{c}}}(\mathbf{w}_{\mathrm{pho}}) &= \vartheta_{\mathrm{c}} \mathbf{w}_{\mathrm{pho}} + (1 - \vartheta_{\mathrm{c}}) \mu(\ell(\mathbf{w}_{\mathrm{pho}})) , \\
	f_{\vartheta_{\mathrm{h}}}(\mathbf{w}_{\mathrm{pho}}) &= (\hat{r}, \hat{g}, \hat{b}) \mapsfrom \left((h+\vartheta_{\mathrm{h}}) \bmod 1,\, s,\, v\right), \\
	f_{\vartheta_{\mathrm{s}}}(\mathbf{w}_{\mathrm{pho}}) &= \vartheta_{\mathrm{s}} \mathbf{w}_{\mathrm{pho}} + (1 - \vartheta_{\mathrm{s}}) \ell(\mathbf{w}_{\mathrm{pho}}) ,
	\end{split}
\end{equation}
where $\mathbf{0}$ denotes the all-zero matrix, $\mu$ denotes the mean operator, $\ell$ denotes the luminance from grayscale conversion, $(h, s, v)$ are the HSV components and $(\hat{r}, \hat{g}, \hat{b})$ are the resulting RGB values. Specifically, brightness adjustment scales intensities relative to black, contrast adjustment varies intensities relative to their mean luminance, hue adjustment shifts the hue while keeping saturation and value fixed, and saturation adjustment interpolates between the colour image and its grayscale version.

\paragraph*{Semantic Reasoning}
Semantic reasoning aims to infer the spatial regions of content editing, taking the form of a matrix $\boldsymbol{\vartheta}_{\mathrm{sem}}$, in contrast to geometric and photometric transformations, which estimate finite-dimensional parameter vectors $\boldsymbol{\vartheta}_{\mathrm{geo}}$ and $\boldsymbol{\vartheta}_{\mathrm{pho}}$. The edited regions are directly revealed by the extracted semantic watermark, without requiring additional computation; that is, $\boldsymbol{\vartheta}_{\mathrm{sem}} = \hat{\mathbf{w}}_{\mathrm{sem}}$. During the learning phase, the encoder and decoder are jointly calibrated to minimise the discrepancy between the extracted semantic watermark and the supervision ground truth, expressed as $\min \| \hat{\mathbf{w}}_{\mathrm{sem}} - \tilde{\mathbf{w}}_{\mathrm{sem}} \|_1$, where the ground truth $\tilde{\mathbf{w}}_{\mathrm{sem}}$ corresponds to the editing mask after geometric transformation $\mathbf{f}_{\mathrm{geo}} ( \mathbf{m} )$. Therefore, the extracted semantic watermark provides a direct indication of the regions subject to content editing.

\section{Evaluation}\label{sec:eval}
The validity of tell-tale watermarking is evaluated through three aspects: watermark insertion assessed by fidelity, watermark extraction assessed by synchronicity, and explanatory reasoning assessed by traceability. Beyond the primary assessments, we further analyse the impact of reasoning dimensionality on reasoning time and accuracy, and provide a comparative study of synthetic media detection performance against state-of-the-art baselines.

\begin{figure}[t!]
    \centering
    \subfloat[semantic and photometric transformations]{
    	\includegraphics[width=0.99\columnwidth]{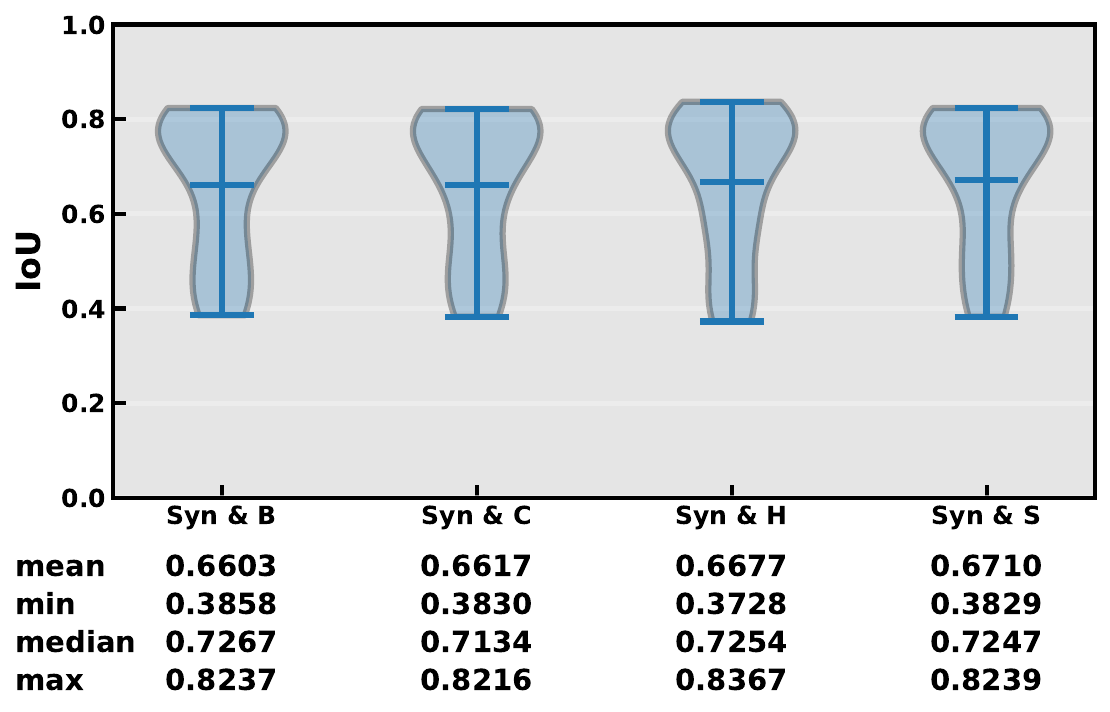}
    }
    \\
    \subfloat[semantic and geometric transformations]{
        \includegraphics[width=0.99\columnwidth]{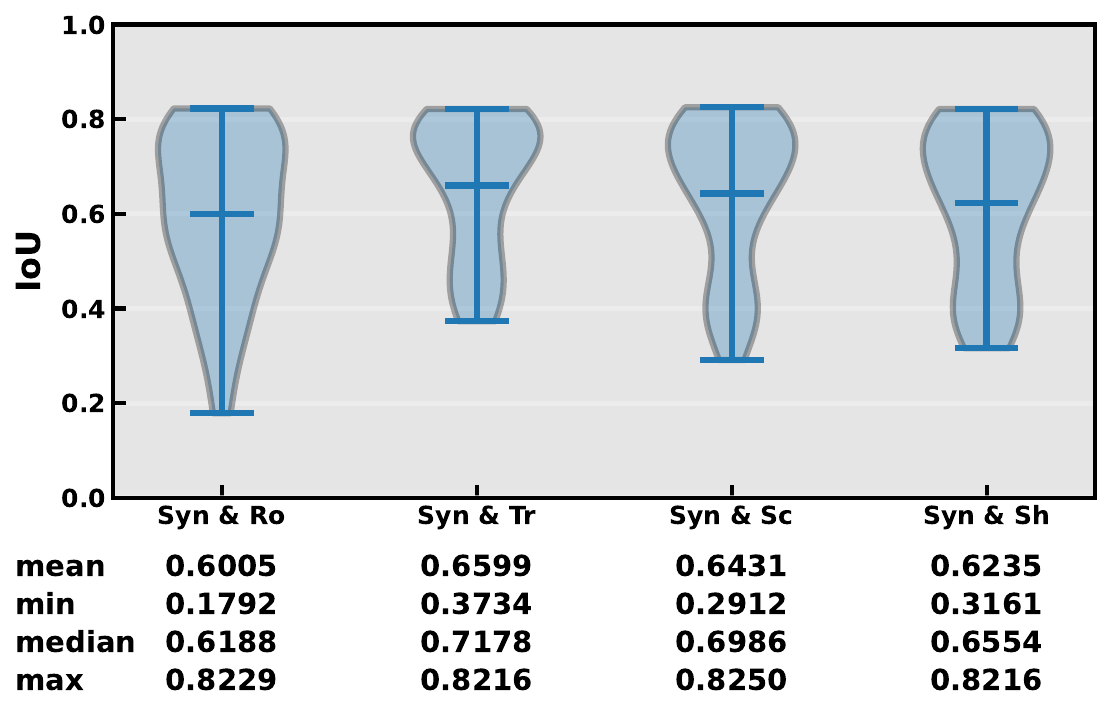}
    }
    \\
    \subfloat[composite transformations]{
        \includegraphics[width=0.99\columnwidth]{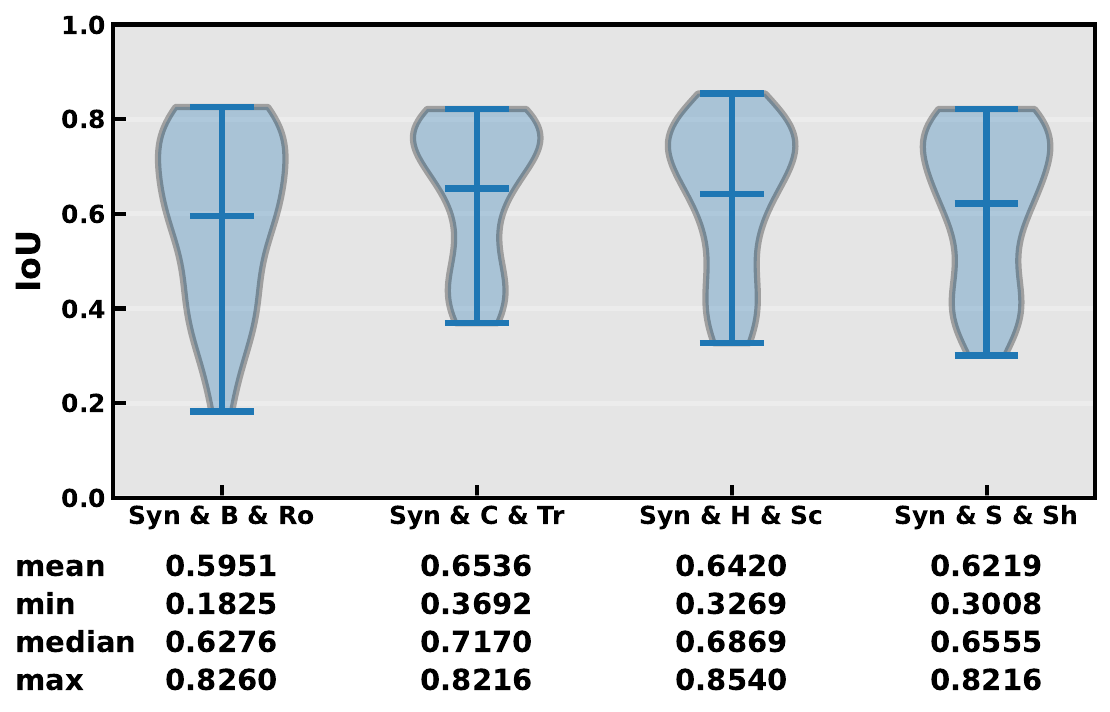}
    }
    \caption{Traceability evaluation of semantic reasoning based on IoU between estimated and true editing regions.}
    \label{fig:sem_iou}
\end{figure}

\subsection{Experimental Setup}
Experiments are conducted on the COCO dataset (common objects in context)~\cite{Lin:2014aa}, partitioned for learning and inference, with all images resampled to a consistent resolution of $128 \times 128$ pixels. The encoder and decoder are based on a U-Net architecture with residual and attention modules~\cite{Ronneberger:2015aa, 7780459, Woo:2018aa}, where the number of base latent channels is configured as 32, as illustrated in Figure~\ref{fig:neural_networks}. Explanatory reasoning of photometric and geometric parameters is performed with a maximum of $100$ iterations, using an update rate of $0.1$.

Semantic editing is performed with a stable diffusion model prompted to synthesise realistic content in masked regions~\cite{9878449}. Photometric transformations are defined with brightness, contrast and saturation varied in $[0.75, 1.25]$ (no adjustment at $1$) and hue in $[-0.35, 0.35]$ (no shift at $0$). Geometric transformations include rotation in $[-30^{\circ}, 30^{\circ}]$ (unchanged at $0^{\circ}$), translation in $[-0.2, 0.2]$ along both axes (unchanged at $0$), scaling in $[0.8, 1.2]$ (unchanged at $1$) and shearing in $[-15^{\circ}, 15^{\circ}]$ along both axes (unchanged at $0^{\circ}$). Results in degrees are normalised to fractions of a full cycle. For notational convenience, the experimental results are presented using the following abbreviations: semantic (sem), photometric (pho), geometric (geo), synthesis (Syn), brightness (B), contrast (C), saturation (S), hue (H), rotation (Ro), translation (Tr), scaling (Sc) and shearing (Sh). A visual demonstration of images and watermarks under composite transformations is shown in Figure~\ref{fig:demo}.

\begin{figure*}[t!]
    \centering
    \subfloat[brightness\\(Syn \& B)]{
    	\includegraphics[width=0.45\columnwidth]{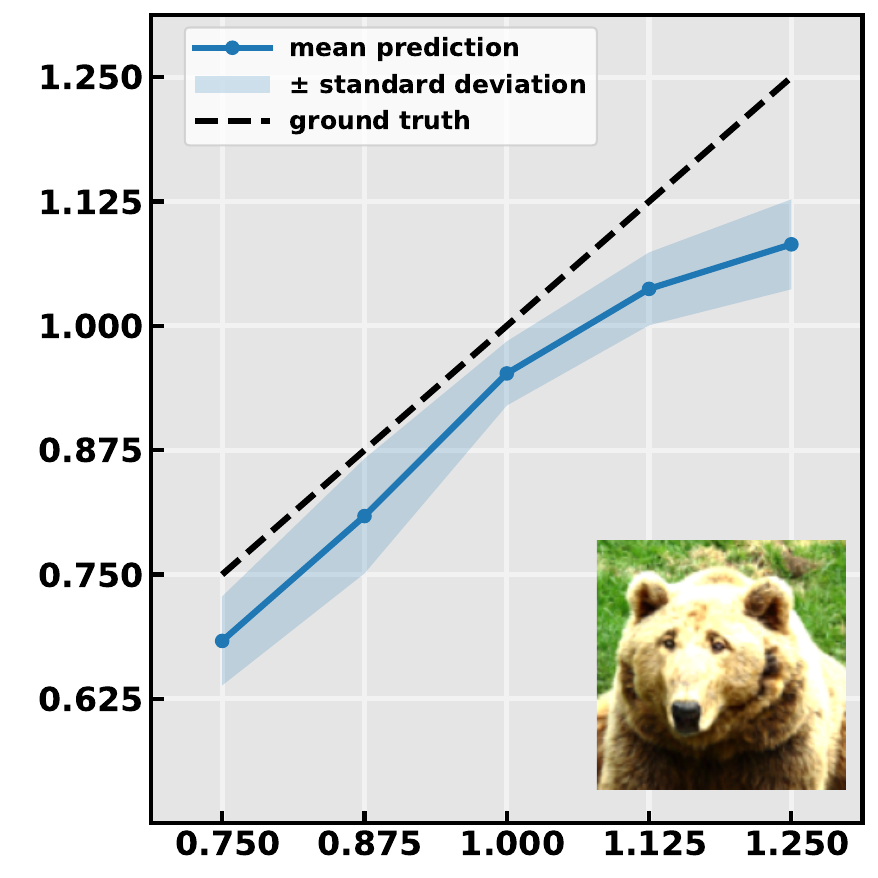}
    }
    \hfil
    \subfloat[contrast\\(Syn \& C)]{
        \includegraphics[width=0.45\columnwidth]{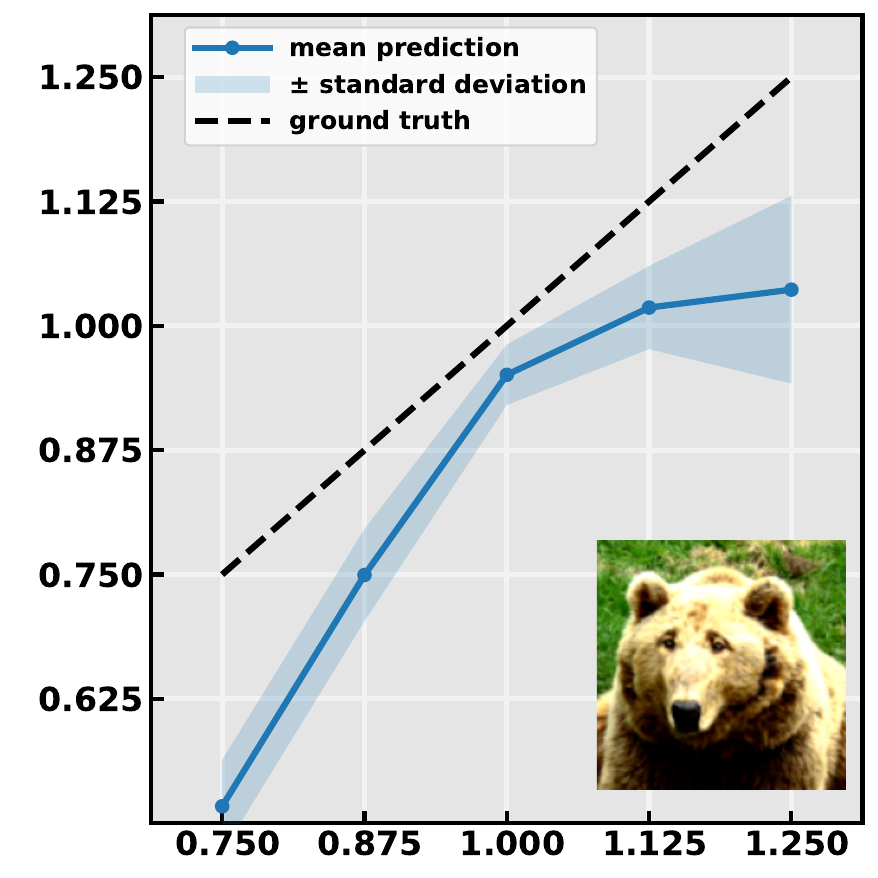}
    }
    \hfil
    \subfloat[hue\\(Syn \& H)]{
        \includegraphics[width=0.45\columnwidth]{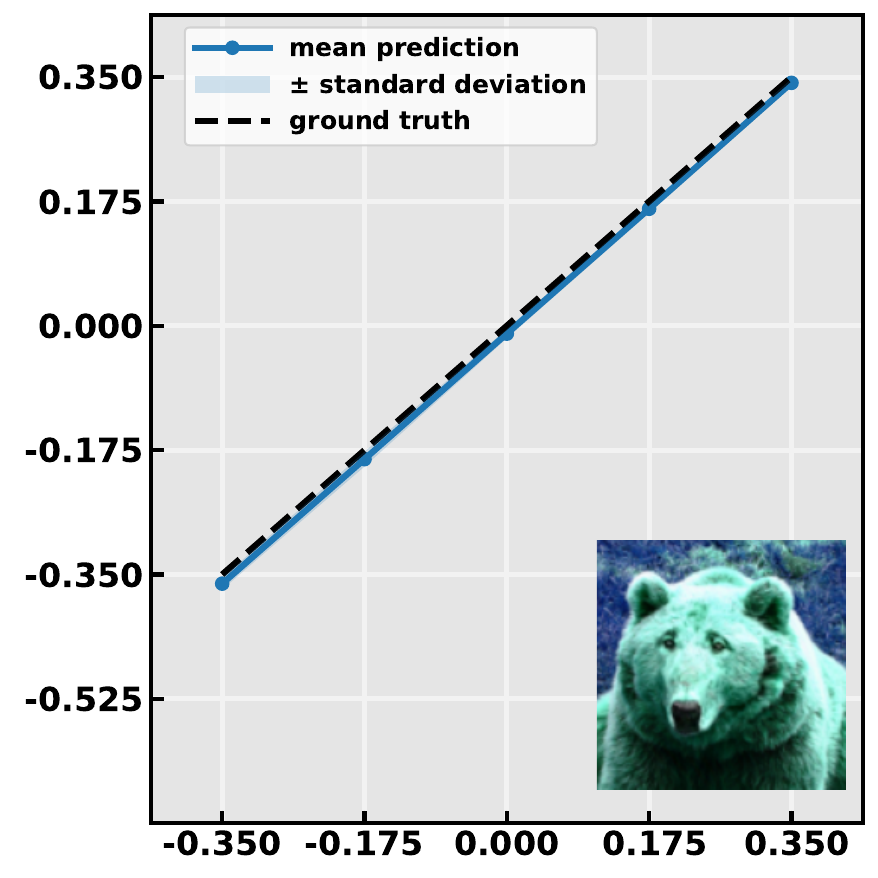}
    }
    \hfil
    \subfloat[saturation\\(Syn \& S)]{
        \includegraphics[width=0.45\columnwidth]{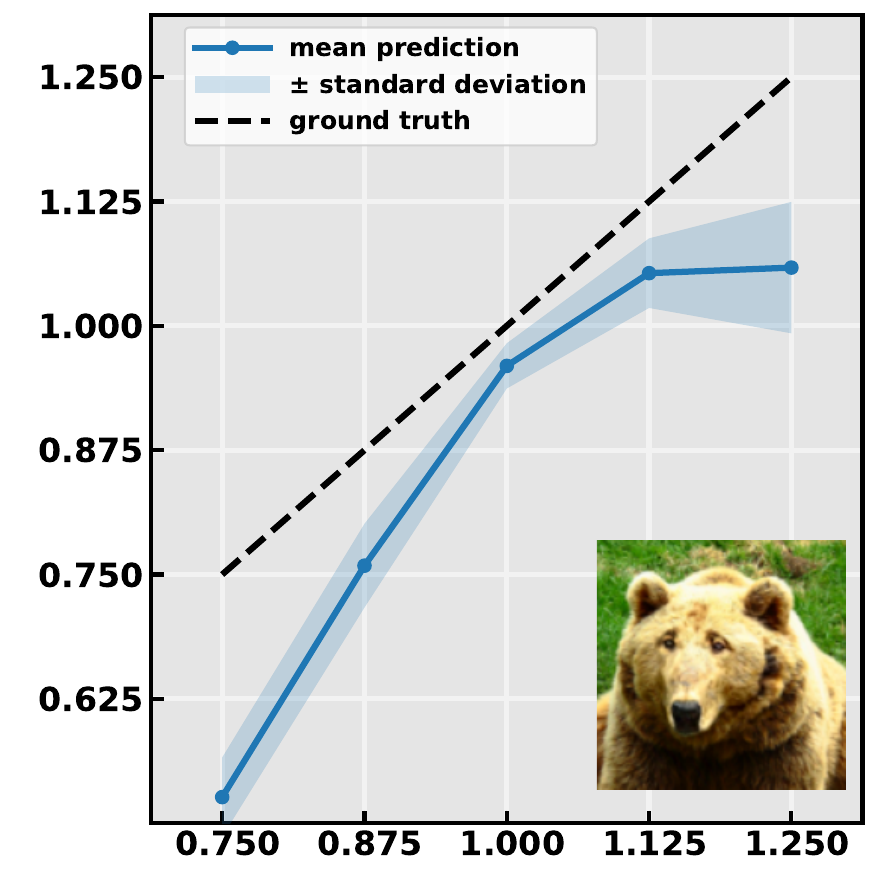}
    }
    \\
    \subfloat[rotation\\(Syn \& Ro)]{
        \includegraphics[width=0.45\columnwidth]{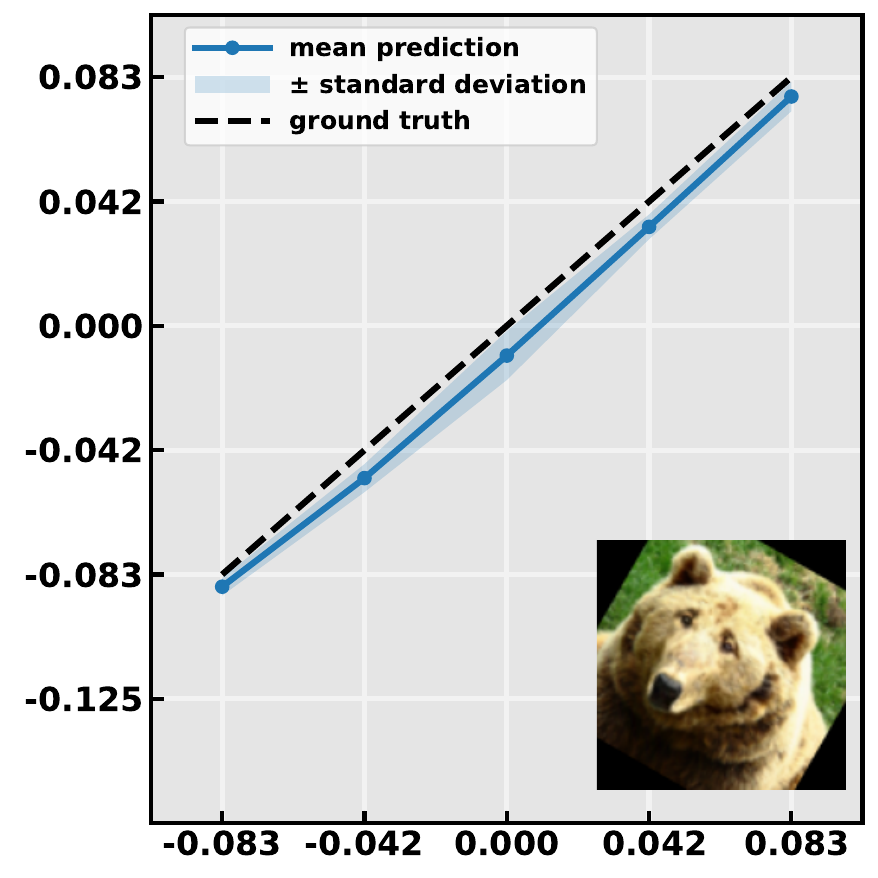}
    }
    \hfil
    \subfloat[translation\\(Syn \& Tr)]{
        \includegraphics[width=0.45\columnwidth]{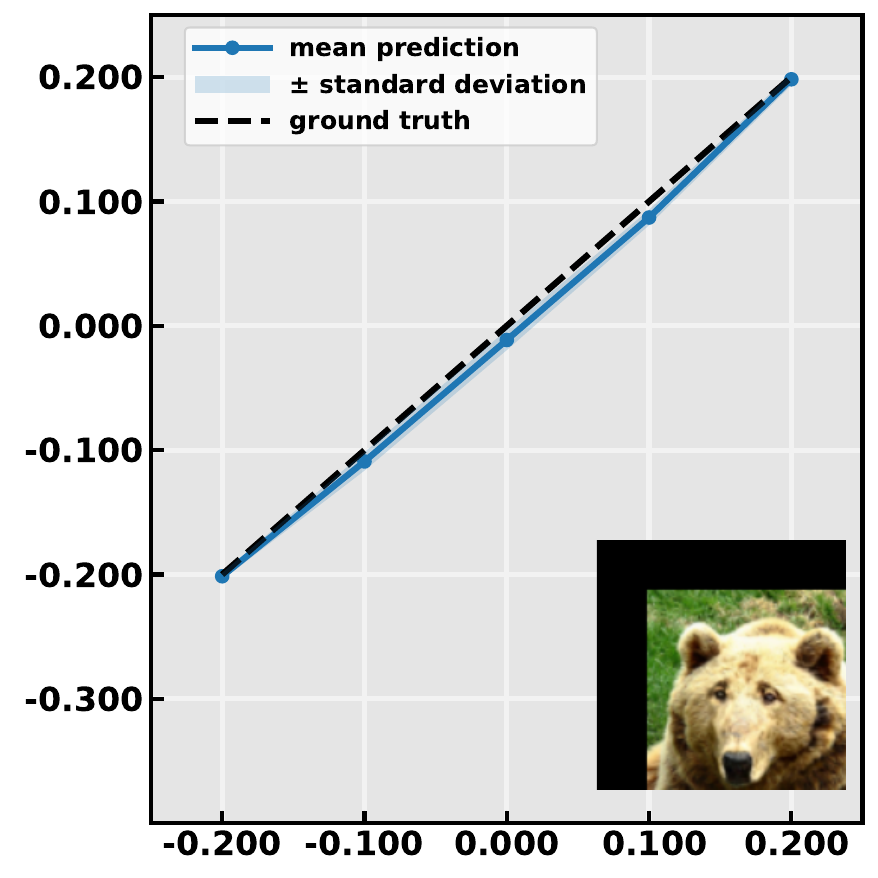}
    }
    \hfil
    \subfloat[scaling\\(Syn \& Sc)]{
        \includegraphics[width=0.45\columnwidth]{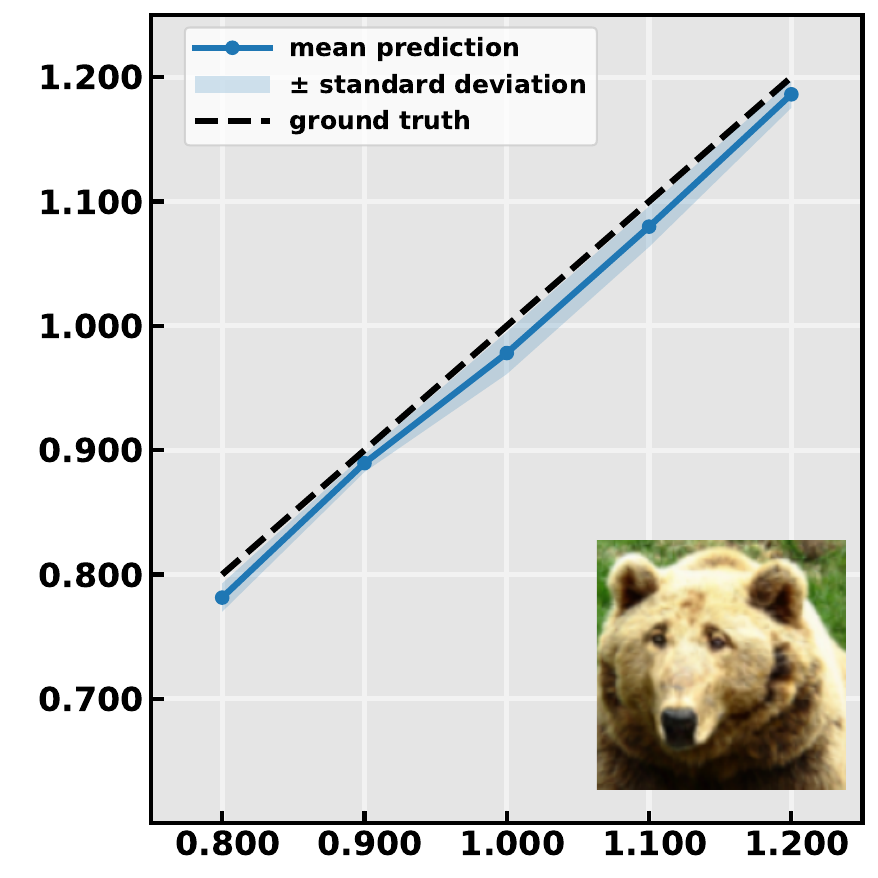}
    }
    \hfil
    \subfloat[shearing\\(Syn \& Sh)]{
        \includegraphics[width=0.45\columnwidth]{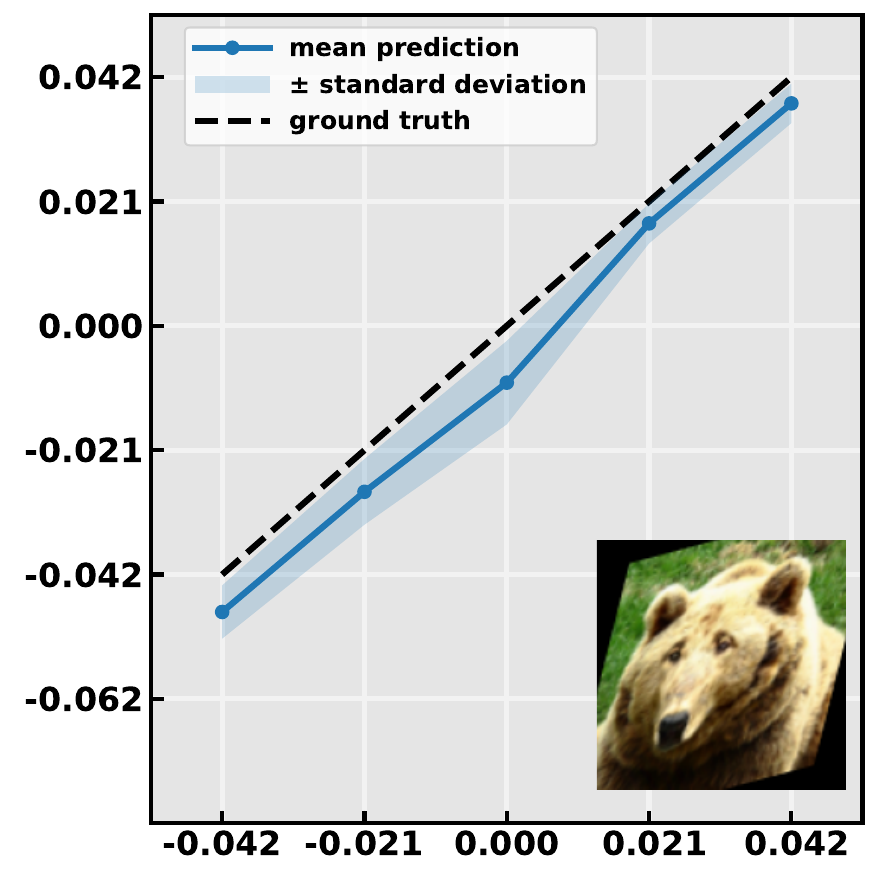}
    }
    \\
    \subfloat[brightness\\(Syn \& B \& Ro)]{
    	\includegraphics[width=0.45\columnwidth]{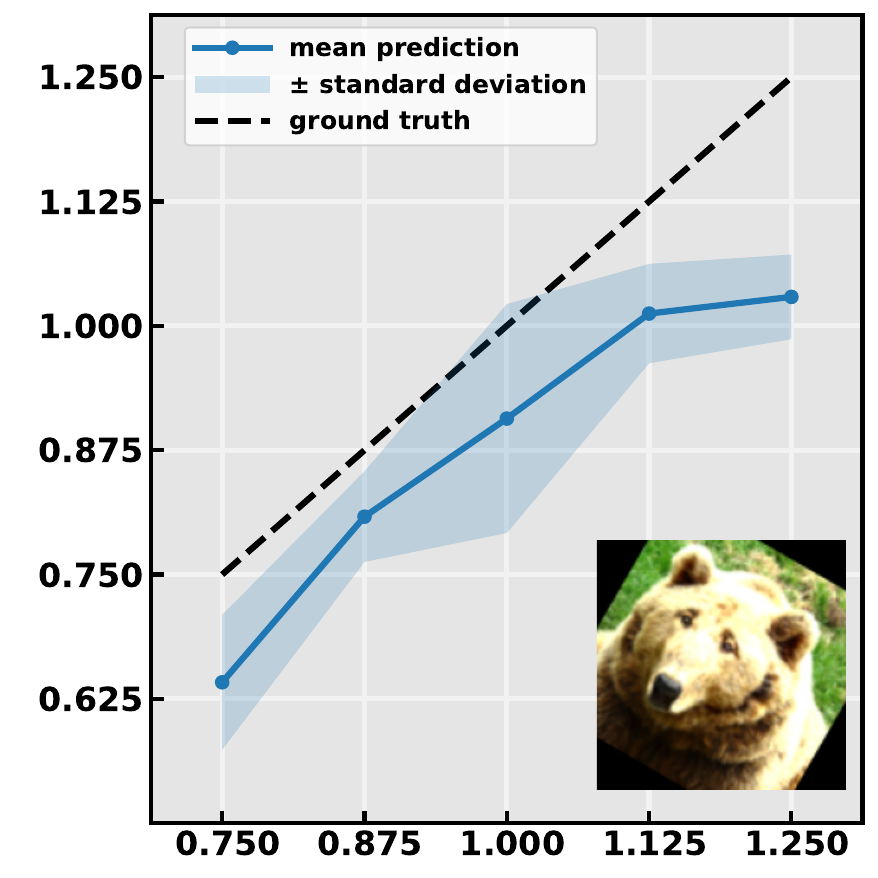}
    }
    \hfil
    \subfloat[contrast\\(Syn \& C \& Tr)]{
        \includegraphics[width=0.45\columnwidth]{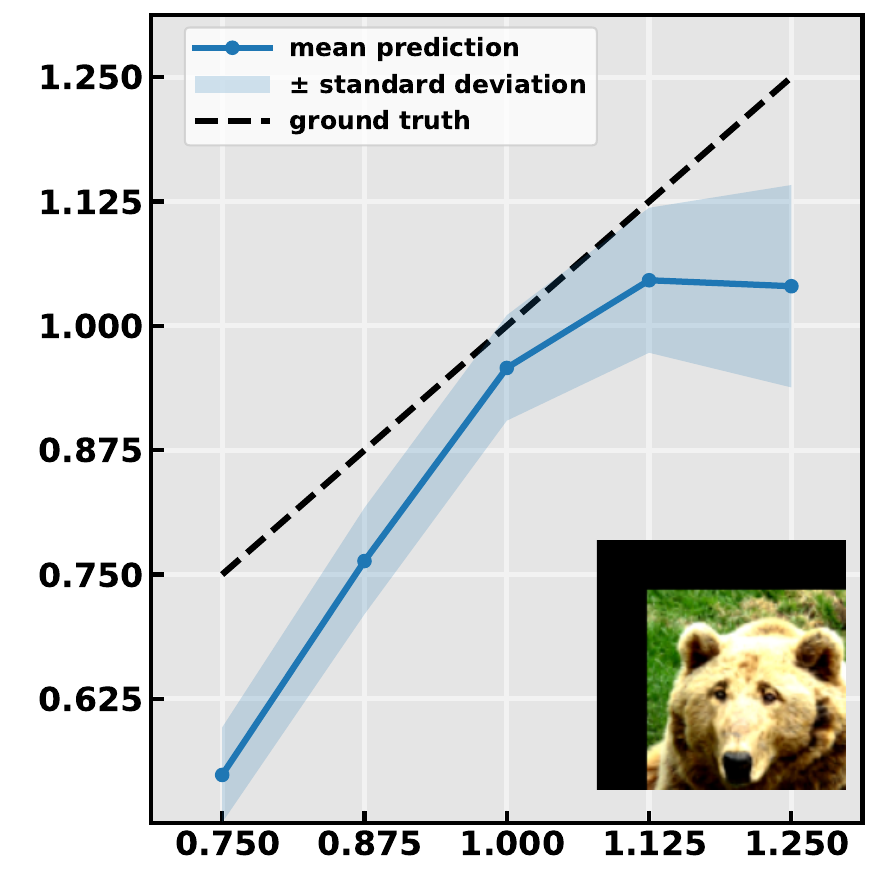}
    }
    \hfil
    \subfloat[hue\\(Syn \& H \& Sc)]{
        \includegraphics[width=0.45\columnwidth]{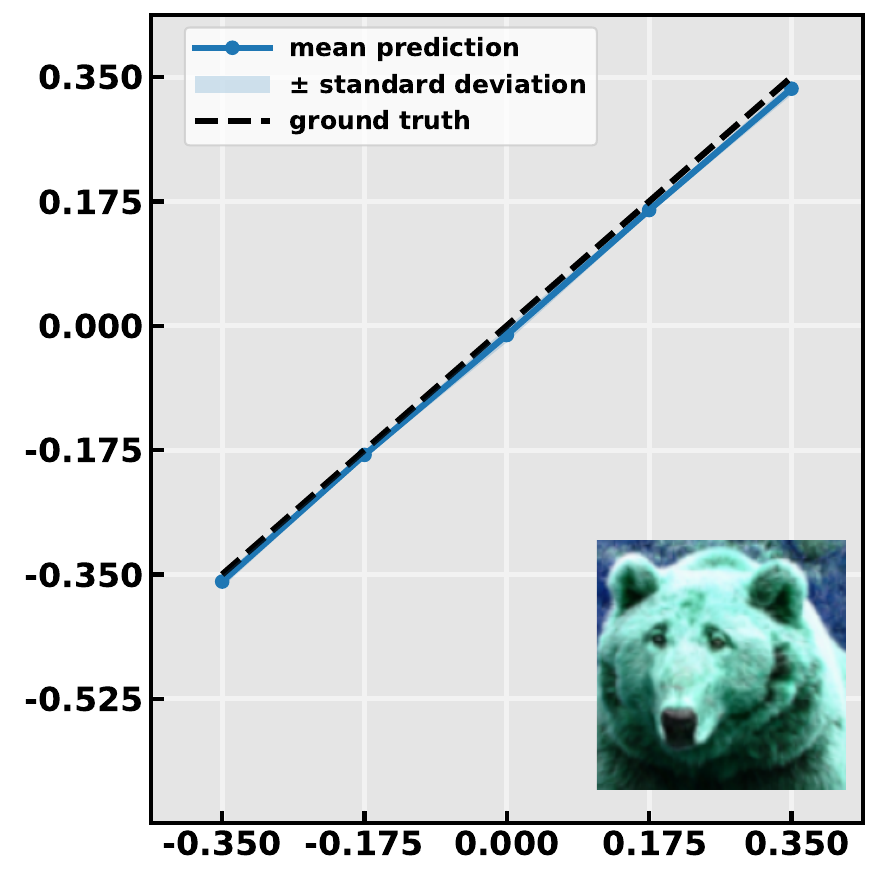}
    }
    \hfil
    \subfloat[saturation\\(Syn \& S \& Sh)]{
        \includegraphics[width=0.45\columnwidth]{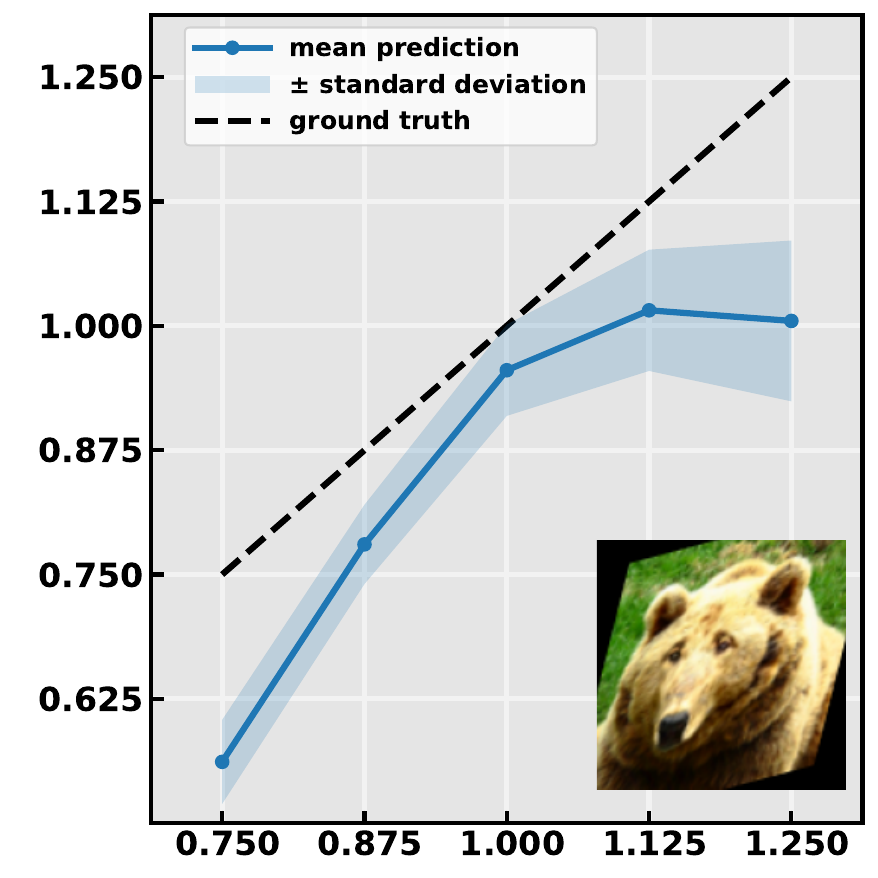}
    }
    \\
    \subfloat[rotation\\(Syn \& B \& Ro)]{
        \includegraphics[width=0.45\columnwidth]{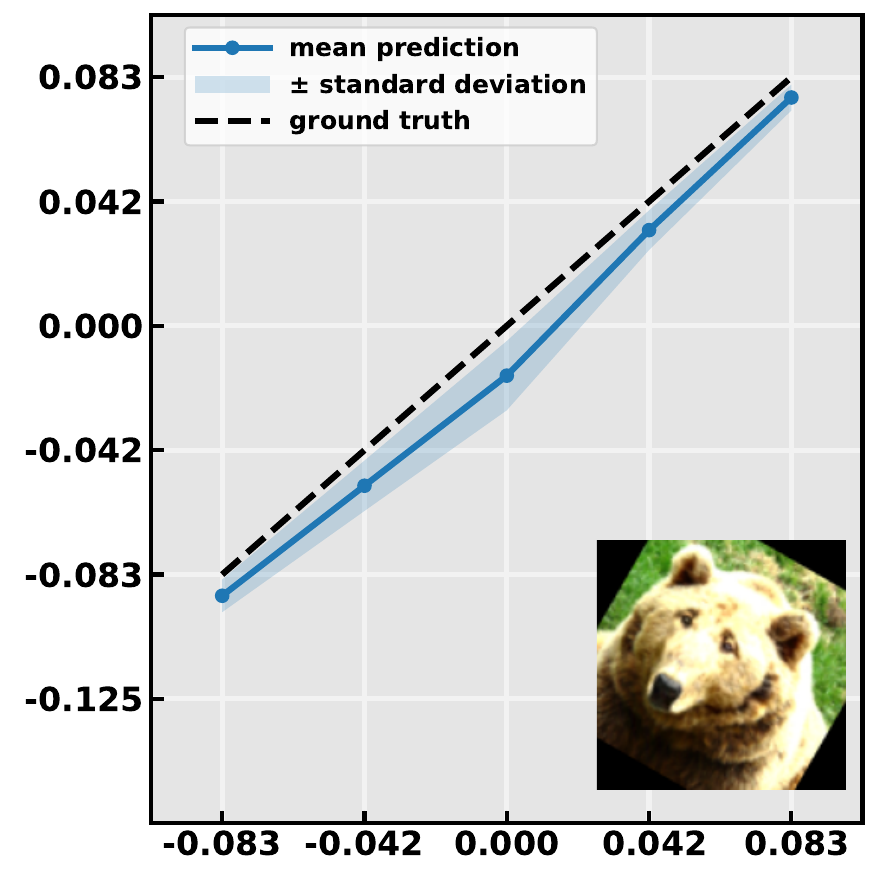}
    }
    \hfil
    \subfloat[translation\\(Syn \& C \& Tr)]{
        \includegraphics[width=0.45\columnwidth]{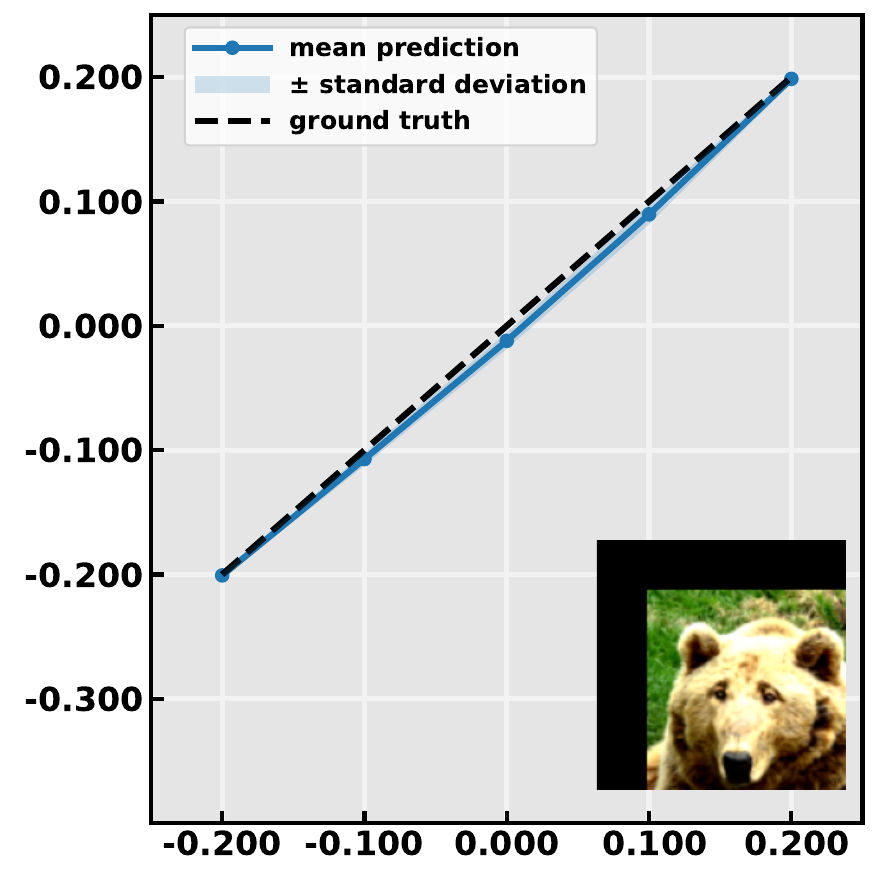}
    }
    \hfil
    \subfloat[scaling\\(Syn \& H \& Sc)]{
        \includegraphics[width=0.45\columnwidth]{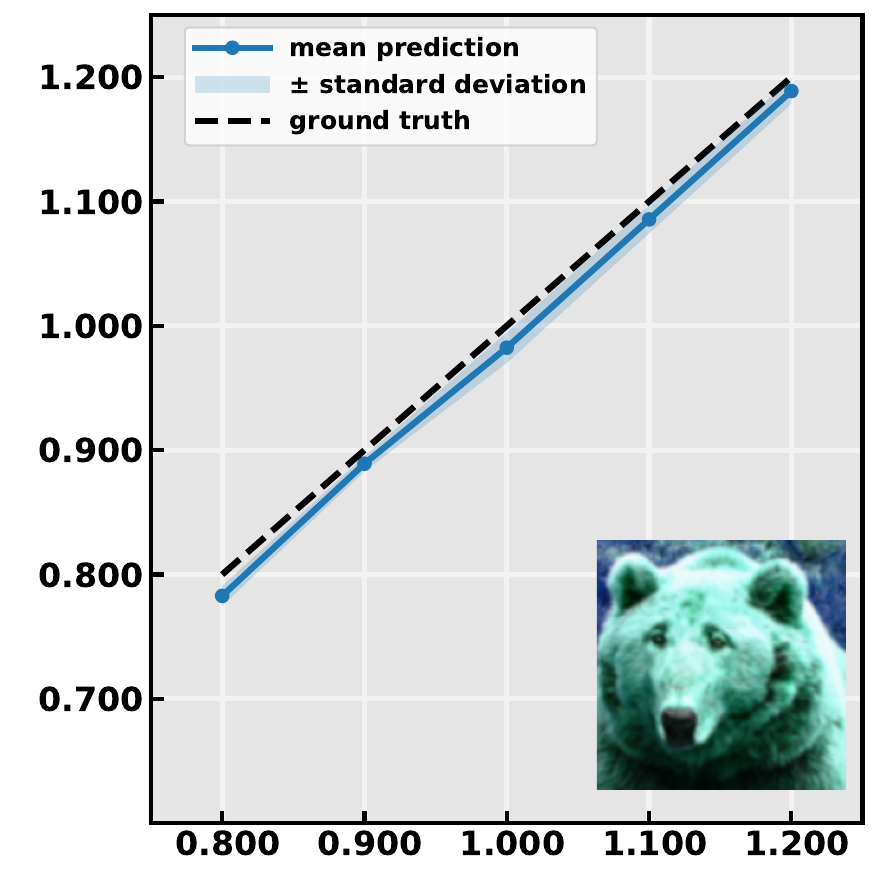}
    }
    \hfil
    \subfloat[shearing\\(Syn \& S \& Sh)]{
        \includegraphics[width=0.45\columnwidth]{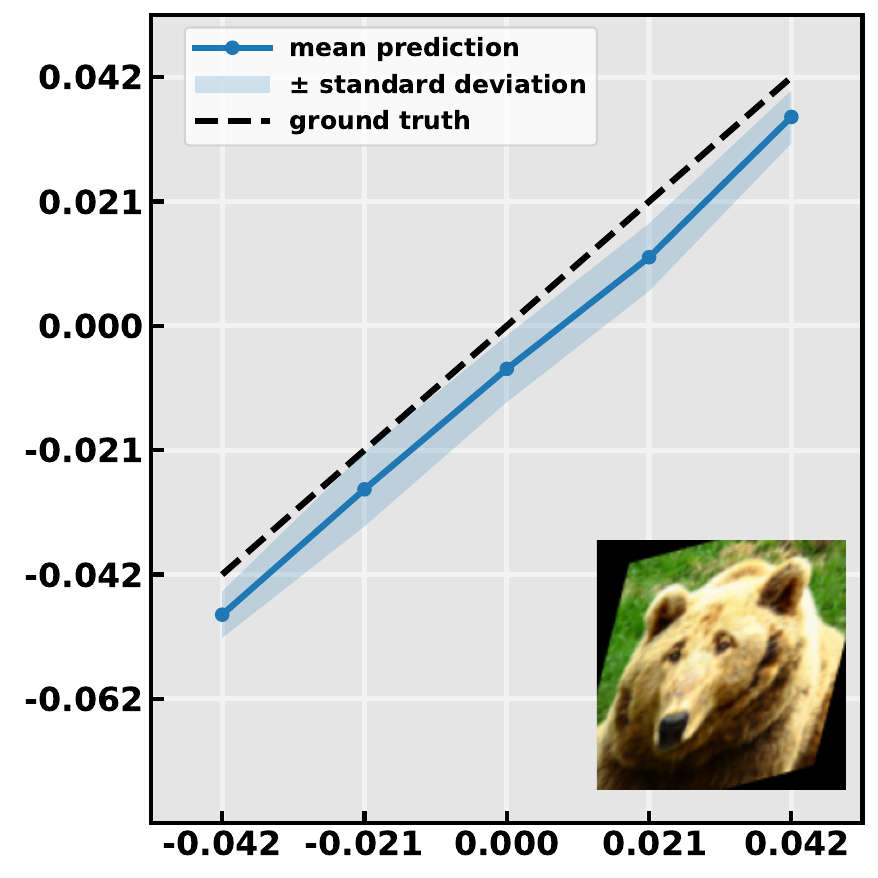}
    }
    \caption{Traceability evaluation of photometric and geometric reasoning based on deviations between estimated and true parameter values.}
    \label{fig:phogeo_trace}
\end{figure*}

\subsection{Fidelity}
Fidelity measures perceptual similarity between carrier and marked images. We evaluate fidelity via the following quality metrics: $\ell_1$ norm, $\ell_{\infty}$ norm, peak signal-to-noise ratio (PSNR), structural similarity index measure (SSIM), and learnt perceptual image patch similarity (LPIPS), as shown in Figure~\ref{fig:fidelity}. The $\ell_{1}$ norm represents the average absolute pixel difference and the $\ell_{\infty}$ norm captures the maximum absolute pixel deviation, where values close to zero signify negligible distortion. PSNR is defined via the ratio of the maximum possible pixel value to the noise, with satisfactory quality values typically between 30 and 50 dB. SSIM compares structural similarity in terms of luminance, contrast and texture, with a score of one denoting identical images. LPIPS is a learning–based perceptual similarity measure for approximating human visual perception, with values close to zero regarded as nearly indistinguishable to human observers. The statistical results across these fidelity metrics demonstrate that watermark insertion introduces only negligible perceptual distortion to the carrier media.

\subsection{Synchronicity}
Synchronicity characterises how closely the extracted watermarks follow the effects of transformations applied to the media, as reflected by their agreement with directly transformed ground-truth counterparts. We quantify synchronicity by the mean absolute error (MAE) between the extracted watermarks $\hat{\mathbf{w}}$ and their ground-truth counterparts $\tilde{\mathbf{w}}$, as shown in Figure~\ref{fig:wm_sync}. For each marked image, we first apply a semantic transformation to synthesise content in a random region, followed by a photometric transformation for colour adjustment, a geometric transformation for perspective projection, or a composite transformation combining the two. Watermarks are then extracted from the transformed image and compared against their ground-truth counterparts, which are obtained by applying the same sequence of transformations directly to the initial reference watermarks. It is observed that composite transformations lead to greater degradation in synchronicity compared with applying a semantic transformation followed by either a photometric or a geometric transformation. The photometric watermark reflects changes in hue more faithfully than changes in brightness, contrast or saturation. The geometric watermark demonstrates relatively stable alignment with its ground-truth counterpart across rotation, translation, scaling, and shearing. The geometric watermark achieves the highest synchronicity, followed by the photometric watermark and then the semantic watermark.

\subsection{Traceability}
Traceability refers to the extent to which explanatory reasoning can correctly infer the parameters of transformations applied to the media from traces left within the extracted watermarks. A semantic transformation is first applied to synthesise content in a random region, followed by a photometric transformation, a geometric transformation, or a composite transformation combining the two. For semantic reasoning, we compare the extracted semantic watermark against the ground-truth mask through the intersection over union (IoU), as shown in Figure~\ref{fig:sem_iou}. The average IoU lies within the range of 0.6 to 0.7 across various transformations, indicating a moderate to high degree of alignment between the estimated and true edited regions. For photometric and geometric reasoning, we first analyse the parameter estimates of each individual transformation across corresponding parameter spectrums and then report traceability in terms of the average prediction error. Figure~\ref{fig:phogeo_trace} presents the mean estimates with standard deviations over the parameter spectrum, where the diagonal line denotes perfect agreement with the ground truth. In photometric reasoning, the estimated hue parameters align closely with their ground-truth counterparts, whereas brightness, contrast and saturation exhibit larger deviations, particularly under stronger adjustments at both extremes. In geometric reasoning, the parameter estimations for rotation, translation, scaling and shearing are nearly perfect. Traceability is slightly degraded under composite transformations, where both photometric and geometric distortions are imposed following semantic synthesis. Table~\ref{tab:phogeo_trace} summarises the average prediction errors for each class of transformation parameter under different transformation chains. It is observed that geometric parameters are estimated with high accuracy, regardless of whether they are applied or not. Among the photometric parameters, hue exhibits the lowest and stablest errors, while brightness, contrast and saturation show larger deviations, particularly when the parameter under evaluation is the one applied in the transformation chain. The results confirm the validity of tracing applied transformations through tell-tale watermarks.

\begin{table*}[ht]
\centering
\caption{Traceability Across Transformations}
\label{tab:phogeo_trace}
\begin{tabular}{@{} l  R{18mm}  C{12mm} C{12mm} C{12mm} C{12mm} C{12mm} C{12mm} C{12mm} C{12mm} @{}}
\toprule
 & & \multicolumn{8}{c}{\vspace{1.0mm}\textbf{Average Prediction Error Per Transform}} \\
 & & Brightness & Contrast & Hue & Saturation & Rotation & Translation & Scaling & Shearing \\
\midrule
\multirow{12}{*}{\rotatebox{90}{\textbf{Applied Transform Chain}}}
& Syn \& B  & \textbf{0.0874} & 0.0504 & 0.0120 & 0.0657 & 0.0108 & 0.0112 & 0.0246 & 0.0117 \\
& Syn \& C  & 0.0487 & \textbf{0.1456} & 0.0106 & 0.0458 & 0.0082 & 0.0107 & 0.0266 & 0.0087 \\          
& Syn \& H  & 0.0479 & 0.0509 & \textbf{0.0110} & 0.0448 & 0.0096 & 0.0130 & 0.0248 & 0.0115 \\
& Syn \& S  & 0.0479 & 0.0420 & 0.0110 & \textbf{0.1287} & 0.0098 & 0.0101 & 0.0272 & 0.0096 \\
& Syn \& Ro & 0.0424 & 0.0516 & 0.0278 & 0.0415 & \textbf{0.0077} & 0.0080 & 0.0128 & 0.0072 \\
& Syn \& Tr & 0.0600 & 0.0645 & 0.0113 & 0.0665 & 0.0082 & \textbf{0.0073} & 0.0132 & 0.0093 \\
& Syn \& Sc & 0.0430 & 0.0438 & 0.0126 & 0.0334 & 0.0097 & 0.0073 & \textbf{0.0170} & 0.0067 \\
& Syn \& Sh & 0.0426 & 0.0445 & 0.0112 & 0.0378 & 0.0083 & 0.0087 & 0.0222 & \textbf{0.0062} \\
& Syn \& B \& Ro  & \textbf{0.1204} & 0.0653 & 0.0378 & 0.0568 & \textbf{0.0104} & 0.0091 & 0.0180 & 0.0082 \\
& Syn \& C \& Tr  & 0.0674 & \textbf{0.1289} & 0.0136 & 0.0577 & 0.0100 & \textbf{0.0063} & 0.0161 & 0.0077 \\
& Syn \& H \& Sc & 0.0406 & 0.0488 & \textbf{0.0115} & 0.0437 & 0.0083 & 0.0093 & \textbf{0.0143} & 0.0078 \\
& Syn \& S \& Sh  & 0.0532 & 0.0513 & 0.0108 & \textbf{0.1364} & 0.0084 & 0.0104 & 0.0264 & \textbf{0.0073}\\
\bottomrule
\end{tabular}
\end{table*}

\subsection{Reasoning Complexity}
\begin{figure*}[t!]
    \centering
    \subfloat[brightness]{
    	\includegraphics[width=0.48\columnwidth]{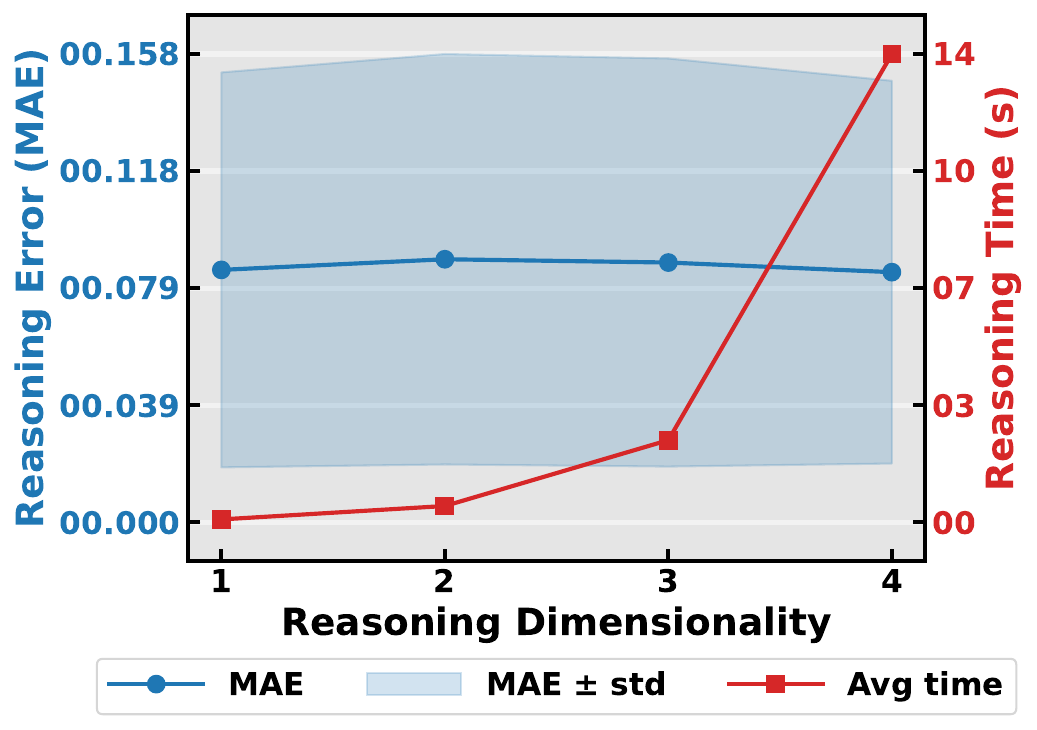}
    }
    \hfil
    \subfloat[contrast]{
        \includegraphics[width=0.48\columnwidth]{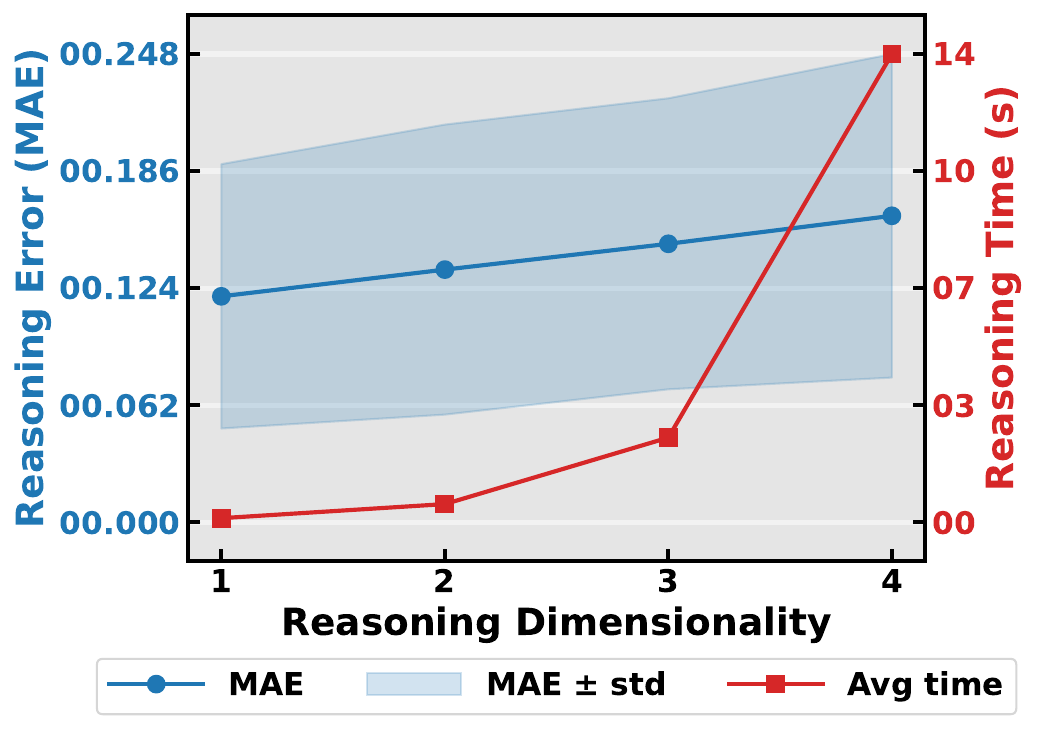}
    }
    \hfil
    \subfloat[hue]{
        \includegraphics[width=0.48\columnwidth]{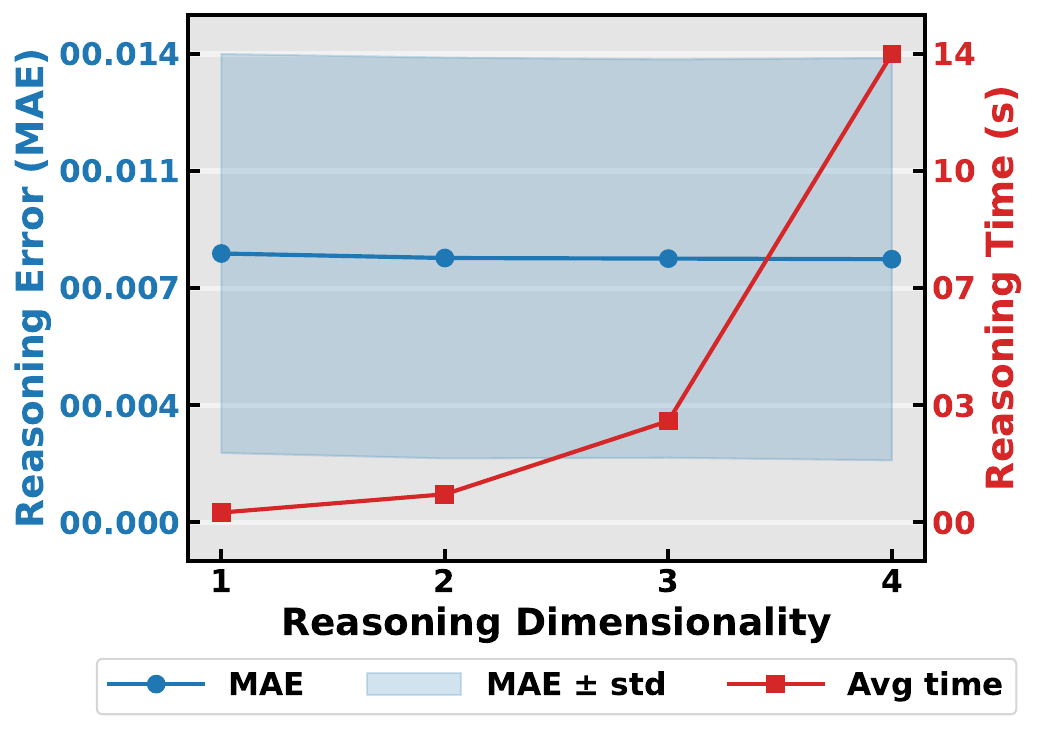}
    }
    \hfil
    \subfloat[saturation]{
        \includegraphics[width=0.48\columnwidth]{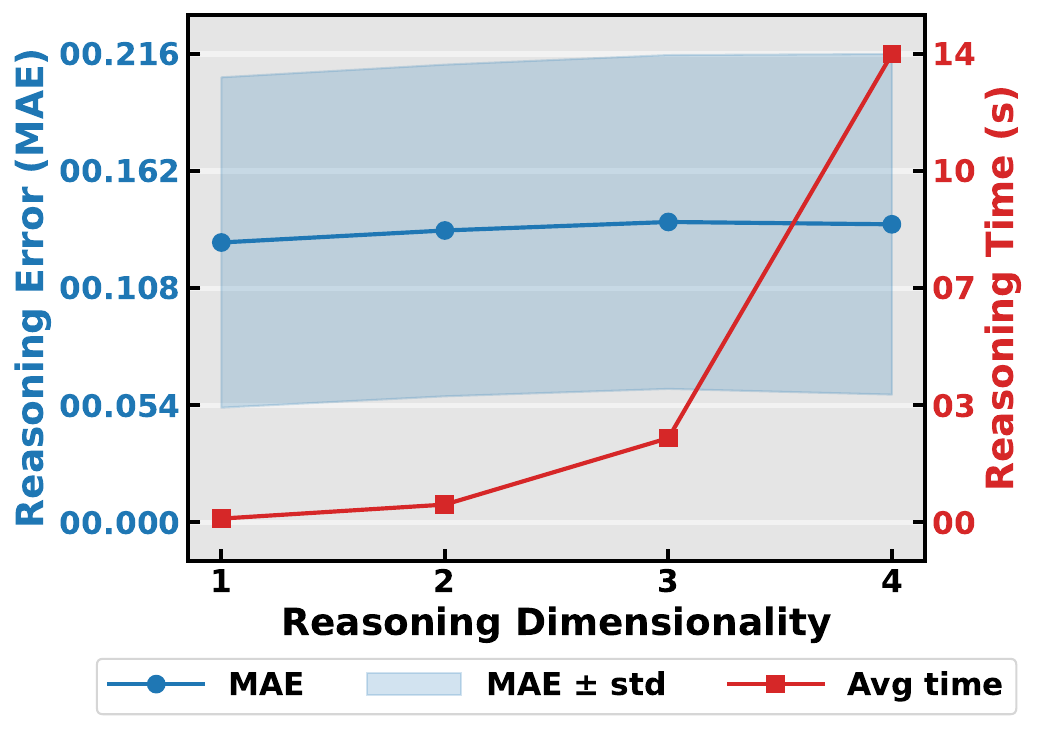}
    }
    \\
    \subfloat[rotation]{
        \includegraphics[width=0.48\columnwidth]{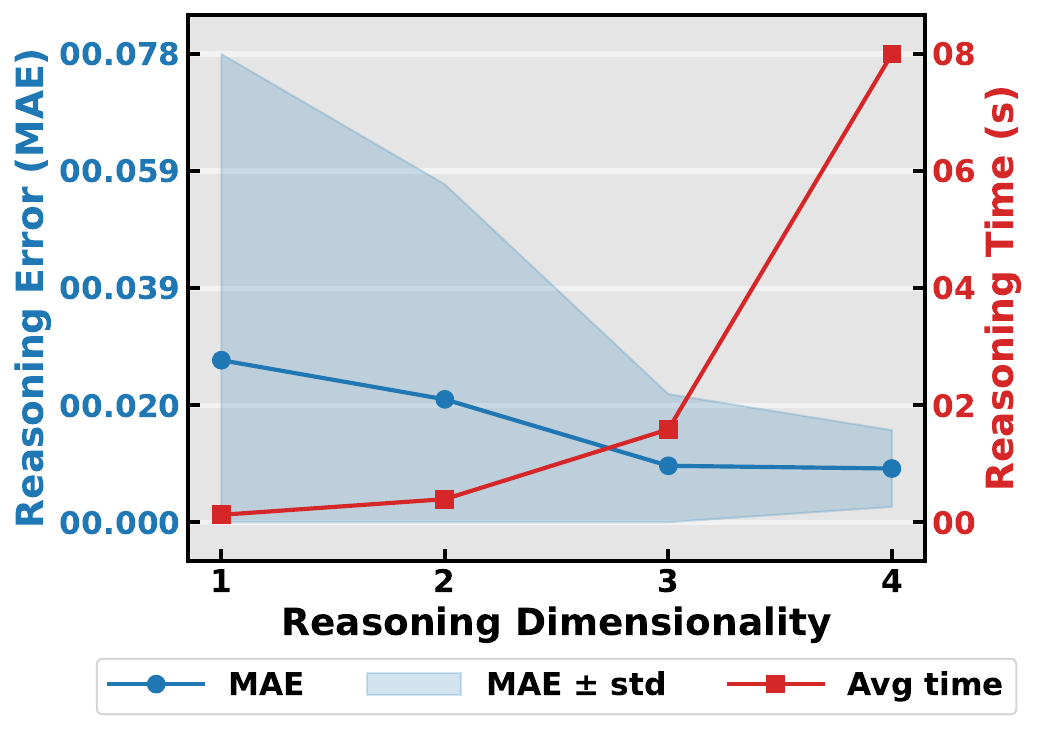}
    }
    \hfil
    \subfloat[translation]{
        \includegraphics[width=0.48\columnwidth]{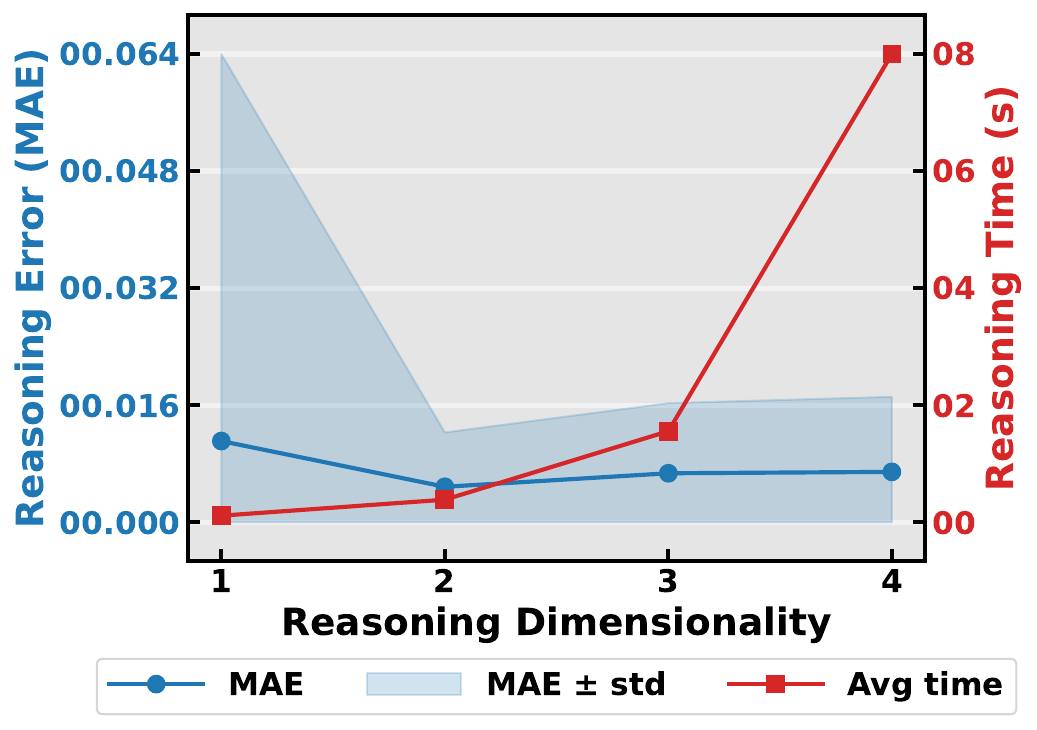}
    }
    \hfil
    \subfloat[scaling]{
        \includegraphics[width=0.48\columnwidth]{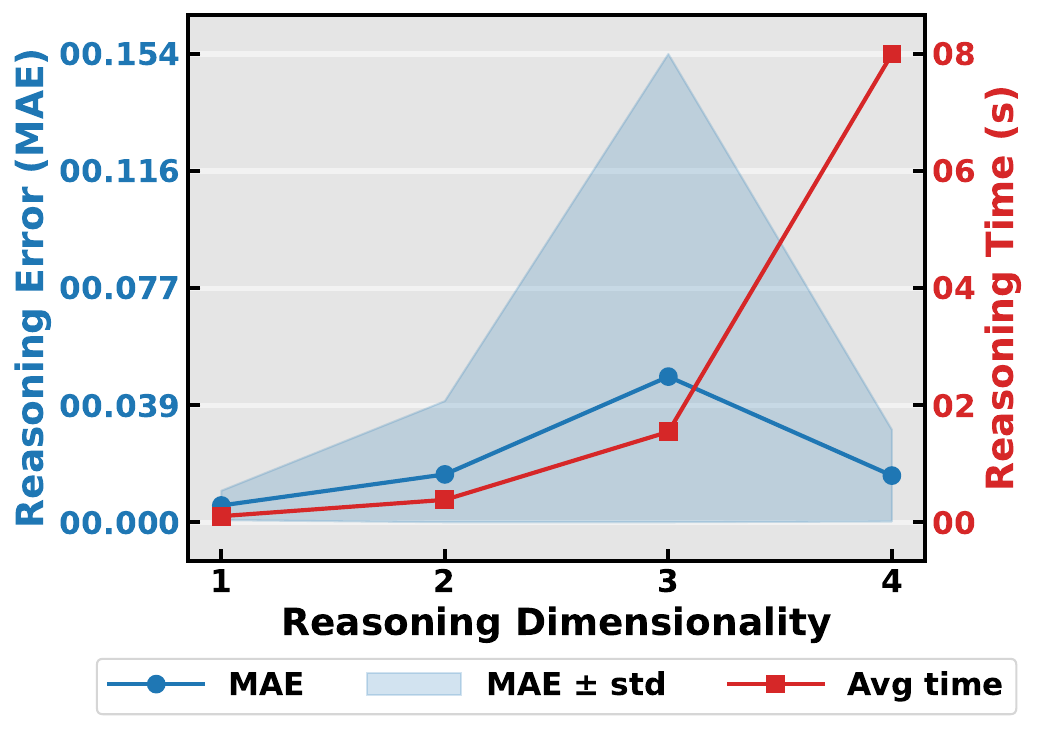}
    }
    \hfil
    \subfloat[shearing]{
        \includegraphics[width=0.48\columnwidth]{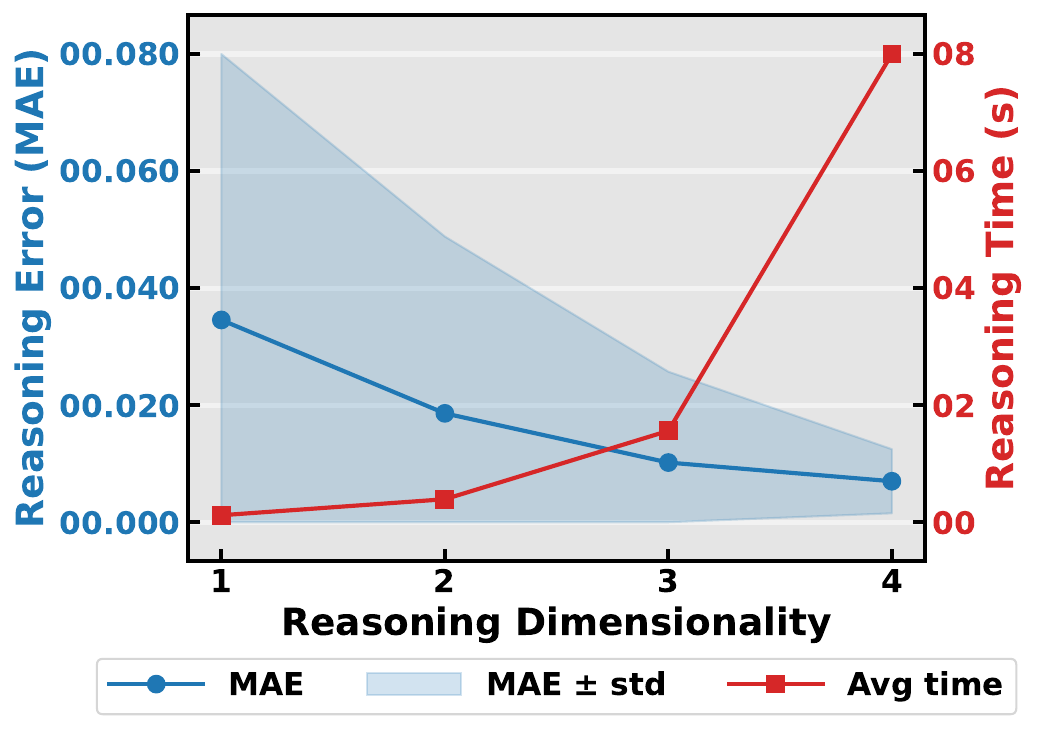}
    }
    \caption{Reasoning complexity analysis of how reasoning dimensionality affects reasoning time and reasoning accuracy.}
    \label{fig:exp_complexity}
\end{figure*}

An analysis of reasoning complexity is conducted to examine how the combinatorial nature of the parameter space affects both reasoning time and reasoning accuracy. For each photometric and geometric transformation, we apply four ground-truth parameter values and attempt to reason about between one and four types of transformation parameters. When reasoning about a single parameter type, it is aligned with the ground-truth transformation. When reasoning about two to four parameters, the additional types are randomly selected from the remaining types within the same family (photometric or geometric). Figure~\ref{fig:exp_complexity} reports the reasoning time and accuracy with respect to the number of parameters to be inferred. As expected, the reasoning time grows exponentially with the number of parameters considered. However, the accuracy does not exhibit a monotonic trend with increasing dimensionality. Under our experimental conditions, we observe no significant impact of reasoning dimensionality on accuracy.

\subsection{Synthetic Media Detection}
\begin{figure}[t!]
    \centering
    \subfloat[none]{
    	\includegraphics[width=0.9\columnwidth]{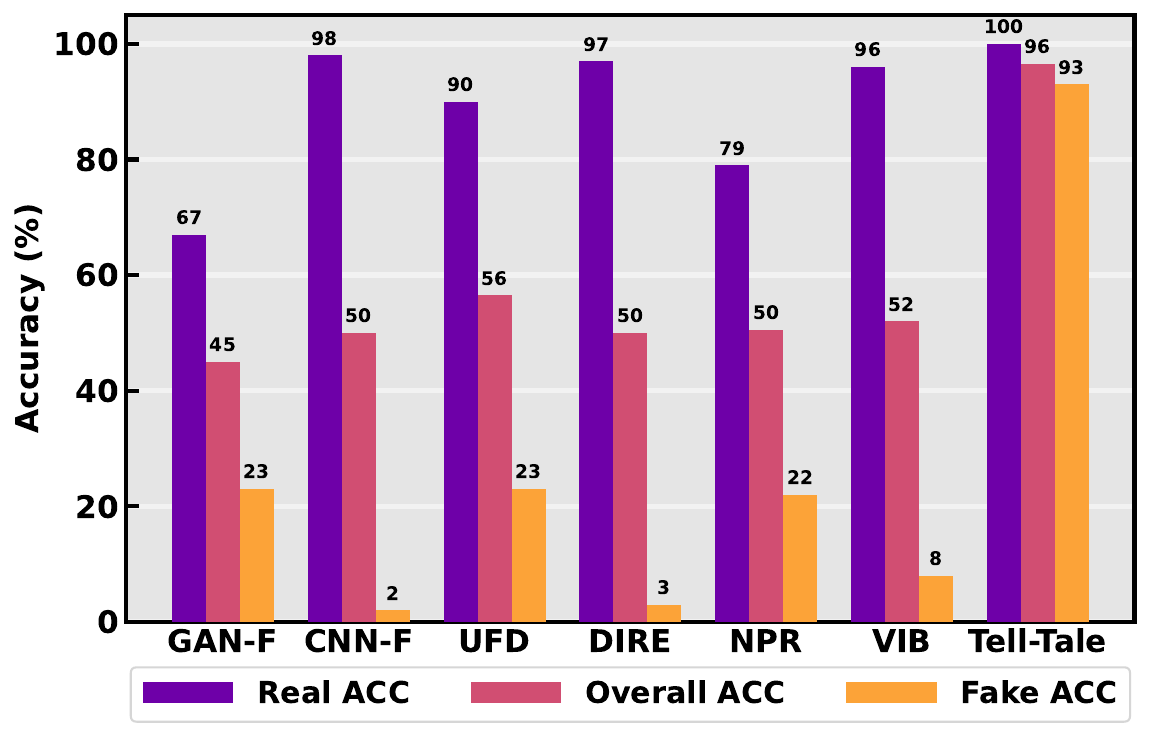}
    }
    \\
    \subfloat[brightness]{
    	\includegraphics[width=0.47\columnwidth]{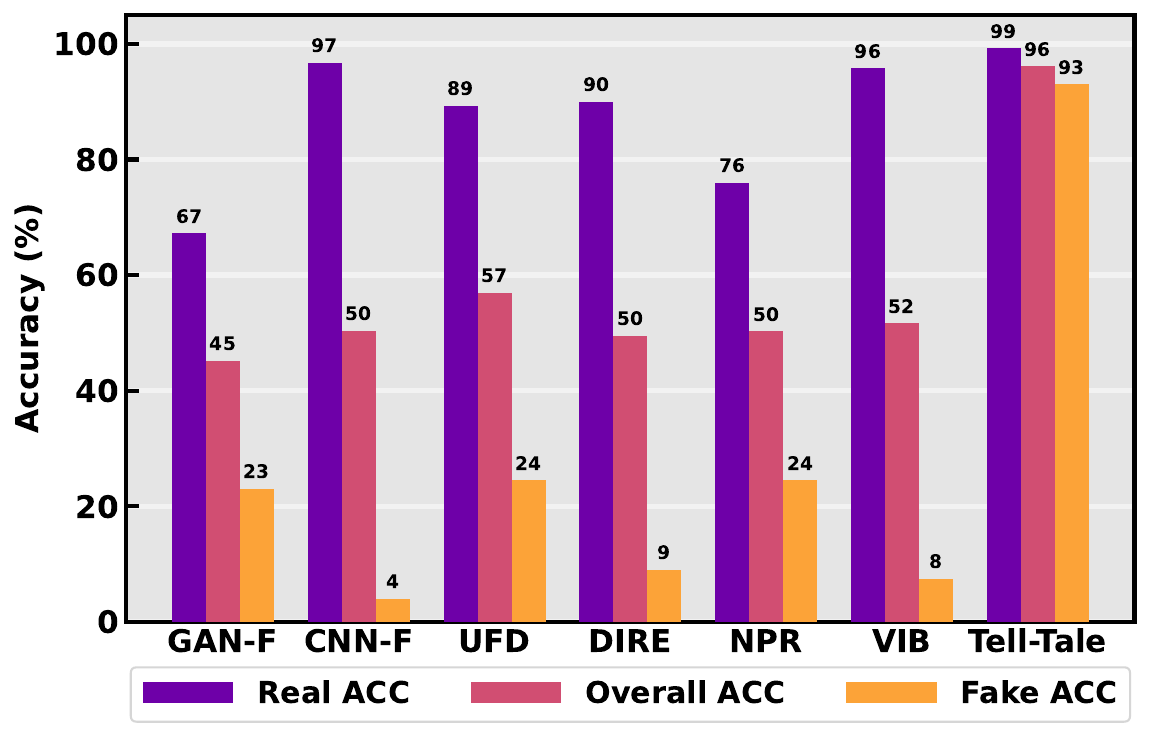}
    }
    \hfil
    \subfloat[contrast]{
        \includegraphics[width=0.47\columnwidth]{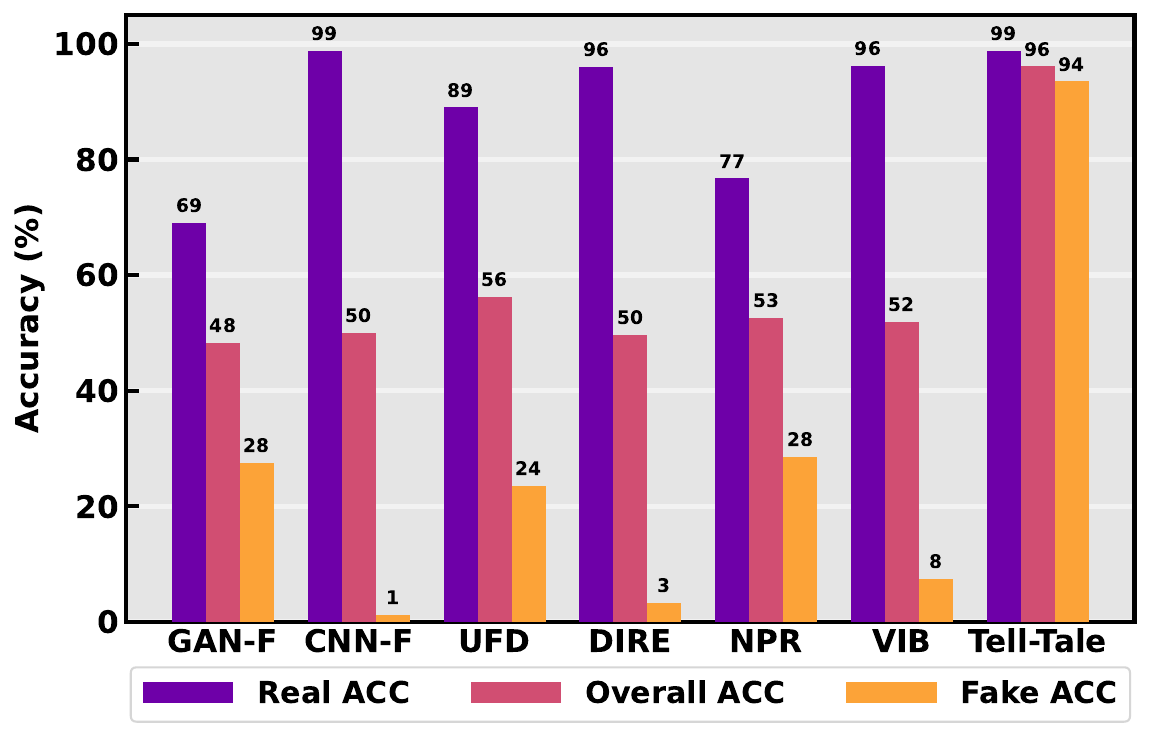}
    }
    \\
    \subfloat[hue]{
        \includegraphics[width=0.47\columnwidth]{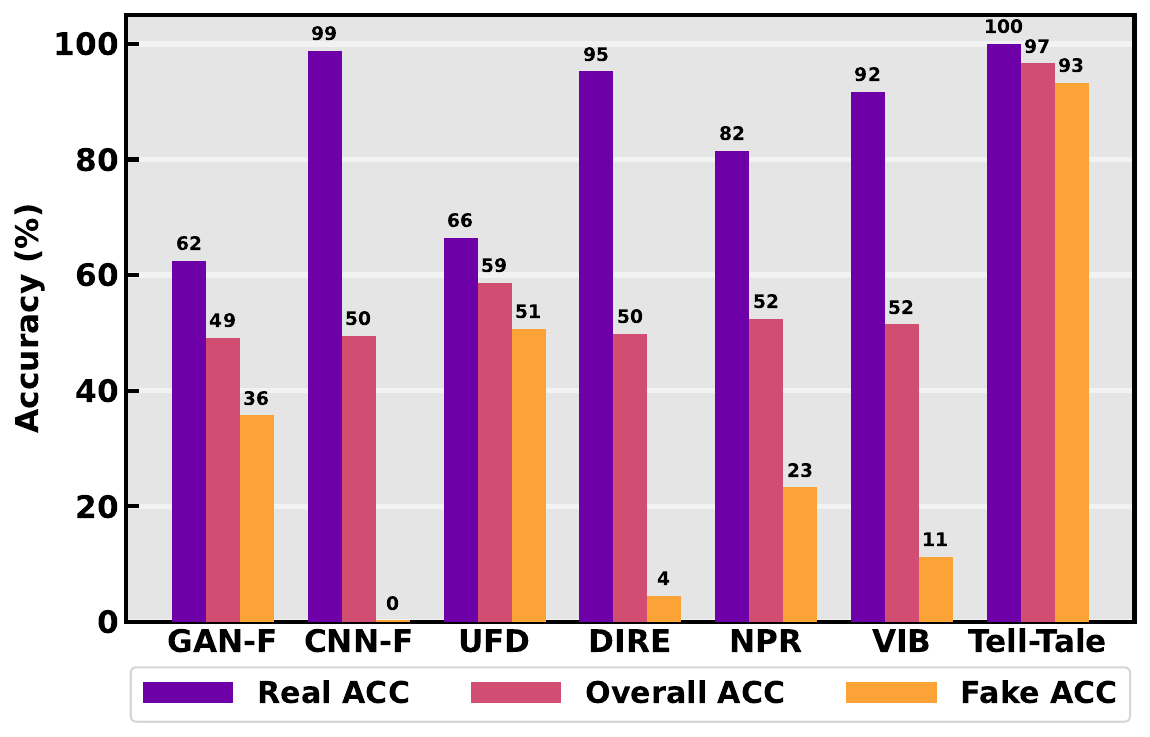}
    }
    \hfil
    \subfloat[saturation]{
        \includegraphics[width=0.47\columnwidth]{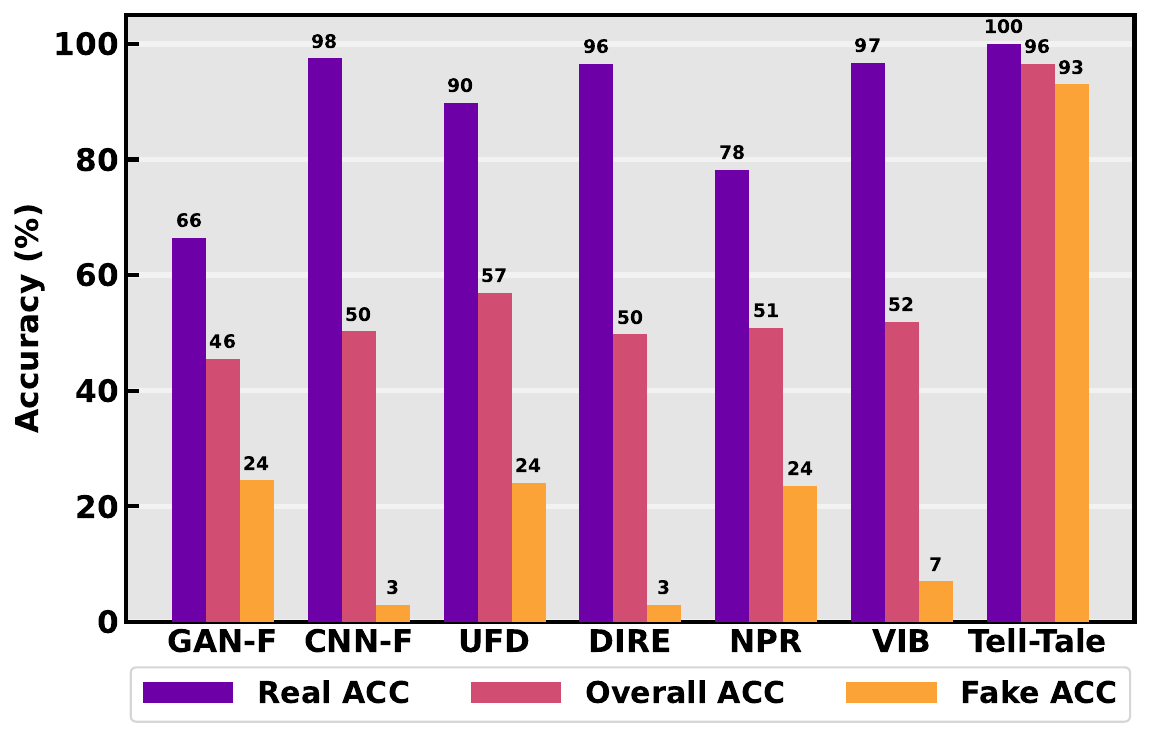}
    }
    \\
    \subfloat[rotation]{
        \includegraphics[width=0.47\columnwidth]{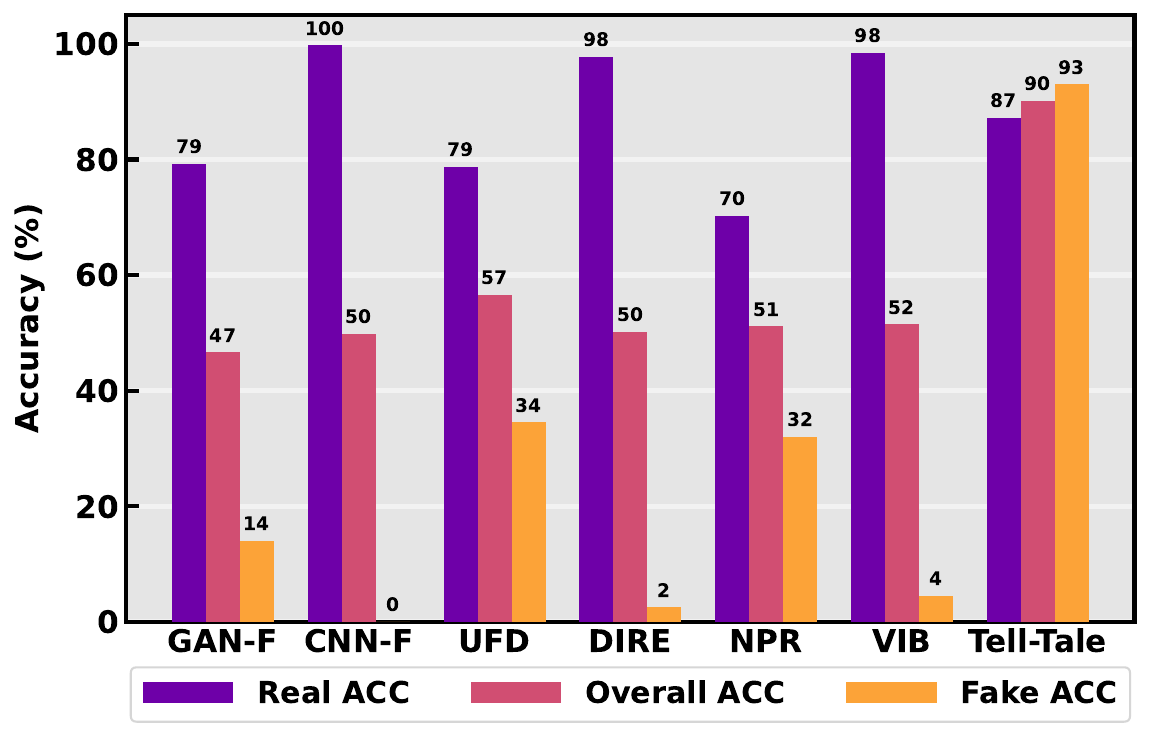}
    }
    \hfil
    \subfloat[translation]{
        \includegraphics[width=0.47\columnwidth]{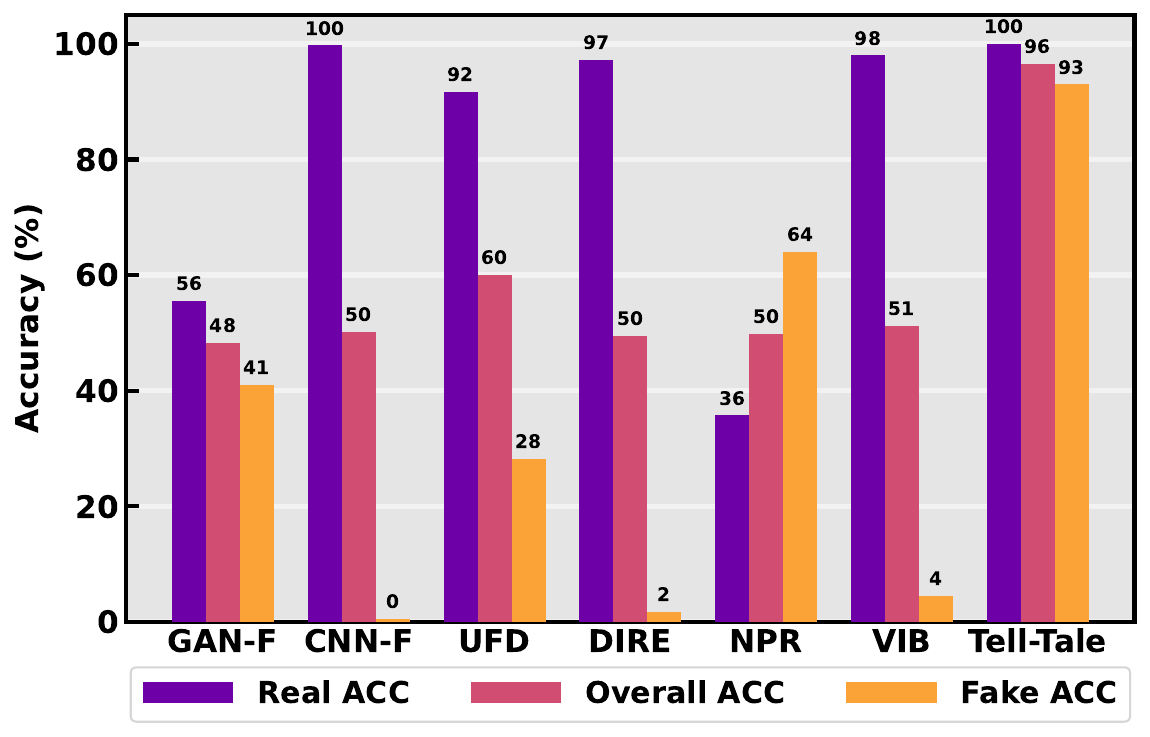}
    }
    \\
    \subfloat[scaling]{
        \includegraphics[width=0.47\columnwidth]{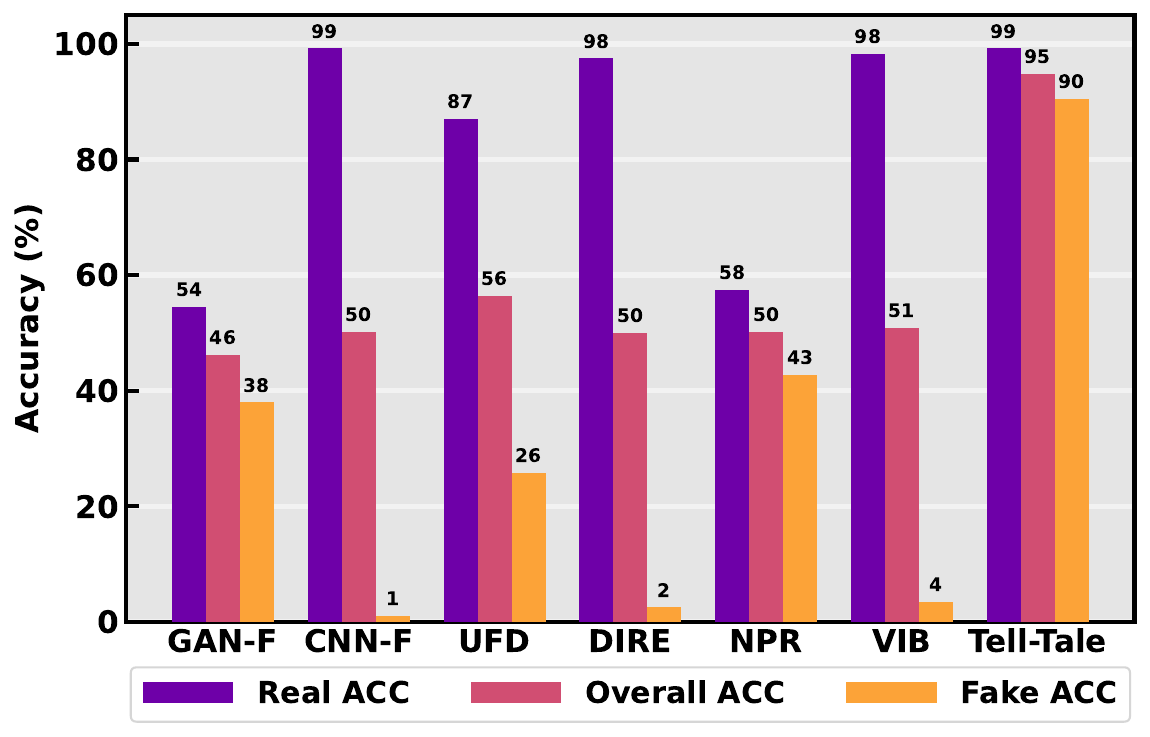}
    }
    \hfil
    \subfloat[shearing]{
        \includegraphics[width=0.47\columnwidth]{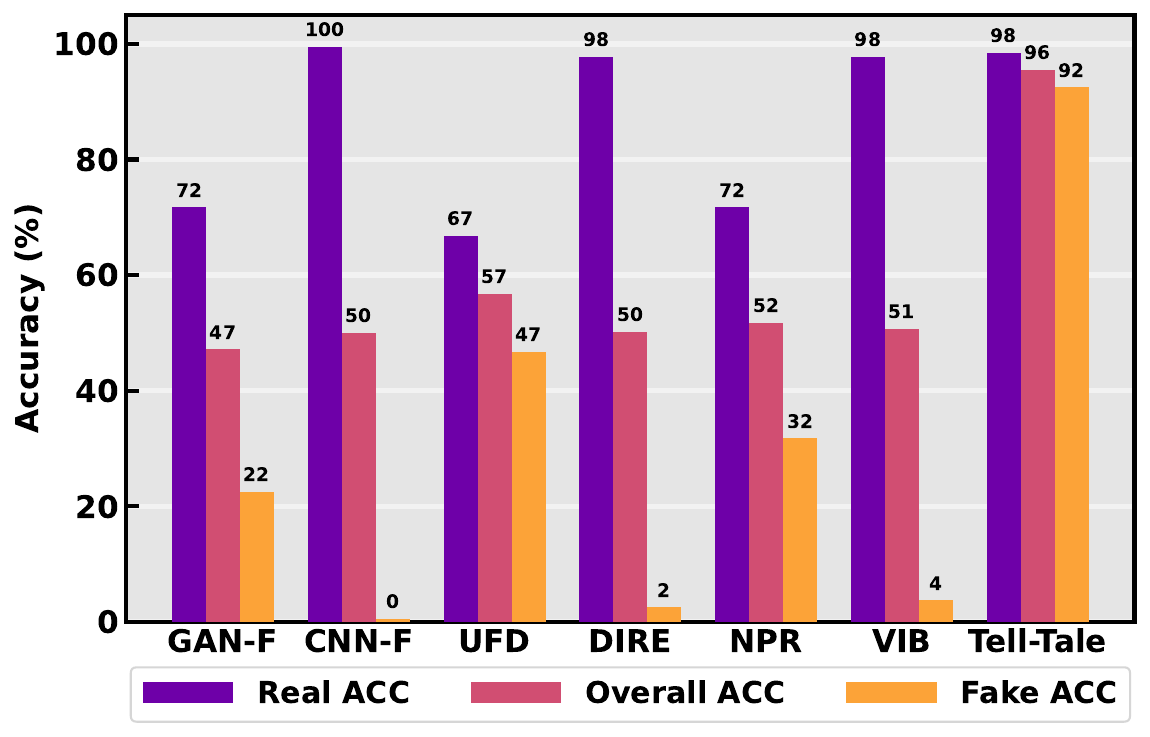}
    }
    \caption{Synthetic media detection accuracy in comparison with state-of-the-art baselines.}
    \label{fig:exp_detection}
\end{figure}

\begin{figure}[t!]
    \centering
    \includegraphics[width=0.72\columnwidth]{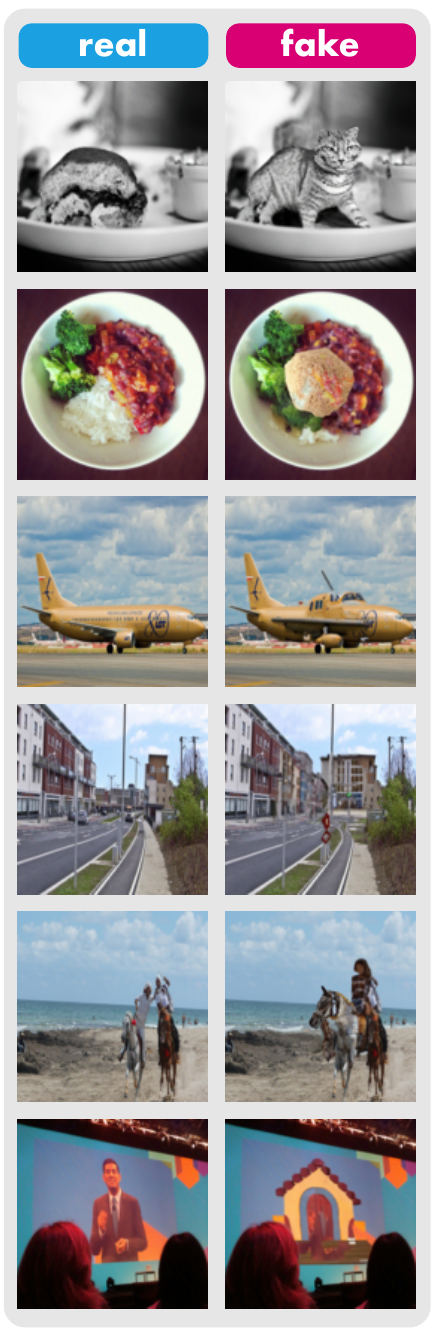}
    \caption{Examples of real and fake images where only part of the content is synthetic.}
    \label{fig:real_fake}
\end{figure}

A comparative study on synthetic media detection is conducted by benchmarking our system against 6 state-of-the-art baselines: generative adversarial network fingerprints (GAN-F)~\cite{9010964}, convolutional neural network fingerprints (CNN-F)~\cite{9156876}, universal fake detectors (UFD)~\cite{10204883}, diffusion reconstruction error (DIRE)~\cite{10377654}, neighbouring pixel relationships (NPR)~\cite{10658459} and variational information bottleneck (VIB)~\cite{11092328}. For our system, we adopt a simple decision rule to determine whether an image is synthetic. An image is deemed synthetic when the proportion of tampered area indicated by the semantic watermark exceeds an empirical threshold of 5\%. Figure~\ref{fig:exp_detection} summarises the classification accuracies on real, fake and overall samples under a range of photometric and geometric transformations. Each transformation type is evaluated using four parameter settings applied to 200 real and 200 fake images. Most baseline methods attain reasonable accuracy on real images but struggle to recognise fake ones, leading to overall accuracies of approximately 40 to 60\%. This reflects a fundamental limitation when applied to our dataset. Beyond the inherent issue of limited generalisability to unseen data, we attribute this failure in part to the partially fake nature of our synthetic media, which are not generated from scratch, but are produced by completing masked regions of real images, as exemplified in Figure~\ref{fig:real_fake}. In comparison, our system, albeit not flawless, achieves over 90\% overall accuracy across all experimental conditions. This performance gap is explained by the fundamental difference in defence assumptions. Our system constitutes a proactive defence that embeds watermarks prior to content synthesis. The baselines, by contrast, embody reactive detection, assuming no preventative mechanism is applied at creation time and relying solely on post hoc analysis of the observed image samples.

\section{Conclusion}\label{sec:con}
This study addressed the forensic challenge of reconstructing the latent transformation chain underlying synthetic media. Tell-tale watermarks are developed to remain synchronised with the transformation dynamics applied to their carrier media, permitting explanatory reasoning from the interpretable traces left behind. The reasoning process, however, is inherently constrained by the combinatorial explosion of possible transformation sequences, which limits scalability in unconstrained environments. The restriction to sequential order leaves open the broader problem of reconstructing the complete timeline of editing history. Future research may explore tell-tale watermark designs tailored or generalised to a wider range of transformations, thereby supporting traceability in the evolving landscape of synthetic media forensics.

\section*{Acknowledgements}
This work was supported in part by the Japan Society for the Promotion of Science (JSPS) under KAKENHI Grants (JP21H04907 and JP24H00732), and in part by the Japan Science and Technology Agency (JST) under CREST Grants (JPMJCR20D3 and JPMJCR2562) including AIP Challenge Program, AIP Acceleration Grant (JPMJCR24U3) and K Program Grant (JPMJKP24C2).

\bibliography{Transactions-Bibliography/bstcontrol, Bib/bib_telltale}

@inproceedings{11092328,
	address = {Nashville, TN, USA},
	author = {Zhang, Haifeng and He, Qinghui and Bi, Xiuli and Li, Weisheng and Liu, Bo and Xiao, Bin},
	booktitle = {Proc. {IEEE/CVF} Conf. Comput. Vis. Pattern Recognit. (CVPR)},
	pages = {23828--23837},
	title = {Towards Universal {AI}-Generated Image Detection by Variational Information Bottleneck Network},
	year = {2025}}

@inproceedings{10658459,
	address = {Seattle, WA, USA},
	author = {Tan, Chuangchuang and Liu, Huan and Zhao, Yao and Wei, Shikui and Gu, Guanghua and Liu, Ping and Wei, Yunchao},
	booktitle = {Proc. {IEEE/CVF} Conf. Comput. Vis. Pattern Recognit. (CVPR)},
	pages = {28130--28139},
	title = {Rethinking the Up-Sampling Operations in {CNN}-Based Generative Network for Generalizable Deepfake Detection},
	year = {2024}}

@inproceedings{10377654,
	address = {Paris, France},
	author = {Wang, Zhendong and Bao, Jianmin and Zhou, Wengang and Wang, Weilun and Hu, Hezhen and Chen, Hong and Li, Houqiang},
	booktitle = {Proc. {IEEE/CVF} Conf. Comput. Vis. Pattern Recognit. (CVPR)},
	pages = {22388--22398},
	title = {{DIRE} for Diffusion-Generated Image Detection},
	year = {2023}}

@inproceedings{10204883,
	address = {Vancouver, BC, Canada},
	author = {Ojha, Utkarsh and Li, Yuheng and Lee, Yong Jae},
	booktitle = {Proc. {IEEE/CVF} Conf. Comput. Vis. Pattern Recognit. (CVPR)},
	pages = {24480--24489},
	title = {Towards Universal Fake Image Detectors that Generalize Across Generative Models},
	year = {2023}}

@inproceedings{9156876,
	address = {Virtual Event},
	author = {Wang, Sheng-Yu and Wang, Oliver and Zhang, Richard and Owens, Andrew and Efros, Alexei A.},
	booktitle = {Proc. {IEEE/CVF} Conf. Comput. Vis. Pattern Recognit. (CVPR)},
	pages = {8692--8701},
	title = {{CNN}-Generated Images Are Surprisingly Easy to Spot{\ldots} for Now},
	year = {2020}}

@article{Chen_Xie_Lin_Liu_Wang_2025,
	author = {Chen, Geng and Xie, Wuyuan and Lin, Di and Liu, Ye and Wang, Miaohui},
	journal = {Proc. AAAI Conf. Artif. Intell. (AAAI)},
	number = {1},
	pages = {58--66},
	title = {{mmFAS}: {M}ultimodal Face Anti-Spoofing Using Multi-Level Alignment and Switch-Attention Fusion},
	volume = {39},
	year = {2025}}

@article{11128946,
	author = {Yan, Pei and Tan, Shunquan and Wang, Miaohui and Huang, Jiwu},
	journal = {IEEE Trans. Dependable Secure Comput.},
	number = {6},
	pages = {7712--7728},
	title = {Prompt Engineering-Assisted Malware Dynamic Analysis Using {GPT-4}},
	volume = {22},
	year = {2025}}

@article{Jin:2022ab,
	author = {Jin, Xiongnan and Lee, Yooyoung and Fiscus, Jonathan and Guan, Haiying and Yates, Amy N. and Delgado, Andrew and Zhou, Daniel F.},
	journal = {Neurocomputing},
	pages = {76--88},
	title = {{MFC-Prov}: {M}edia forensics challenge image provenance evaluation and data analysis on large-scale datasets},
	volume = {470},
	year = {2022}}

@article{9107445,
	author = {Zhang, Xu and Sun, Zhaohui H. and Karaman, Svebor and Chang, Shih-Fu},
	journal = {IEEE J. Sel. Top. Signal Process.},
	number = {5},
	pages = {1012--1023},
	title = {Discovering Image Manipulation History by Pairwise Relation and Forensics Tools},
	volume = {14},
	year = {2020}}

@article{Breitinger:2025aa,
	author = {Breitinger, Frank and Studiawan, Hudan and Hargreaves, Chris},
	journal = {Forensic Sci. Int. Digit. Investig.},
	pages = {1--15},
	title = {{SoK}: {T}imeline based event reconstruction for digital forensics: {T}erminology, methodology, and current challenges},
	volume = {53},
	year = {2025}}

@article{10.1145/1978802.1978805,
	author = {Rocha, Anderson and Scheirer, Walter and Boult, Terrance and Goldenstein, Siome},
	journal = {ACM Comput. Surv.},
	number = {4},
	pages = {1--42},
	title = {Vision of the unseen: {C}urrent trends and challenges in digital image and video forensics},
	volume = {43},
	year = {2011}}

@inproceedings{Lin:2014aa,
	address = {Zurich, Switzerland},
	author = {Lin, Tsung-Yi and Maire, Michael and Belongie, Serge and Hays, James and Perona, Pietro and Ramanan, Deva and Doll{\'a}r, Piotr and Zitnick, C. Lawrence},
	booktitle = {Proc. Eur. Conf. Comput. Vis. (ECCV)},
	pages = {740--755},
	title = {{Microsoft} {COCO}: {C}ommon Objects in Context},
	year = {2014}}

@inproceedings{10205460,
	address = {Vancouver, BC, Canada},
	author = {Guillaro, Fabrizio and Cozzolino, Davide and Sud, Avneesh and Dufour, Nicholas and Verdoliva, Luisa},
	booktitle = {Proc. {IEEE/CVF} Conf. Comput. Vis. Pattern Recognit. (CVPR)},
	pages = {20606--20615},
	title = {{TruFor}: {L}everaging All-Round Clues for Trustworthy Image Forgery Detection and Localization},
	year = {2023}}

@article{9786832,
	author = {Korus, Pawe{\l} and Memon, Nasir},
	journal = {IEEE Trans. Inf. Forensics Secur.},
	pages = {2508--2523},
	title = {Computational Sensor Fingerprints},
	volume = {17},
	year = {2022}}

@article{5487389,
	author = {Stamm, Matthew C. and Liu, K.J. Ray},
	journal = {IEEE Trans. Inf. Forensics Secur.},
	number = {3},
	pages = {492--506},
	title = {Forensic detection of image manipulation using statistical intrinsic fingerprints},
	volume = {5},
	year = {2010}}

@article{Nightingale:2017aa,
	author = {Nightingale, Sophie J. and Wade, Kimberley A. and Watson, Derrick G.},
	journal = {Cogn. Res. Princ. Implic.},
	number = {1},
	pages = {1--21},
	title = {Can people identify original and manipulated photos of real-world scenes?},
	volume = {2},
	year = {2017}}

@article{Harman:1965aa,
	author = {Harman, Gilbert H.},
	journal = {Philos. Rev.},
	number = {1},
	pages = {88--95},
	title = {The Inference to the Best Explanation},
	volume = {74},
	year = {1965}}

@article{Niiniluoto:2011aa,
	author = {Niiniluoto, Ilkka},
	journal = {Stud. Hist. Philos. Sci. A},
	number = {1},
	pages = {135--139},
	title = {Abduction, tomography, and other inverse problems},
	volume = {42},
	year = {2011}}

@article{Lazer:2018aa,
	author = {Lazer, David M. J. and Baum, Matthew A. and Benkler, Yochai and Berinsky, Adam J. and Greenhill, Kelly M. and Menczer, Filippo and Metzger, Miriam J. and Nyhan, Brendan and Pennycook, Gordon and Rothschild, David and Schudson, Michael and Sloman, Steven A. and Sunstein, Cass R. and Thorson, Emily A. and Watts, Duncan J. and Zittrain, Jonathan L.},
	journal = {Science},
	number = {6380},
	pages = {1094--1096},
	title = {The science of fake news},
	volume = {359},
	year = {2018}}

@article{4806202,
	author = {Farid, Hany},
	journal = {IEEE Signal Process. Mag.},
	number = {2},
	pages = {16--25},
	title = {Image forgery detection},
	volume = {26},
	year = {2009}}

@article{10.1145/1113034.1113074,
	author = {Richard, Golden G. and Roussev, Vassil},
	journal = {Commun. ACM},
	number = {2},
	pages = {76--80},
	title = {Next-generation digital forensics},
	volume = {49},
	year = {2006}}

@article{Del-Vicario:2016aa,
	author = {Del Vicario, Michela and Bessi, Alessandro and Zollo, Fabiana and Petroni, Fabio and Scala, Antonio and Caldarelli, Guido and Stanley, H. Eugene and Quattrociocchi, Walter},
	journal = {Proc. Natl. Acad. Sci. USA},
	number = {3},
	pages = {554--559},
	title = {The spreading of misinformation online},
	volume = {113},
	year = {2016}}

@article{Vosoughi:2018aa,
	author = {Vosoughi, Soroush and Roy, Deb and Aral, Sinan},
	journal = {Science},
	number = {6380},
	pages = {1146--1151},
	title = {The spread of true and false news online},
	volume = {359},
	year = {2018}}

@article{Lewandowsky:2017aa,
	author = {Lewandowsky, Stephan and Ecker, Ullrich K. H. and Cook, John},
	journal = {J. Appl. Res. Mem. Cogn.},
	number = {4},
	pages = {353--369},
	title = {Beyond Misinformation: {U}nderstanding and Coping with the ``Post-Truth'' Era},
	volume = {6},
	year = {2017}}

@article{4806203,
	author = {Fridrich, Jessica},
	journal = {IEEE Signal Process. Mag.},
	number = {2},
	pages = {26--37},
	title = {Digital image forensics},
	volume = {26},
	year = {2009}}

@article{Zhao:2024aa,
	author = {Zhao, Yuan and Liu, Bo and Zhu, Tianqing and Ding, Ming and Yu, Xin and Zhou, Wanlei},
	journal = {Neurocomputing},
	pages = {1--14},
	title = {Proactive image manipulation detection via deep semi-fragile watermark},
	volume = {585},
	year = {2024}}

@article{10.1145/3640466,
	author = {Neekhara, Paarth and Hussain, Shehzeen and Zhang, Xinqiao and Huang, Ke and McAuley, Julian and Koushanfar, Farinaz},
	journal = {ACM Trans. Multimedia Comput. Commun. Appl.},
	number = {11},
	pages = {1--21},
	title = {{FaceSigns}: {S}emi-fragile Watermarks for Media Authentication},
	volume = {20},
	year = {2024}}

@inproceedings{10030248,
	address = {Waikoloa, HI, USA},
	author = {Zhao, Yuan and Liu, Bo and Ding, Ming and Liu, Baoping and Zhu, Tianqing and Yu, Xin},
	booktitle = {Proc. IEEE/CVF Winter Conf. Appl. Comput. Vis. (WACV)},
	pages = {4591--4600},
	title = {Proactive Deepfake Defence via Identity Watermarking},
	year = {2023}}

@inproceedings{723413,
	address = {Chicago, IL, USA},
	author = {Min Wu and Bede Liu},
	booktitle = {Proc. Int. Conf. Image Process. (ICIP)},
	pages = {437--441},
	title = {Watermarking for image authentication},
	volume = {2},
	year = {1998}}

@inproceedings{638587,
	address = {Santa Barbara, CA, USA},
	author = {Yeung, M.M. and Mintzer, F.},
	booktitle = {Proc. Int. Conf. Image Process. (ICIP)},
	pages = {680--683},
	title = {An invisible watermarking technique for image verification},
	volume = {2},
	year = {1997}}

@inproceedings{nijkamp2023codegen,
	address = {Kigali, Rwanda},
	author = {Erik Nijkamp and Bo Pang and Hiroaki Hayashi and Lifu Tu and Huan Wang and Yingbo Zhou and Silvio Savarese and Caiming Xiong},
	booktitle = {Proc. Int. Conf. Learn. Represent. (ICLR)},
	pages = {1--25},
	title = {{CodeGen}: {A}n Open Large Language Model for Code with Multi-Turn Program Synthesis},
	year = {2023}}

@inproceedings{10.1145/3520312.3534862,
	address = {San Diego, CA, USA},
	author = {Xu, Frank F. and Alon, Uri and Neubig, Graham and Hellendoorn, Vincent Josua},
	booktitle = {Proc. ACM SIGPLAN Int. Symp. Mach. Program. (MAPS)},
	pages = {1--10},
	title = {A systematic evaluation of large language models of code},
	year = {2022}}

@article{Li:2022aa,
	author = {Li, Yujia and Choi, David and Chung, Junyoung and Kushman, Nate and Schrittwieser, Julian and Leblond, R{\'e}mi and Eccles, Tom and Keeling, James and Gimeno, Felix and Dal Lago, Agustin and Hubert, Thomas and Choy, Peter and de Masson d'Autume, Cyprien and Babuschkin, Igor and Chen, Xinyun and Huang, Po-Sen and Welbl, Johannes and Gowal, Sven and Cherepanov, Alexey and Molloy, James and Mankowitz, Daniel J. and Sutherland Robson, Esme and Kohli, Pushmeet and de Freitas, Nando and Kavukcuoglu, Koray and Vinyals, Oriol},
	journal = {Science},
	number = {6624},
	pages = {1092--1097},
	title = {Competition-level code generation with {AlphaCode}},
	volume = {378},
	year = {2022}}

@inproceedings{Woo:2018aa,
	address = {Munich, Germany},
	author = {Woo, Sanghyun and Park, Jongchan and Lee, Joon-Young and Kweon, In So},
	booktitle = {Proc. Eur. Conf. Comput. Vis. (ECCV)},
	pages = {3--19},
	title = {{CBAM}: {C}onvolutional Block Attention Module},
	year = {2018}}

@inproceedings{10.5555/3692070.3693829,
	address = {Vienna, Austria},
	author = {Roman, Robin San and Fernandez, Pierre and Elsahar, Hady and D\'{e}fossez, Alexandre and Furon, Teddy and Tran, Tuan},
	booktitle = {Proc. Int. Conf. Mach. Learn. (ICML)},
	pages = {43180--43196},
	title = {Proactive detection of voice cloning with localized watermarking},
	year = {2024}}

@article{Dathathri:2024aa,
	author = {Dathathri, Sumanth and See, Abigail and Ghaisas, Sumedh and Huang, Po-Sen and McAdam, Rob and Welbl, Johannes and Bachani, Vandana and Kaskasoli, Alex and Stanforth, Robert and Matejovicova, Tatiana and Hayes, Jamie and Vyas, Nidhi and Merey, Majd Al and Brown-Cohen, Jonah and Bunel, Rudy and Balle, Borja and Cemgil, Taylan and Ahmed, Zahra and Stacpoole, Kitty and Shumailov, Ilia and Baetu, Ciprian and Gowal, Sven and Hassabis, Demis and Kohli, Pushmeet},
	journal = {Nature},
	number = {8035},
	pages = {818--823},
	title = {Scalable watermarking for identifying large language model outputs},
	volume = {634},
	year = {2024}}

@inproceedings{10377226,
	address = {Paris, France},
	author = {Fernandez, Pierre and Couairon, Guillaume and J{\'e}gou, Herv{\'e} and Douze, Matthijs and Furon, Teddy},
	booktitle = {Proc. IEEE/CVF Int. Conf. Comput. Vis. (ICCV)},
	pages = {22409--22420},
	title = {The Stable Signature: Rooting Watermarks in Latent Diffusion Models},
	year = {2023}}

@article{267415,
	author = {Friedman, G.L.},
	journal = {IEEE Trans. Consum. Electron.},
	number = {4},
	pages = {905--910},
	title = {The trustworthy digital camera: {R}estoring credibility to the photographic image},
	volume = {39},
	year = {1993}}

@article{10.1145/3422622,
	author = {Goodfellow, Ian and Pouget-Abadie, Jean and Mirza, Mehdi and Xu, Bing and Warde-Farley, David and Ozair, Sherjil and Courville, Aaron and Bengio, Yoshua},
	journal = {Commun. ACM},
	number = {11},
	pages = {139--144},
	title = {Generative adversarial networks},
	volume = {63},
	year = {2020}}

@article{Chesney:2019ab,
	author = {Chesney, Robert and Citron, Danielle},
	journal = {Foreign Aff.},
	number = {1},
	pages = {147--155},
	title = {Deepfakes and the New Disinformation War},
	volume = {98},
	year = {2019}}

@inproceedings{pmlr-v202-mitchell23a,
	address = {Honolulu, HI, USA},
	author = {Mitchell, Eric and Lee, Yoonho and Khazatsky, Alexander and Manning, Christopher D and Finn, Chelsea},
	booktitle = {Proc. Int. Conf. Mach. Learn. (ICML)},
	pages = {24950--24962},
	title = {{DetectGPT}: {Z}ero-Shot Machine-Generated Text Detection using Probability Curvature},
	volume = {202},
	year = {2023}}

@inproceedings{NEURIPS2019_3e9f0fc9,
	address = {Vancouver, BC, Canada},
	author = {Zellers, Rowan and Holtzman, Ari and Rashkin, Hannah and Bisk, Yonatan and Farhadi, Ali and Roesner, Franziska and Choi, Yejin},
	booktitle = {Proc. Int. Conf. Neural Inf. Process. Syst. (NeurIPS)},
	pages = {9054--9065},
	title = {Defending Against Neural Fake News},
	volume = {32},
	year = {2019}}

@article{9141516,
	author = {Ciftci, Umur Aybars and Demir, Ilke and Yin, Lijun},
	journal = {IEEE Trans. Pattern Anal. Mach. Intell.},
	pages = {1-1},
	title = {{FakeCatcher}: {D}etection of Synthetic Portrait Videos using Biological Signals},
	year = {2020}}

@inproceedings{pmlr-v202-kirchenbauer23a,
	address = {Honolulu, HI, USA},
	author = {Kirchenbauer, John and Geiping, Jonas and Wen, Yuxin and Katz, Jonathan and Miers, Ian and Goldstein, Tom},
	booktitle = {Proc. Int. Conf. Mach. Learn. (ICML)},
	pages = {17061--17084},
	title = {A Watermark for Large Language Models},
	volume = {202},
	year = {2023}}

@inproceedings{NEURIPS2022_ec795aea,
	address = {New Orleans, LA, USA},
	author = {Saharia, Chitwan and Chan, William and Saxena, Saurabh and Li, Lala and Whang, Jay and Denton, Emily L and Ghasemipour, Kamyar and Gontijo Lopes, Raphael and Karagol Ayan, Burcu and Salimans, Tim and Ho, Jonathan and Fleet, David J and Norouzi, Mohammad},
	booktitle = {Proc. Int. Conf. Neural Inf. Process. Syst. (NeurIPS)},
	pages = {36479--36494},
	title = {Photorealistic Text-to-Image Diffusion Models with Deep Language Understanding},
	volume = {35},
	year = {2022}}

@article{10238689,
	author = {Chang, Ching-Chun and Nguyen, Huy H. and Yamagishi, Junichi and Echizen, Isao},
	journal = {IEEE Access},
	pages = {105027--105039},
	title = {Cyber Vaccine for Deepfake Immunity},
	volume = {11},
	year = {2023}}

@inproceedings{Luo:2020aa,
	address = {Seattle, WA, USA},
	author = {X. Luo and R. Zhan and H. Chang and F. Yang and P. Milanfar},
	booktitle = {Proc. {IEEE/CVF} Conf. Comput. Vis. Pattern Recognit. (CVPR)},
	pages = {13545--13554},
	title = {Distortion Agnostic Deep Watermarking},
	year = {2020}}

@article{771072,
	author = {Voyatzis, G. and Pitas, I.},
	journal = {Proc. IEEE},
	number = {7},
	pages = {1197--1207},
	title = {The use of watermarks in the protection of digital multimedia products},
	volume = {87},
	year = {1999}}

@article{951543,
	author = {P. W. Wong and N. Memon},
	journal = {IEEE Trans. Image Process.},
	number = {10},
	pages = {1593--1601},
	title = {Secret and public key image watermarking schemes for image authentication and ownership verification},
	volume = {10},
	year = {2001}}

@article{771068,
	author = {I. J. Cox and M. L. Miller and A. L. McKellips},
	journal = {Proc. IEEE},
	number = {7},
	pages = {1127--1141},
	title = {Watermarking as communications with side information},
	volume = {87},
	year = {1999}}

@article{687830,
	author = {M. D. Swanson and M. Kobayashi and A. H. Tewfik},
	journal = {Proc. IEEE},
	number = {6},
	pages = {1064--1087},
	title = {Multimedia data-embedding and watermarking technologies},
	volume = {86},
	year = {1998}}

@article{650120,
	author = {I. J. Cox and J. Kilian and F. T. Leighton and T. Shamoon},
	journal = {IEEE Trans. Image Process.},
	number = {12},
	pages = {1673--1687},
	title = {Secure spread spectrum watermarking for multimedia},
	volume = {6},
	year = {1997}}

@article{985560,
	author = {Ruizhen Liu and Tieniu Tan},
	journal = {IEEE Trans. Multimed.},
	number = {1},
	pages = {121--128},
	title = {An {SVD}-based watermarking scheme for protecting rightful ownership},
	volume = {4},
	year = {2002}}

@article{771066,
	author = {F. Hartung and M. Kutter},
	journal = {Proc. IEEE},
	number = {7},
	pages = {1079--1107},
	title = {Multimedia watermarking techniques},
	volume = {87},
	year = {1999}}

@article{BARNI1998357,
	author = {Mauro Barni and Franco Bartolini and Vito Cappellini and Alessandro Piva},
	journal = {Signal Process.},
	number = {3},
	pages = {357--372},
	title = {A {DCT}-domain system for robust image watermarking},
	volume = {66},
	year = {1998}}

@article{771065,
	author = {Petitcolas, F.A.P. and Anderson, R.J. and Kuhn, M.G.},
	journal = {Proc. IEEE},
	number = {7},
	pages = {1062--1078},
	title = {Information hiding{\textemdash}{A} survey},
	volume = {87},
	year = {1999}}

@inproceedings{9880195,
	address = {New Orleans, LA, USA},
	author = {Shiohara, Kaede and Yamasaki, Toshihiko},
	booktitle = {Proc. IEEE/CVF Conf. Comput. Vis. Pattern Recognit. (CVPR)},
	pages = {18699-18708},
	title = {Detecting Deepfakes with Self-Blended Images},
	year = {2022}}

@inproceedings{9578592,
	address = {Nashville, TN, USA},
	author = {Zhou, Tianfei and Wang, Wenguan and Liang, Zhiyuan and Shen, Jianbing},
	booktitle = {Proc. IEEE/CVF Conf. Comput. Vis. Pattern Recognit. (CVPR)},
	pages = {5774--5784},
	title = {Face Forensics in the Wild},
	year = {2021}}

@inproceedings{9578910,
	address = {Nashville, TN, USA},
	author = {Haliassos, Alexandros and Vougioukas, Konstantinos and Petridis, Stavros and Pantic, Maja},
	booktitle = {Proc. IEEE/CVF Conf. Comput. Vis. Pattern Recognit. (CVPR)},
	pages = {5037--5047},
	title = {Lips Don't Lie: {A} Generalisable and Robust Approach to Face Forgery Detection},
	year = {2021}}

@inproceedings{8683164,
	address = {Brighton, UK},
	author = {Yang, Xin and Li, Yuezun and Lyu, Siwei},
	booktitle = {Proc. IEEE Int. Conf. Acoust. Speech Signal Process. (ICASSP)},
	pages = {8261--8265},
	title = {Exposing Deep Fakes Using Inconsistent Head Poses},
	year = {2019}}

@inproceedings{9010964,
	address = {Seoul, Korea},
	author = {Yu, Ning and Davis, Larry and Fritz, Mario},
	booktitle = {Proc. IEEE/CVF Int. Conf. Comput. Vis. (ICCV)},
	pages = {7555--7565},
	title = {Attributing Fake Images to {GANs}: {L}earning and Analyzing {GAN} Fingerprints},
	year = {2019}}

@inproceedings{10.1007/978-3-030-01261-8_41,
	address = {Munich, Germany},
	author = {Wiles, Olivia and Koepke, A. Sophia and Zisserman, Andrew},
	booktitle = {Proc. Eur. Conf. Comput. Vis. (ECCV)},
	pages = {690--706},
	title = {{X2Face}: {A} Network for Controlling Face Generation Using Images, Audio, and Pose Codes},
	year = {2018}}

@inproceedings{7780459,
	address = {Las Vegas, NV, USA},
	author = {He, Kaiming and Zhang, Xiangyu and Ren, Shaoqing and Sun, Jian},
	booktitle = {Proc. IEEE Conf. Comput. Vis. Pattern Recognit. (CVPR)},
	pages = {770--778},
	title = {Deep Residual Learning for Image Recognition},
	year = {2016}}

@inproceedings{9878449,
	address = {New Orleans, LA, USA},
	author = {Rombach, Robin and Blattmann, Andreas and Lorenz, Dominik and Esser, Patrick and Ommer, Bj{\"o}rn},
	booktitle = {Proc. IEEE/CVF Conf. Comput. Vis. Pattern Recognit. (CVPR)},
	pages = {10674--10685},
	title = {High-Resolution Image Synthesis with Latent Diffusion Models},
	year = {2022}}

@inproceedings{9010912,
	address = {Seoul, Korea},
	author = {R{\"o}ssler, Andreas and Cozzolino, Davide and Verdoliva, Luisa and Riess, Christian and Thies, Justus and Niessner, Matthias},
	booktitle = {Proc. IEEE/CVF Int. Conf. Comput. Vis. (ICCV)},
	pages = {1--11},
	title = {{FaceForensics++}: {L}earning to Detect Manipulated Facial Images},
	year = {2019}}

@inproceedings{Ronneberger:2015aa,
	address = {Cham},
	author = {Ronneberger, Olaf and Fischer, Philipp and Brox, Thomas},
	booktitle = {Proc. Int. Conf. Med. Image Comput. Comput.-Assist. Interv. (MICCAI)},
	pages = {234--241},
	title = {{U-Net}: {C}onvolutional Networks for Biomedical Image Segmentation},
	year = {2015}}

@inproceedings{Brock:2017aa,
	address = {Toulon, France},
	author = {Brock, Andrew and Lim, Theodore and Ritchie, James M. and Weston, Nick},
	booktitle = {Proc. Int. Conf. Learn. Represent. (ICLR)},
	pages = {1--15},
	title = {Neural Photo Editing with Introspective Adversarial Networks},
	year = {2017}}

@article{10.1145/3425780,
	author = {Mirsky, Yisroel and Lee, Wenke},
	journal = {ACM Comput. Surv.},
	number = {1},
	pages = {1--41},
	title = {The Creation and Detection of Deepfakes: A Survey},
	volume = {54},
	year = {2021}}

@article{10.1145/3292039,
	author = {Thies, Justus and Zollh\"{o}fer, Michael and Stamminger, Marc and Theobalt, Christian and Nie\ss{}ner, Matthias},
	journal = {Commun. ACM},
	number = {1},
	pages = {96--104},
	title = {{Face2Face}: {R}eal-Time Face Capture and Reenactment of {RGB} Videos},
	volume = {62},
	year = {2018}}

@article{1276112,
	author = {Zhu, B.B. and Swanson, M.D. and Tewfik, A.H.},
	journal = {IEEE Signal Process. Mag.},
	number = {2},
	pages = {40--49},
	title = {When seeing isn't believing [multimedia authentication technologies]},
	volume = {21},
	year = {2004}}

@article{771070,
	author = {Kundur, D. and Hatzinakos, D.},
	journal = {Proc. IEEE},
	number = {7},
	pages = {1167--1180},
	title = {Digital watermarking for telltale tamper proofing and authentication},
	volume = {87},
	year = {1999}}

@inproceedings{723401,
	address = {Chicago, IL, USA},
	author = {Fridrich, J.},
	booktitle = {Proc. IEEE Int. Conf. Image Process. (ICIP)},
	pages = {404--408},
	title = {Image watermarking for tamper detection},
	volume = {2},
	year = {1998}}

@inproceedings{8578214,
	address = {Salt Lake City, UT, USA},
	author = {Zhou, Peng and Han, Xintong and Morariu, Vlad I. and Davis, Larry S.},
	booktitle = {Proc. IEEE/CVF Conf. Comput. Vis. Pattern Recognit. (CVPR)},
	pages = {1053--1061},
	title = {Learning Rich Features for Image Manipulation Detection},
	year = {2018}}

@inproceedings{8695364,
	address = {San Jose, CA, USA},
	author = {Marra, Francesco and Gragnaniello, Diego and Verdoliva, Luisa and Poggi, Giovanni},
	booktitle = {Proc. IEEE Conf. Multimedia Inf. Process. Retr. (MIPR)},
	pages = {506--511},
	title = {Do {GANs} Leave Artificial Fingerprints?},
	year = {2019}}

@inproceedings{8630787,
	address = {Hong Kong, China},
	author = {Li, Yuezun and Chang, Ming-Ching and Lyu, Siwei},
	booktitle = {Proc. IEEE Int. Workshop Inf. Forensics Secur. (WIFS)},
	pages = {1--7},
	title = {In Ictu Oculi: {E}xposing {AI} Created Fake Videos by Detecting Eye Blinking},
	year = {2018}}

@article{10.1145/3371409,
	author = {Greengard, Samuel},
	journal = {Commun. ACM},
	number = {1},
	pages = {17--19},
	title = {Will Deepfakes Do Deep Damage?},
	volume = {63},
	year = {2019}}

@inproceedings{9157215,
	address = {Seattle, WA, USA},
	author = {Li, Lingzhi and Bao, Jianmin and Zhang, Ting and Yang, Hao and Chen, Dong and Wen, Fang and Guo, Baining},
	booktitle = {Proc. IEEE/CVF Conf. Comput. Vis. Pattern Recognit. (CVPR)},
	pages = {5000--5009},
	title = {Face {X}-Ray for More General Face Forgery Detection},
	year = {2020}}

@article{10.1145/3333002,
	articleno = {139},
	author = {Liu, Lingjie and Xu, Weipeng and Zollh\"{o}fer, Michael and Kim, Hyeongwoo and Bernard, Florian and Habermann, Marc and Wang, Wenping and Theobalt, Christian},
	journal = {ACM Trans. Graph.},
	note = {{{A}rt. no. 139}},
	number = {5},
	pages = {1--14},
	title = {Neural Rendering and Reenactment of Human Actor Videos},
	volume = {38},
	year = {2019}}

@inproceedings{10.1007/978-3-030-58517-4_42,
	address = {Glasgow, UK},
	author = {Thies, Justus and Elgharib, Mohamed and Tewari, Ayush and Theobalt, Christian and Nie\ss{}ner, Matthias},
	booktitle = {Proc. Eur. Conf. Comput. Vis. (ECCV)},
	pages = {716--731},
	title = {Neural Voice Puppetry: {A}udio-Driven Facial Reenactment},
	year = {2020}}

@article{10.1145/3197517.3201283,
	author = {Kim, Hyeongwoo and Garrido, Pablo and Tewari, Ayush and Xu, Weipeng and Thies, Justus and Niessner, Matthias and P\'{e}rez, Patrick and Richardt, Christian and Zollh\"{o}fer, Michael and Theobalt, Christian},
	journal = {ACM Trans. Graph.},
	number = {4},
	pages = {1--14},
	title = {Deep Video Portraits},
	volume = {37},
	year = {2018}}

@article{10.1145/2816795.2818056,
	author = {Thies, Justus and Zollh\"{o}fer, Michael and Nie\ss{}ner, Matthias and Valgaerts, Levi and Stamminger, Marc and Theobalt, Christian},
	journal = {ACM Trans. Graph.},
	number = {6},
	pages = {1--14},
	title = {Real-Time Expression Transfer for Facial Reenactment},
	volume = {34},
	year = {2015}}

@IEEEtranBSTCTL{IEEEexample:BSTcontrol,
  CTLuse_forced_etal = "yes",
  CTLmax_names_forced_etal = "6",
  CTLnames_show_etal = "1",
  CTLuse_url = "no",
  CTLdash_repeated_names  = "no"
}
\bibliographystyle{Transactions-Bibliography/IEEEtran}

\vspace{-3em}
\begin{IEEEbiography}[{\includegraphics[width=1in,height=1.25in,clip,keepaspectratio]{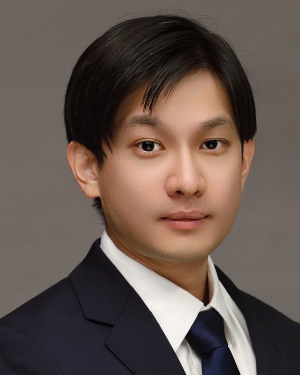}}]{Ching-Chun Chang} received the PhD in Computer Science from the University of Warwick, UK, in 2019. He is currently affiliated with the National Institute of Informatics, Japan, as a Project Assistant Professor. He also serves as a Visiting Researcher at Peking University, China, and a Distinguished Professor at Hangzhou Dianzi University, China. He participated in the Short-Term Scientific Mission supported by European Cooperation in Science and Technology Actions at the Faculty of Computer Science, Otto von Guericke University of Magdeburg, Germany, in 2016. He was granted the Marie-Curie Fellowship and participated in the Research and Innovation Staff Exchange supported by Marie Skłodowska-Curie Actions at the Department of Electrical and Computer Engineering, New Jersey Institute of Technology, USA, in 2017. He was a Visiting Scholar at the School of Computing and Mathematics, Charles Sturt University, Australia, in 2018, and at the School of Information Technology, Deakin University, Australia, in 2019. He was a Research Fellow at the Department of Electronic Engineering, Tsinghua University, China, in 2020. His research interests include artificial intelligence, biometrics, cryptography, cybersecurity, evolutionary computation, forensics, information theory, steganography, and watermarking.
\end{IEEEbiography}

\vspace{-3em}

\begin{IEEEbiography}[{\includegraphics[width=1in,height=1.25in,clip,keepaspectratio]{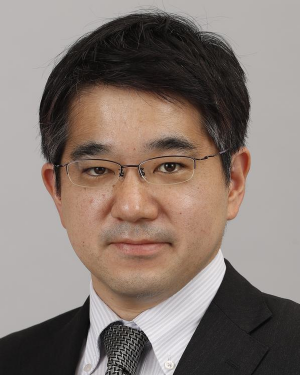}}]{Isao Echizen} received BS, MS, and DE degrees from the Tokyo Institute of Technology, Japan, in 1995, 1997 and 2003, respectively. He joined Hitachi, Ltd. in 1997 and until 2007 was a Research Engineer in the company's systems development laboratory. He is currently a Director and Professor of the Information and Society Research Division, as well as a Director of the Global Research Center for Synthetic Media, at the National Institute of Informatics; a Professor in the Department of Information and Communication Engineering, Graduate School of Information Science and Technology, the University of Tokyo; and a Professor in the Graduate University for Advanced Studies (SOKENDAI), Japan. He was a Visiting Professor at the Tsuda University, Japan; at the University of Freiburg, Germany; and at the University of Halle-Wittenberg, Germany. He is currently engaged in research on AI security, multimedia security and multimedia forensics, serving as a Research Director for the CREST FakeMedia project and the K Program SYNTHETIQ X project of the Japan Science and Technology Agency (JST). He received the Commendation for Science and Technology by the Minister of Education, Culture, Sports, Science and Technology (Research Category) in 2025. He also received the IEICE Best Paper Award in 2023; the IPSJ Best Paper Awards in 2005 and 2014; the IPSJ Nagao Special Researcher Award in 2011; the DOCOMO Mobile Science Award in 2014; the IISEC Information Security Cultural Award in 2016; and the IEEE WIFS Best Paper Award in 2017. He is an IEICE Fellow, an IPSJ Fellow, an IEEE Senior Member, an IFIP Japanese Representative, and an APSIPA Vice President.
\end{IEEEbiography}

\end{document}